\documentclass[%
reprint,
superscriptaddress,
 amsmath,amssymb,
 aps,
]{revtex4-2}

\usepackage{graphicx}
\usepackage{dcolumn}
\usepackage{bm}
\usepackage{braket}
\usepackage[dvipsnames]{xcolor}
\usepackage{hyperref}
\hypersetup{
    colorlinks=true,
    linkcolor=blue,
    citecolor=blue,
    filecolor=blue,
    urlcolor=blue,
    pdftitle={Your Document Title},
    pdfauthor={Your Name},
    pdfsubject={},
    pdfkeywords={},
}
\usepackage[normalem]{ulem}

\newcommand{\blue}[1]{\textcolor{black}{#1}}

\newcommand{\red}[1]{\textcolor{black}{#1}}
\newcommand{\green}[1]{\textcolor{black}{#1}}

\begin{document}
%

\title{\blue{Crossover in the Ordered Phase in the Non-Mermin-Wagner-Hohenberg Regime of Spin Models with Long-Range Coupling }}

\author{Jiewei Ding}
\affiliation{
Department of Physics, City University of Hong Kong, Kowloon, Hong Kong
}%
\author{Jiahao Su}
\affiliation{%
  School of Science, Harbin Institute of Technology, Shenzhen, 518055, China
}%
\affiliation{Shenzhen Key Laboratory of Advanced Functional Carbon Materials Research and Comprehensive Application, Shenzhen 518055, China.}
\author{Ho-Kin Tang}
\email{denghaojian@hit.edu.cn}
\affiliation{%
  School of Science, Harbin Institute of Technology, Shenzhen, 518055, China
}%
\affiliation{Shenzhen Key Laboratory of Advanced Functional Carbon Materials Research and Comprehensive Application, Shenzhen 518055, China.}
\author{Wing Chi Yu}
\email{wingcyu@cityu.edu.hk}
\affiliation{
Department of Physics, City University of Hong Kong, Kowloon, Hong Kong
}
%

\date{\today}

\begin{abstract}
Continuous spin models with long-range interactions \blue{of the form \( r^{-\sigma} \), where \( r \) is the distance between two spins and \( \sigma \) controls the decay of the interaction,} exhibit enhanced order that competes with thermal disturbances, leading to a richer variety of phases and types of phase transitions. In-depth research in this area not only aids in comprehending the complex behaviors in theoretical models but also provides valuable insights into the diverse phase transitions observed in real materials. Here, we identify that the true long-range ordered phase encompasses distinct scaling regimes, which we term Enhance  Long-Range Ordered (EnLRO) and Reduce Long-Range Ordered (ReLRO) regimes. In the one-dimensional XY model, the crossover from EnLRO to ReLRO regimes occurs around \(\sigma \approx 1.575\), while in two dimensions, the crossover happens near \(\sigma \approx 3.2\). Applying finite-size scaling analysis, we extract the critical exponents that characterize the \blue{order-to-disorder phase transitions in the} EnLRO and ReLRO regimes, constructing comprehensive phase diagrams. The analysis is further extended to the 1D and 2D long-range Heisenberg models, where we find the EnLRO-ReLRO crossover at \(\sigma \approx 1.575\) and \(\sigma \approx 3.22\), respectively. The similar crossover points suggest that the distinction between EnLRO and ReLRO regime is a generic feature in continuous spin models with long-range interactions.
\end{abstract}

\maketitle


\section{Introduction}
Classical spin models such as the Ising, Potts, and XY models have been extensively studied over the past three decades, yielding significant insights. For example, the Potts model has successfully explained alloy behavior and cell sorting \cite{radhakrishnan1998modeling,adam20183d,graner1992simulation,allena2016cellular,rozikov2021gibbs}, and the XY model has been effective in describing systems like superfluids, superconductors, and liquid crystals \cite{ohta1979xy,rosenblatt1980temperature,gingras1996topological}. 
Recent studies have revealed intricate phase transitions in models with complex interactions \cite{song2022phase,zhang2022surface} and interesting behaviors in long-range interacting models, such as the aging patterns in the Ising model \cite{christiansen2020aging} and phase transitions in the non-Mermin-Wagner-Hohenberg (non-MWH) regime of the XY model\cite{giachetti2022berezinskii}.~\\

\blue{The Mermin-Wagner-Hohenberg  theorem states that continuous symmetries cannot be spontaneously broken at finite temperatures in one- and two-dimensional systems with sufficiently short-range interactions, making true long-range order impossible \cite{ma2018modern, jenkins2022breaking}. However, when long-range interactions are present, the system can evade the constraints of the MWH theorem and exhibit true long-range order even in low dimensions. The region where true long-range order can exist is defined as the non-MWH regime, characterized by \red{\( \sigma < 2D \)}, where $\sigma$ is the exponent governing the decay of the interactions, and \( D \) is the spatial dimensionality of the system \cite{fisher1972critical,kunz1976first}. Conversely, the region where only quasi-long-range order \red{or short-range order} exist is termed the MWH regime, with \( \sigma > 2D \) \cite{bruno2001absence}. Determining the critical thresholds of \( \sigma \) that delineate these regimes is crucial for comprehensively characterizing phase transitions and scaling behaviors in spin models.}~\\

In recent years, advancements in exploring fundamental spin models using Rydberg atom experiments have been substantial. Researchers have successfully implemented long-range interactions in Rydberg atom arrays, thereby deepening our understanding of spin models under long-range interactions \cite{browaeys2020many,chen2023continuous}. In nano-systems, the strong system size dependence exhibited by some nanoparticles and nano-thin films also implies that these systems are related to spin systems with long-range interactions \cite{brede2014long,ellis2015switching,zhao2016high}. Additionally, in some rare-earth magnetic materials, interactions between rare-earth elements can span multiple atomic distances, such as in NdFeB \cite{samin2015monte,li2023coercivity}. This resurgence in interest has reignited our study of classical spin models with long-range interactions.~\\

Prior Monte Carlo studies 
have explored some properties of spin models with long range coupling of the form $r^{- \sigma}$, where $r$ is the distance between two spins. In one study, evidence of Berezhinskiǐ-Kosterlitz-Thouless-like phase transitions in the non-MWH regime was found by investigating the magnetization and specific heat of one-dimensional XY and Heisenberg models\cite{romano1997computer}. In another study, a new Monte Carlo algorithm proposed by the researchers was used to study the phase transition at the boundary between the non-MWH regime and the MWH regime in the two-dimensional XY model\cite{muller2023fast}. However, previous works often focused on the model behavior at specific values of \(\sigma\) and a comprehensive investigation of the system's behavior over the entire non-MWH region is not well-addressed. Moreover, the main focus of the studies was on the phase transition between the ordered and the disordered phases, and the scaling behaviors between different ordered regimes have been overlooked.~\\

To fill this gap, we provide a detail investigation in the non-MWH region of the long-range spin models using Monte Carlo simulations in this work. By analyzing the variation of magnetization with the system size, \blue{we discover that the true long-range ordered phase of continuous spin models can be further divided into distinct scaling regimes}\blue{, which we called the \red{Enhanced Long-Range Order (EnLRO) and the Reduced Long-Range Order (ReLRO)}, with different characteristics.} In the EnLRO, the magnetization increases with system size, while in the ReLRO, the magnetization decreases as the system size increases. Near the EnLRO-ReLRO crossover, the magnetization exhibits a non-monotonic behavior, first decreasing and then increasing with system size. By analyzing the spin-spin correlation functions, we found that in the ReLRO, the correlation functions decay algebraically \red{to a nonzero value even in large systems, whereas the correlation functions decay exponentially to a nonzero value in the EnLRO.} \blue{We determine the value of $\sigma$ at which the crossover between EnLRO and ReLRO occurs, and obtain the temperature-$\sigma$ phase diagram through finite-size scaling.} 


The paper is organised as follows: in Sec. \ref{sec:model_method}, the model concerned in this study and the methodology adopted are discussed. Section \ref{sec:simulation_results} presents our findings in the long-range one-dimensional (1D) XY model, including its critical exponents and the phase diagram in the temperature $T-\sigma$ plane. Similar results obtained in the two-dimensional (2D) XY model are also discussed. Evidence of distinct scaling behaviors in both long-range 1D and 2D Heisenberg models is presented. \red{Section \ref{sec:discuss_results} discusses the possible mechanism behind the transition between EnLRO and ReLRO.} 
Section \ref{sec:conculsion} summarizes our work.

\section{The Models and the Measurements}
\label{sec:model_method}
We consider the long-range interacting spin model defined on a 1D chain and a 2D square lattice with periodic boundary conditions. \blue{When calculating the distances between spins under periodic boundary conditions, the shortest path is always chosen, with \( r \leq L/2 \) in the 1D case and \( r \leq \sqrt{2}L/2 \) in the 2D case, where \( L \) is the linear dimension of the system.} The Hamiltonian of the model reads:
    
\begin{equation}
H = - \sum_{i \neq j}^{}{\frac{J}{{|i - j|}^{\sigma}}\mathbf{S}_{\mathbf{i}} \cdot \mathbf{S}_{\mathbf{j}}},
\end{equation}
    
\noindent where \(\frac{J}{{|i - j|}^{\sigma}}\) denotes the strength of the interaction between spins, and \(\sigma\) is the decay exponent controlling the range of interactions. The sum is taken over all sites \(i \neq j\). \(\mathbf{S}_{\mathbf{i}}\) represents the vector of the classical spin at the site \(i\); for the XY model, \(\mathbf{S}_{\mathbf{i}} = ({S}_{i}^{x},{S}_{i}^{y})\). Since the magnitude of the spins is constant, \(\mathbf{S}_{\mathbf{i}}\) can be represented solely by the polar angle \(\theta_{i}\), and the Hamiltonian simplifies to:
\begin{equation}
H_{\rm{XY}} = -\sum_{i \neq j}^{}{\frac{J}{{|i - j|^{\sigma}}} \cos\left(\theta_{i} - \theta_{j}\right)}.
\end{equation}
    
Similarly, for the Heisenberg model, \(\mathbf{S}_{\mathbf{i}}\) can be represented by the polar angle \(\theta_i\) and the azimuthal angle \(\phi_i\),
\(\mathbf{S}_{i} = ({S}_{i}^{x}, {S}_{i}^{y}, {S}_{i}^{z}) = (\sin\theta_i \cos\phi_i, \sin\theta_i \sin\phi_i, \cos\theta_i)\), and the Hamiltonian is given by:
    
\begin{equation}
\begin{aligned}
& H_x = -\sum_{i \neq j}\frac{J}{|i - j|^{\sigma}}\sin\theta_{i}\cos{\phi}_{i}\sin\theta_{j}\cos{\phi}_{j}, \\
& H_y = -\sum_{i \neq j}\frac{J}{|i - j|^{\sigma}}\sin\theta_{i}\sin{\phi}_{i}\sin\theta_{j}\sin{\phi}_{j}, \\
& H_z = -\sum_{i \neq j}\frac{J}{|i - j|^{\sigma}}\cos\theta_{i}\cos\theta_{j}, \\
& H_{\rm{Heisenberg}} = H_x + H_y + H_z.
\end{aligned}
\end{equation}

\blue{To characterize the thermodynamic properties of the system, we consider the partition function \( Z \) in the canonical ensemble, which is given by
\begin{equation}
Z = \sum_{i} e^{-\beta NE_i},
\label{eq:Z}
\end{equation}
where \( N \) is the number of spin in the system, \( E_i \) is the energy per spin of each possible state, \( \beta = \frac{1}{k_B T} \) is the inverse temperature with \( k_B \) being the Boltzmann constant and \( T \) being the temperature. For simplicity, we set \( J = k_B = 1 \), thereby measuring temperature in units of \( J/k_B \). For any physical quantity \( O \), its expectation value is defined through the partition function as
\begin{equation}
\langle O \rangle = \sum_{i} \frac{O_i e^{-\beta NE_i}}{Z},
\label{eq:O}
\end{equation}
where \( O_i \) is the value of the physical quantity \( O \) in each possible state. In Monte Carlo sampling, the expectation value can be further simplified to \( \langle O \rangle = \frac{1}{I} \sum_{i=1}^{I} O_{i}\), where \( O_i \) represents the quantity measured at the \( i \)-th Monte Carlo sweep, and \( I \) is the total number of measurements.}~\\

In our simulations of the XY and Heisenberg models, both one-dimensional spin chains and two-dimensional spin lattices were modeled using the single-update Metropolis Monte Carlo method. Each simulation began with \(10^{5}\) warm-up sweeps over randomly initialized spin configurations, where one sweep is defined as \(N\) attempted spin flips (\(N\) being the total number of spins in the system). Subsequently, \({5 \times 10}^{6}\) sweeps were employed for sampling, \blue{one sample is recorded for every 10 sweeps. In the results presented below, each data point is an average of \(5 \times 10^{5}\) samples and if the error bars are not visible in the figures, this indicates that they are smaller than the size of the data points. }~\\

For low-temperature simulations \red{of XY model} (\(T < 0.1\)), each proposed spin flip at the \(m\)-th step was defined as \(\theta_{i}(m + 1) = \theta_{i}(m) + \mathcal{U}( - 0.25\pi, 0.25\pi)\), where \(\mathcal{U}\) represents a uniform distribution, to mitigate the likelihood of rejecting large-angle flip proposals and to avoid the system being trapped in metastable states, which could reduce simulation efficiency. For high-temperature simulations, the proposed spin flip was reset to \(\theta_{i}(m + 1) = \theta_{i}(m) + \mathcal{U}( - \pi, \pi)\). A similar approach was applied for \({\theta}_{i}\) \red{in the Heisenberg model, while \({\phi}_{i}\) of the proposed spin is \(\cos^{-1}[2\mathcal{U}(0,1) - 1]\) for any temperatures}. ~\\

\blue{To investigate the scaling behaviors and critical phenomena of the system, we measured various physical quantities including the energy per site \(E\), magnetization per site \(M\), magnetic susceptibility \(\chi\), and \(n\)-th order magnetic cumulants (also known as Binder ratios) \(U_{n}\) \cite{fehske2007computational}. The magnetization per site} in the XY model and the Heisenberg model is defined as  

\begin{equation}
M_{\rm{XY}} = \sqrt{\left(\frac{1}{N}\sum_{i=1}^{N}{\sin\theta_{i}}\right)^{2} + \left(\frac{1}{N}\sum_{i=1}^{N}{\cos\theta_{i}}\right)^{2}},
\end{equation}
    
and 
    
\begin{equation}
M_{\rm{Heisenberg}} = \sqrt{M_{x}^{2} + M_{y}^{2} + M_{z}^{2}},
\end{equation}
where \(M_{x} = \sum_{i}\sin\theta_{i}\cos{\phi}_{i}/N\), \(M_{y} = \sum_{i}\sin\theta_{i}\sin{\phi}_{i}/N\), and \(M_{z} = \sum_{i}\cos\theta_{i}/N\), respectively. The magnetic susceptibility and the \(n\)-th order magnetic cumulants are defined as
    
\begin{equation}
\blue{\chi = \frac{N}{T}(\langle M^{2} \rangle - \langle M \rangle^{2}),}
\end{equation}
and 
{\begin{equation}
\blue{U_{n} = 1 - \frac{\langle M^{n} \rangle}{3\langle M^{n/2} \rangle^{2}},}
\label{eq:Un}
\end{equation} 
respectively. Additionally, we evaluated the first derivatives of the magnetic cumulant with respect to the inverse temperature $\beta$
\begin{equation}
U_{n}' = \frac{dU_{n}}{d\beta} = N(1 - U_{n})\left( \langle E \rangle - 2\frac{\langle M^{n/2}E \rangle}{\langle M^{n/2} \rangle} + \frac{\langle M^{n}E \rangle}{\langle M^{n} \rangle} \right),
\label{eq:dUn}
\end{equation}
and that for the $\ln\langle M^{n} \rangle$
\begin{equation}
(\ln\langle M^{n} \rangle)' = \frac{d\ln(\langle M^{n} \rangle)}{d\beta} = N\left(\langle E \rangle -\frac{\langle M^{n}E \rangle}{\langle M^{n} \rangle} \right).
\label{eq:dlnM^n}
\end{equation}

\blue{The detailed derivation of Eqs.~(\ref{eq:dUn}) and~(\ref{eq:dlnM^n}) is provided in Appendix \ref{sec:Detailed_derivation}. It is noteworthy that we utilize Eqs.~(\ref{eq:dUn}) instead of the magnetic cumulant in Eq. (\ref{eq:Un}) to locate the phase transition points because when using the latter, one needs to identify the intersection points of magnetic cumulant curves for different system sizes. In contrast, the former allows us to determine the phase transition point based on its maximum for a specific system size. This makes our analysis more convenient and the identified phase transition points more reliable. A detailed discussion of using Eq.~(\ref{eq:dUn}) and Eq. (\ref{eq:dlnM^n}) to study the order-disorder phase transition in the 2D Ising model can be found in Ref. \cite{fehske2007computational}.} \\

\blue{In addition to magnetization, we also explore the characteristics of the system under different values of \(\sigma\) in the XY model through the spin-spin correlation function \(G(r)\) defined as
\begin{equation}
G(r) = \langle \cos(\theta_i - \theta_{i+r}). \rangle
\end{equation}
}

In systems with long-range interactions, the physical quantities above can be described by the following scaling relationships \cite{bayong1999effect,bayong1999potts,gonzalez2021finite}:
    
\begin{equation}
\max\lbrack{U_{n}}^{'}(L)\rbrack \propto L^{1/\upsilon},
\label{eq:maxdUn}
\end{equation}
    
\begin{equation}
\max\left[(\ln M^{n}(L))'\right] \propto L^{1/\upsilon}, 
\label{eq:maxlnM}
\end{equation}
    
\begin{equation}
\max\lbrack\chi(L)\rbrack \propto L^{\gamma/\upsilon},
\label{eq:maxchi}
\end{equation}
    
\begin{equation}
T_{c}(L) = T_{c}(\infty) + KL^{- 1/\upsilon}.
\label{eq:Tc}
\end{equation}
where \(L\) is the linear dimension of the system. The parameters \(\upsilon\) and \(\gamma\) are the critical exponents of the correlation length \(\xi\) and the magnetic susceptibility, respectively, while \(K\) is a parameter to be determined. The \(T_{c}(L)\) is the temperature determined by the maximum of \(\chi(L)\), whereas \(T_{c}(\infty)\) is the critical temperature for an infinitely large system. Using Eqs. (\ref{eq:maxdUn}) and (\ref{eq:maxlnM}), we can estimate \(\upsilon\) and employ Eq. (\ref{eq:Tc}) to perform a linear fit between \(T_{c}(L)\) and \blue{\(L^{- 1/\upsilon}\)}, thereby estimating \(T_{c}(\infty)\). ~\\

\section{Results}
\label{sec:simulation_results}

\subsection{Long-range XY model}
\label{subsec:lr_xy_model}

\blue{Figure \ref{fig:fig1}(a) illustrates the three distinctive regimes in the long-range coupling XY model with respect to the decay exponent \(\sigma\) and the system's spatial dimension \(D\).} In the regime where \(0 < \sigma < D\), the energy is non-extensive and diverges in an infinite system, making the system statistically non-meaningful for conventional analytical approaches. Although this issue can be mitigated by introducing the Kac normalization factor \cite{kac1963van}, this region is outside the scope of interest in this work. In the Mermin-Wagner-Hohenberg (MWH) regime where \(\sigma > 2D\), it is found that only a quasi-long-range ordered phase \red{or short-range ordered phase} exist at low temperatures. This implies that a small subsystem of the model exhibits strong order, but this order diminishes gradually as the subsystem size increases. Specifically, the spin-spin correlation function decays as a power-law with increasing distance \(r\) between two spins, with a temperature-dependent exponent. On the other hand, when the system enters the non-MWH regime where \( D < \sigma < 2D \), recent renormalization group studies have shown that the system possesses a true long-range ordered phase \blue{with finite magnetization at low temperatures \cite{giachetti2021berezinskii,giachetti2022berezinskii}.} \red{In Appendix~\ref{sec:Correlation_function_diffL}, we use numerical methods to show evidence for the existence of a true long-range ordered phase when $\sigma < 2D$.}~\\


\begin{figure} [t!]
\centering
\includegraphics[width=0.8\linewidth]{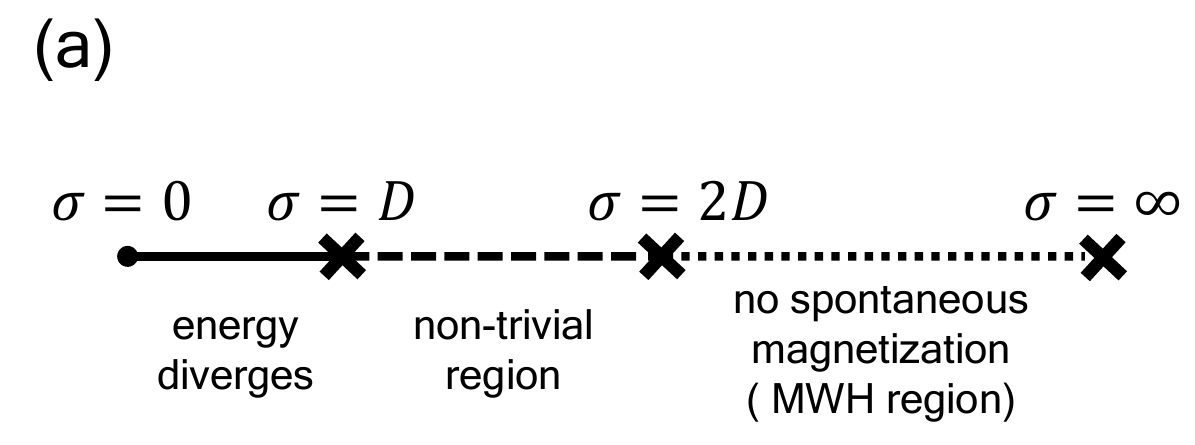}
\includegraphics[width=0.49\linewidth]{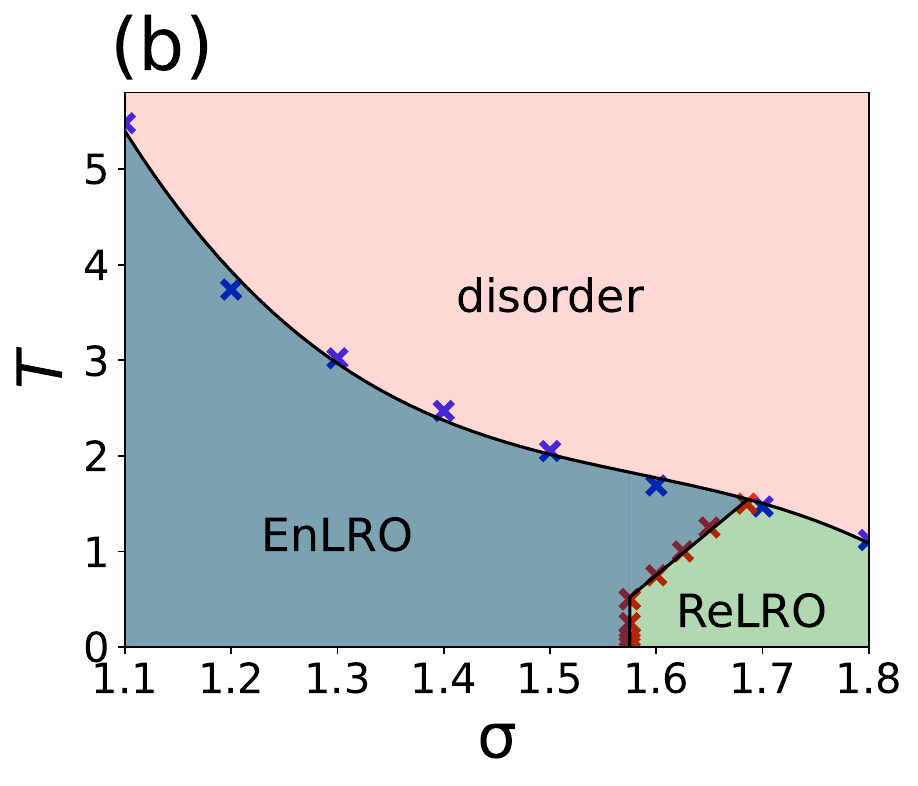}
\includegraphics[width=0.49\linewidth]{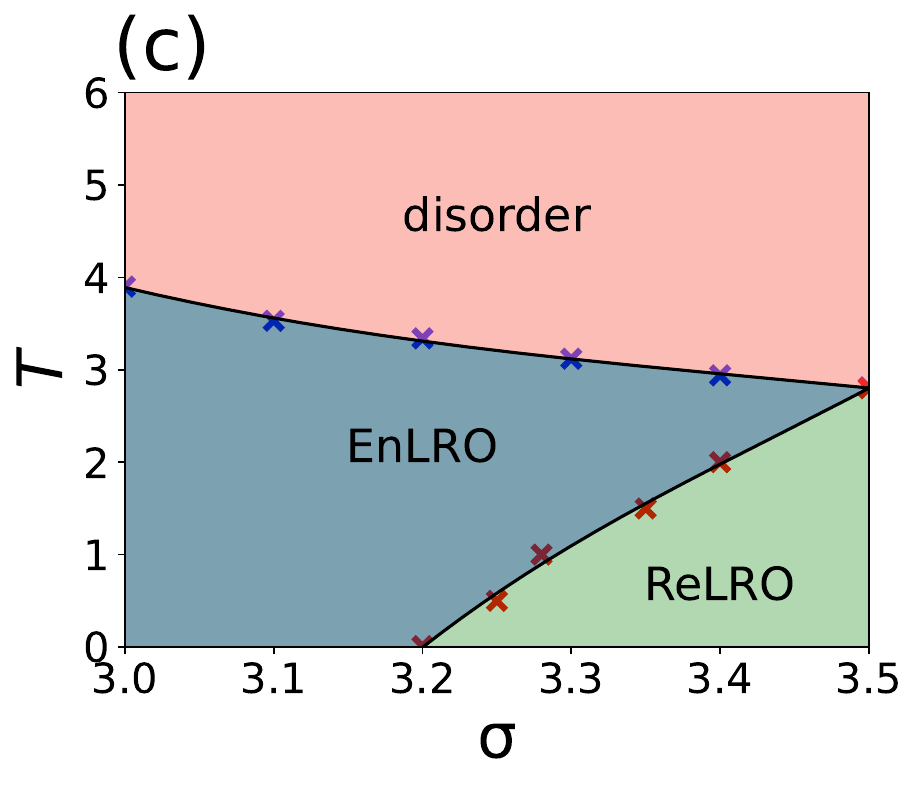}
\caption{(a) The figure shows three characteristic regimes in the LR XY model. The system is in the non-extensive regime for \(0 < \sigma < D\) and the MWH regime for \(\sigma > 2D\), in which its behavior is akin to that of nearest-neighbor interacting models exhibiting quasi-long-range order \red{or short-range order}. At intermediate values of \(D < \sigma < 2D\), \blue{where true long-range orders are found to exists at low temperatures \cite{giachetti2021berezinskii,giachetti2022berezinskii}}, the system exhibits distinct scaling behaviors due to long-range interactions. (b, c) The figures respectively show the \blue{phase diagrams} of the 1D and 2D LR XY models in the non-MWH regime. The red crosses indicate the numerically extracted \blue{boundaries for the crossover between EnLRO and ReLRO regions}, while the blue crosses indicate the ordered-disordered phase transitions. Solid black lines are guides to the eye.} 
\label{fig:fig1}
\end{figure}

Although previous studies have demonstrated the existence of a true long-range ordered phase in the non-MWH regime where a finite magnetization is attained in the system \cite{giachetti2021berezinskii}, the relationship between magnetization and system size was not thoroughly addressed. Specifically, it remains unclear whether the magnetization decreases with increasing \(L\) before converging to a finite value (indicative of a ReLRO scaling behavior), or whether it increases with \(L\) and converges to a finite value indicative of a EnLRO scaling behavior. In this work, we conducted detailed Monte Carlo simulations and analysis on both 1D and 2D LR XY models and found that both EnLRO and ReLRO scaling behavior can coexist in the system depending on the driving parameters \(\sigma\) and \(T\). The crossover between EnLRO and ReLRO does not constitute a true phase transition but rather reflects a change in the scaling behavior of the magnetization with system size. The main results are summarized in the phase diagrams shown in Fig. \ref{fig:fig1}(b) and (c).~\\

\subsubsection{1D LR XY model: EnLRO-ReLRO crossover at low temperatures}
\label{subsubsec:1d_lr_xy_model_lowT}

\begin{figure} [t!]
\includegraphics[width=0.49\linewidth]{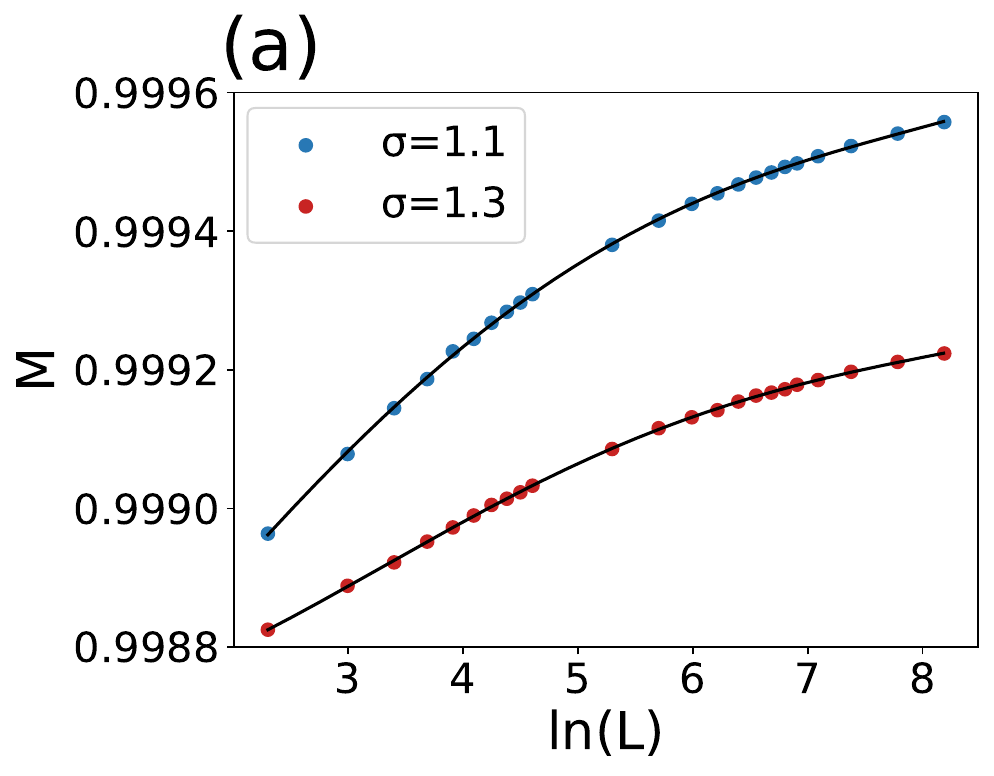}
\includegraphics[width=0.49\linewidth]{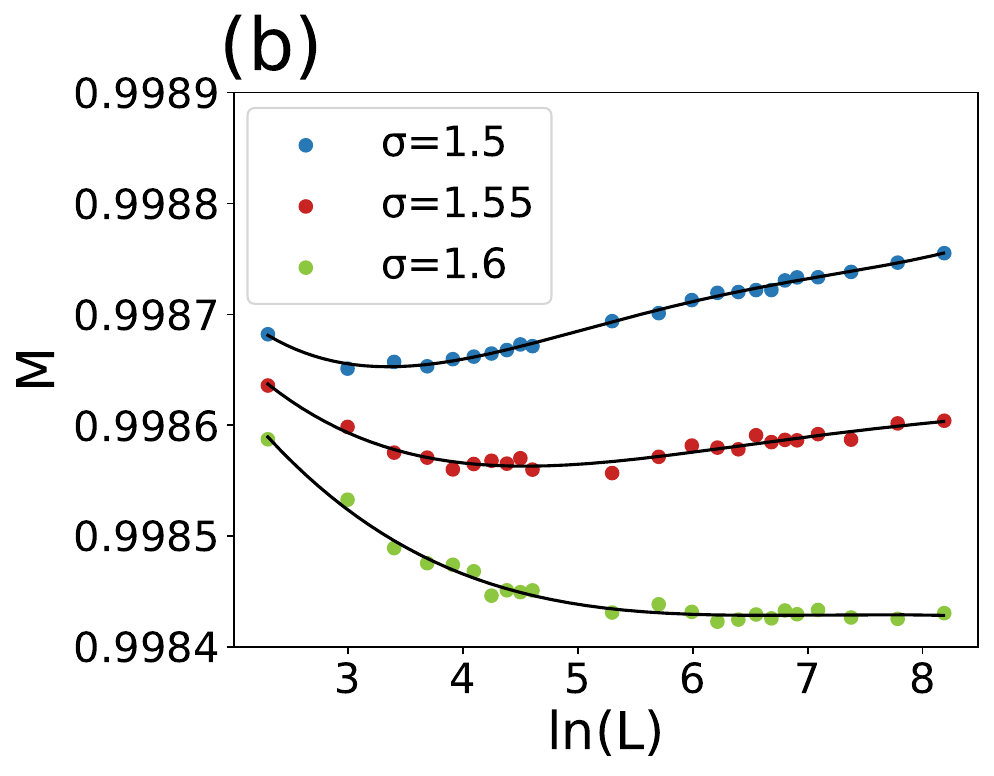}
\includegraphics[width=0.49\linewidth]{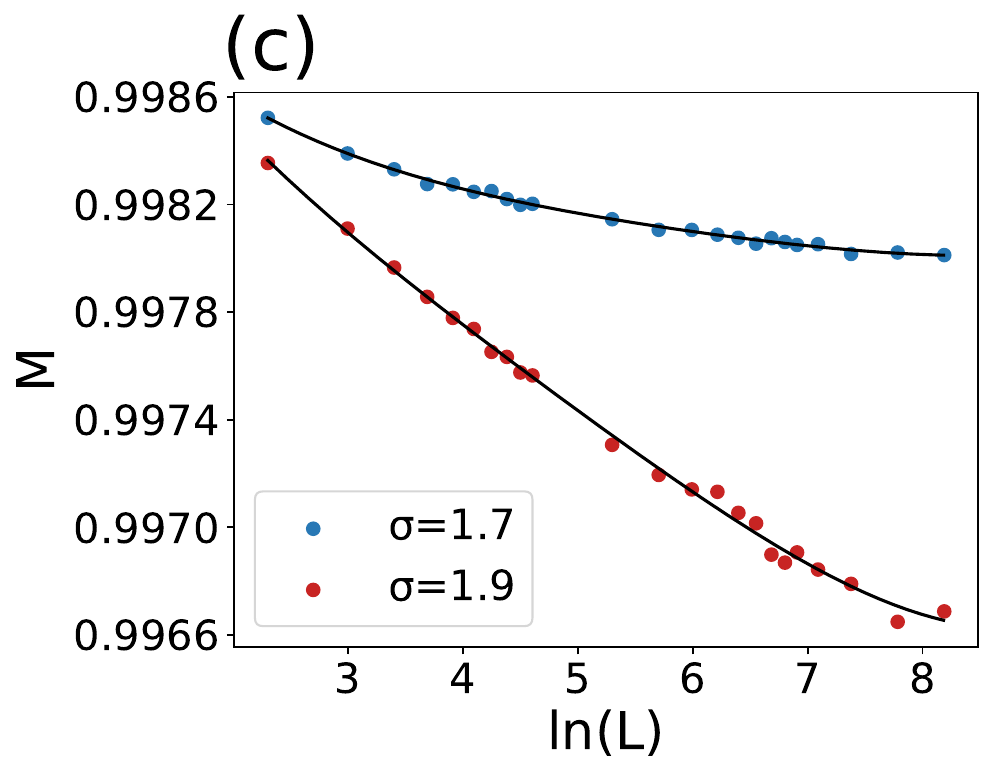}
\includegraphics[width=0.49\linewidth]{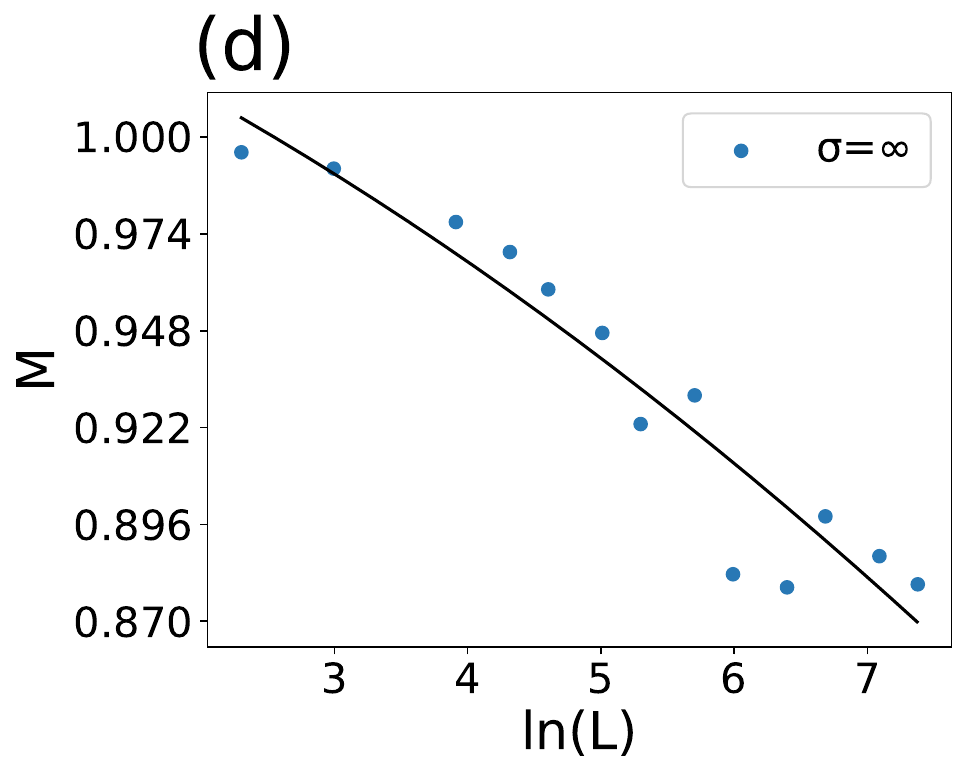}
\includegraphics[width=0.487\linewidth]{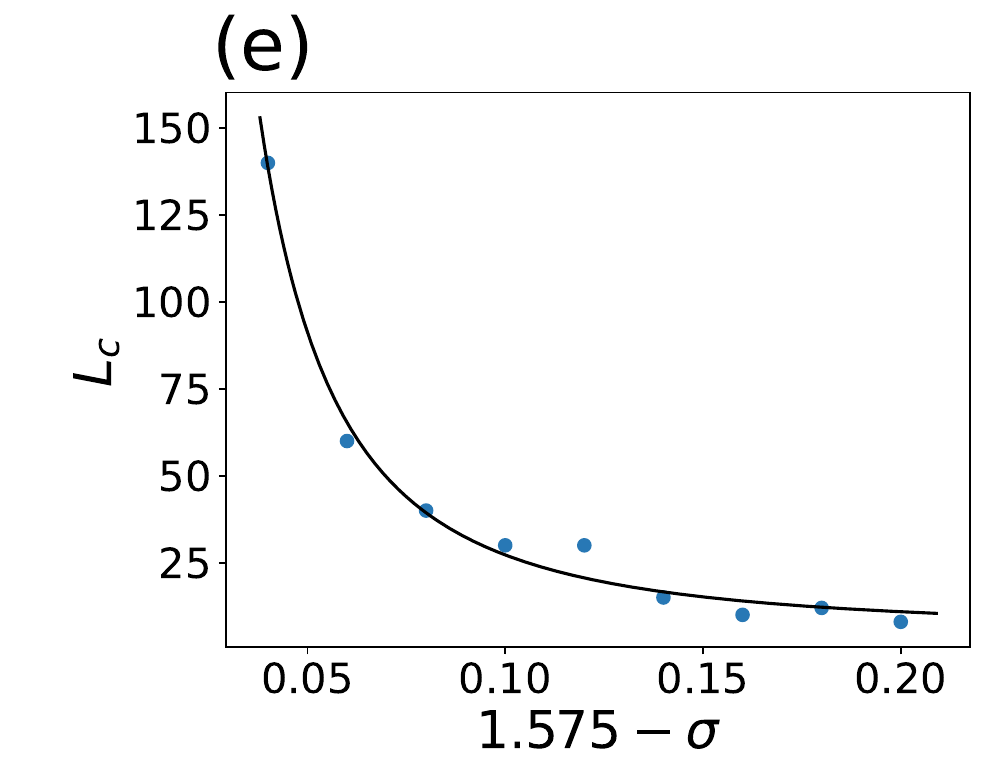}
\includegraphics[width=0.487\linewidth]{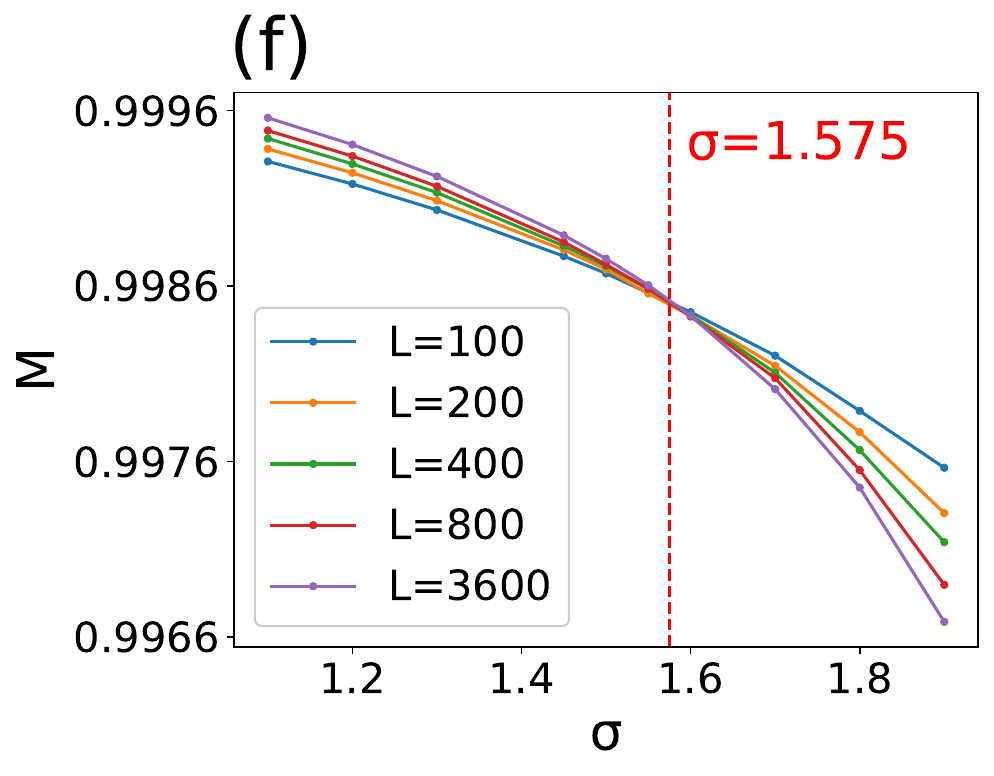}
\caption{(a-d) The figures show variation trends of
magnetization $M$ with $\ln L$ for different values of \(\sigma\) in the 1D LR XY model at $T=0.01$, with the
black lines are polynomial fits (up to fourth order) to the data points \blue{that guide the eye}. In non-MWH regions of small \(\sigma\) in (a) and large \(\sigma\) in
(c), $M$ exhibits diametrically opposed trends, whereas the dependence of $M$ on $\ln L$ in (b) is non-monotonic. \red{In the MWH regime of \(\sigma = \infty\) in (d), \(M\) decays more rapidly as \(\ln L\) increases, in clear contrast to the data trends in (c).} (e) The figure shows \( L_{c} \) as a function of \( 1.575-\sigma \) at \( T = 0.01 \). The blue circles represent $L_c$ at which the magnetization changes its size dependence from decrease to increase. (f) The \( M \)-\( \sigma \) curves for different system sizes at \( T = 0.01 \) cross at a single point \( \sigma = 1.575 \). }
\label{fig:fig2}
\end{figure}

As our goal is to study the true long-range ordered phases, we focused our attention on low temperatures in the non-MWH regime, specifically where \(1 < \sigma < 2\) for the 1D model. \blue{The data points in Fig.~\ref{fig:fig2} (a-d) show the Monte Carlo simulation results for $M$ as a function of $\ln L$ at $T = 0.01$ for different $\sigma$ values, while the black lines are polynomial fits of the data that guide the eye.} It can be clearly observed from Fig. \ref{fig:fig2}(c) that for \(1.7 \leq \sigma \leq 1.9\), $M$ decays with $\ln L$. The decay is much slower than that observed in the nearest-neighbor interacting model (\(\sigma = \infty\)) as shown in Fig. \ref{fig:fig2}(d), and the magnetization tends to attain a finite value. Thus, we anticipate the presence of true long-range order in this region, which is consistent with the conclusions obtained from \red{theoretical analysis}\cite{bruno2001absence}. When \(\sigma\) is small, we observe a sudden change in the trend of the $M-\ln L$ plot, shifting from a decaying trend to an increasing one, as shown in Fig. \ref{fig:fig2}(a). Specifically, for \(\sigma \leq 1.3\), $M$ increases monotonically over the entire $\ln L$ region.~\\ 

For \(\sigma \approx 1.5\), $M$ first undergoes rapid decay, followed by an increase with $\ln L$, as depicted in Fig. \ref{fig:fig2}(b). \blue{One possible explanation for this trend is that, in small systems, thermal fluctuations introduce significant disorder. This disorder exceeds the ordering tendency resulting from the limited number of interacting spins, leading to a decrease in magnetization with increasing system size. When the system size surpasses a threshold \(L_c\), the cumulative effect of long-range interactions begins to dominate, causing magnetization to increase with \(\ln L\).}~\\

\begin{figure} [t!]
\includegraphics[width=0.49\linewidth]{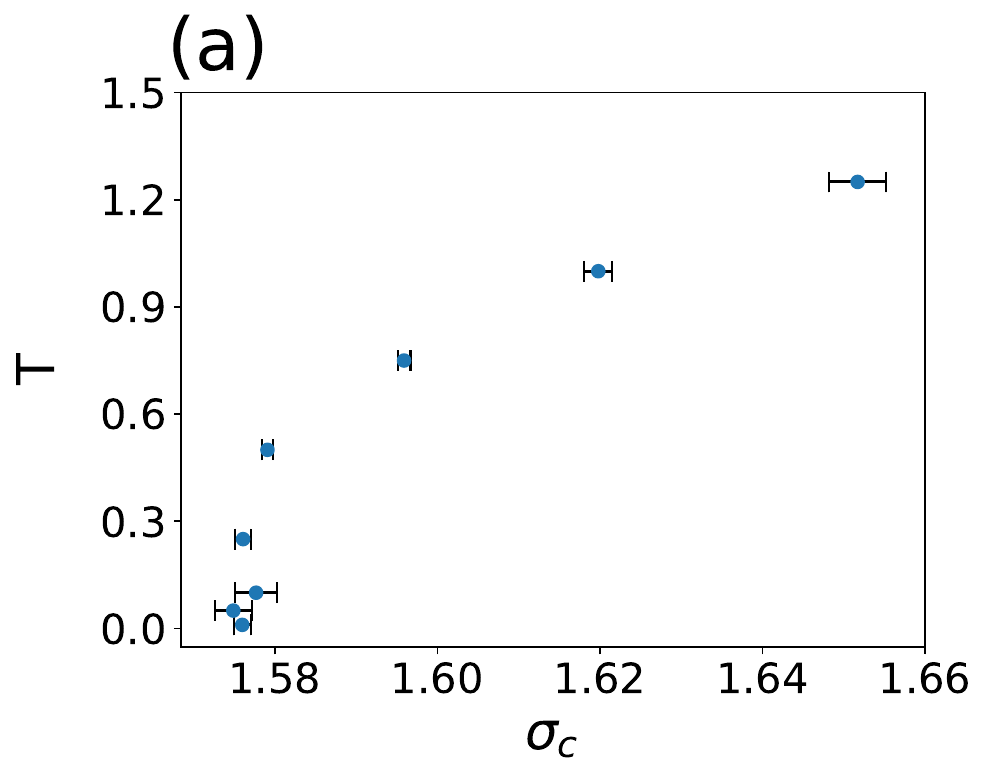}
\includegraphics[width=0.48\linewidth]{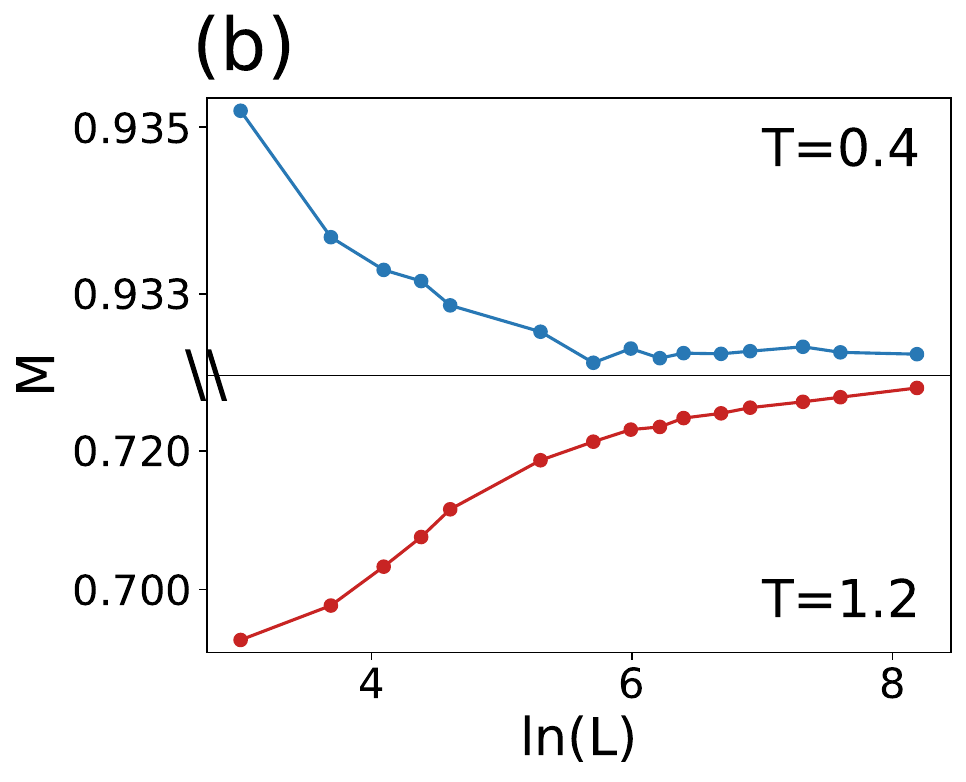}
\includegraphics[width=0.48\linewidth]{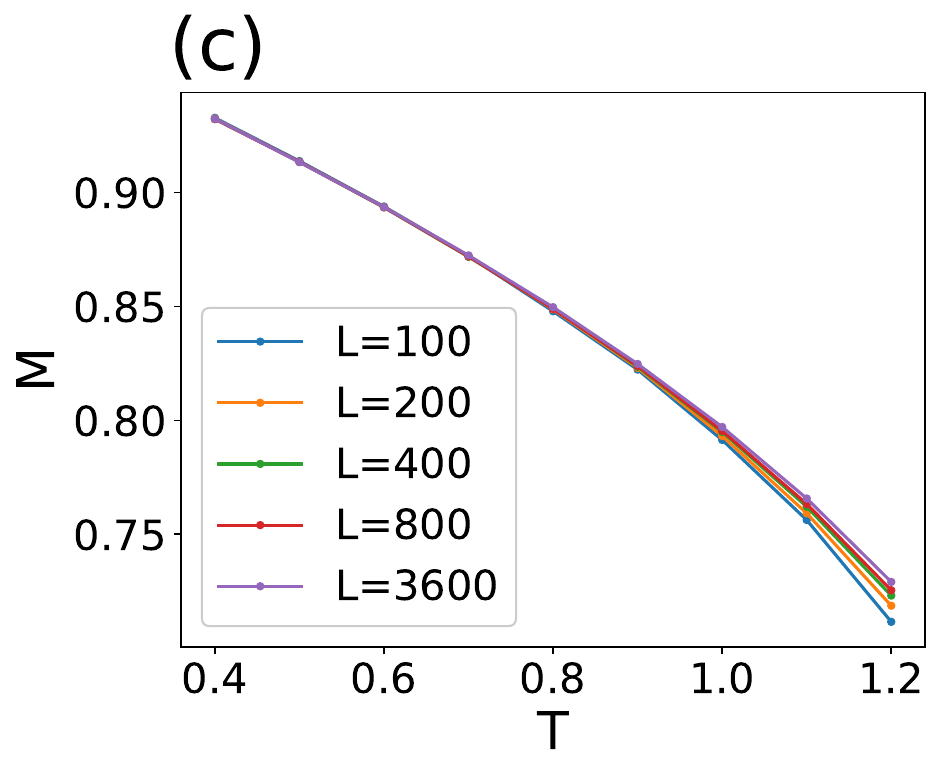}
\includegraphics[width=0.48\linewidth]{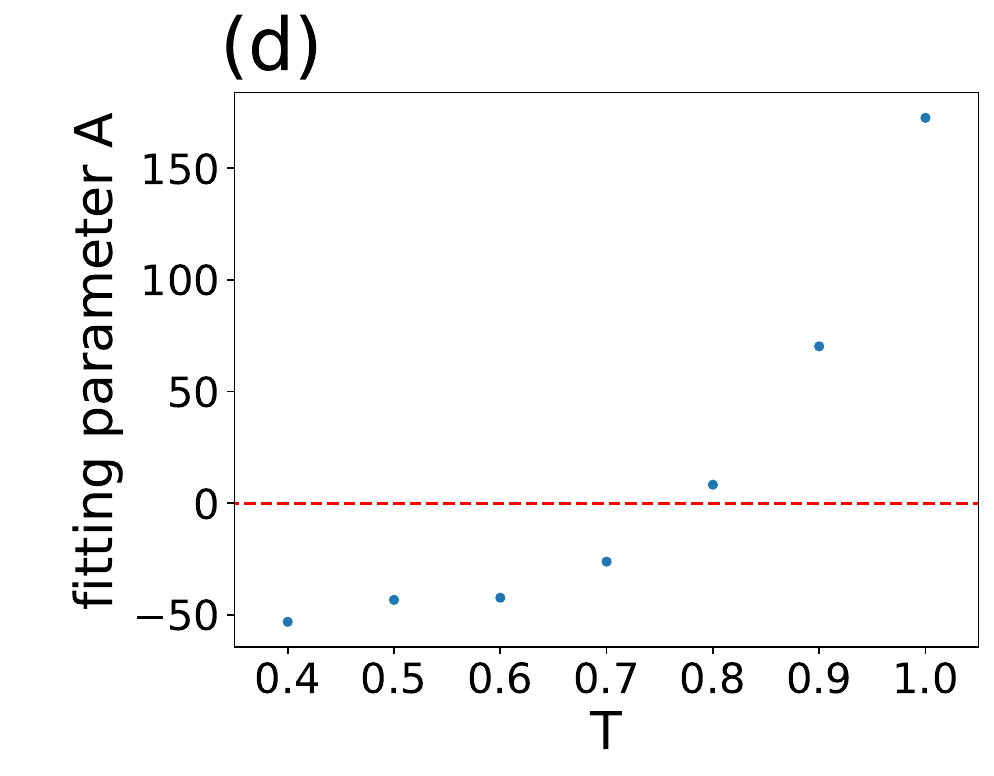}
\caption{Data collected in the 1D LR XY model. (a) The figure shows critical \( \sigma \) where the EnLRO-ReLRO scaling behavior crossovers take place at various temperatures. (b-d) The figures show the ReLRO-EnLRO scaling behavior crossover at \( \sigma=1.6 \) while temperature is the driving parameter. (b) The variation of
magnetization $M$ with $\ln L$ for different values of \(T\). (c) The \( M \)-\( T \) curves for different system sizes. (d) The fitted parameter $A$ in \red{$\frac{\mathrm{d} M}{\mathrm{d} (\ln L)}=A(\ln L)^{-2}$}. The sign change of $A$ indicates the scaling behavior crossovers.}
\label{fig:fig3}
\end{figure}

To investigate the observed decreasing trend in magnetization $M$ with increasing $\ln L$ for \(\sigma = 1.7\) and 1.9, and to determine if the observation is due to insufficient system size, we conducted a detailed analysis for $\sigma\in [1.40,1.56]$ with a step size of \blue{0.02}. \blue{Specifically, we estimated the critical system size \( L_{c} \) for each \( \sigma \) and established a power-law relationship between \( L_{c} \) and \( 1.575 - \sigma \), as shown in Fig.~\ref{fig:fig2}(e). It is evident that as \( 1.575 - \sigma \) approaches zero (i.e., \( \sigma \) approaches 1.575), \( L_{c} \) diverges. This indicates that there is no turning point for the \( M - \ln(L) \) curve when \( \sigma > \sigma_{c} \), which is consistent with the results presented in Fig.~\ref{fig:fig2}(b-d). The divergence of \( L_{c} \) also suggests that a crossover in scaling behavior occurs.}~\\

According to our analysis, a small \(\sigma\) strongly enhances $M$ when
$L$ increases. As \(\sigma\) increases, this upward trend of $M$ gradually diminishes,
and finally, when \(\sigma > \sigma_{c}\), $M$ decreases with increasing
$L$. \blue{Therefore, the \(\sigma\) at which $M$ remains constant as
$L$ increases can be defined as the crossover point \(\sigma_{c}\) where the system transits from EnLRO to ReLRO as $\sigma$ increases.} To locate \(\sigma_{c}\) more accurately, we plotted $M$ versus \(\sigma\) for
different $L> L_{c}$ in Fig. \ref{fig:fig2} 
(f). It is clear that at
$\sigma_c =1.575$, the curves for different $L$ cross,
indicating that $M$ does not change significantly with increasing $L$. As
\(\sigma\) gradually moves away from 1.575, the separation between data
points for different $L$ also increases, which aligns with our definition of $\sigma_c$
above.~\\

\begin{figure} [t!]
\includegraphics[width=0.48\linewidth]{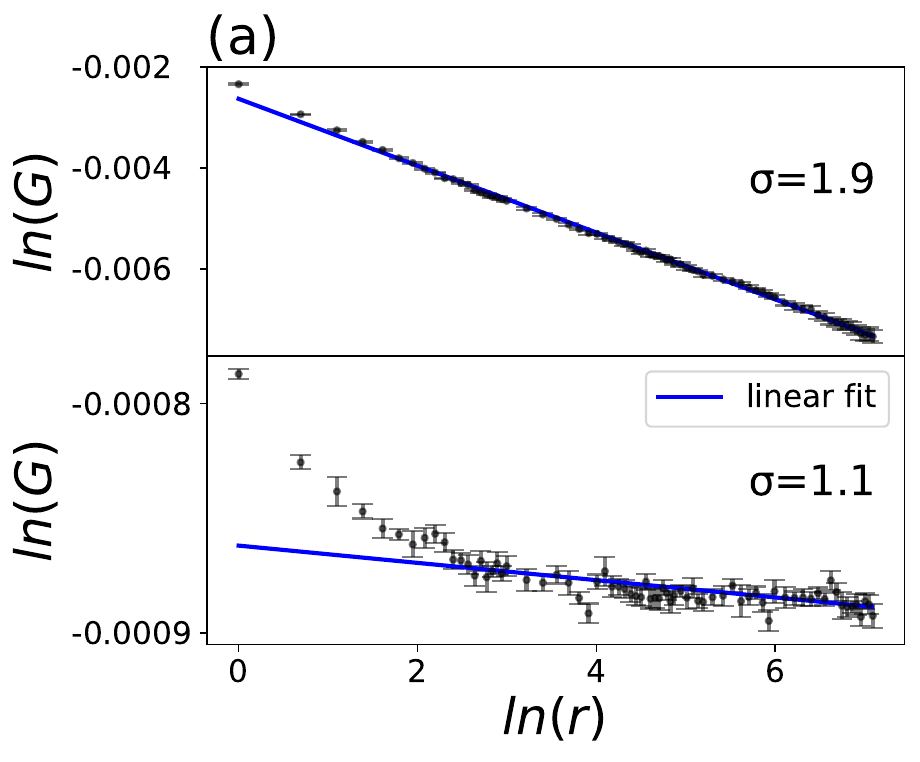}
\includegraphics[width=0.485\linewidth]{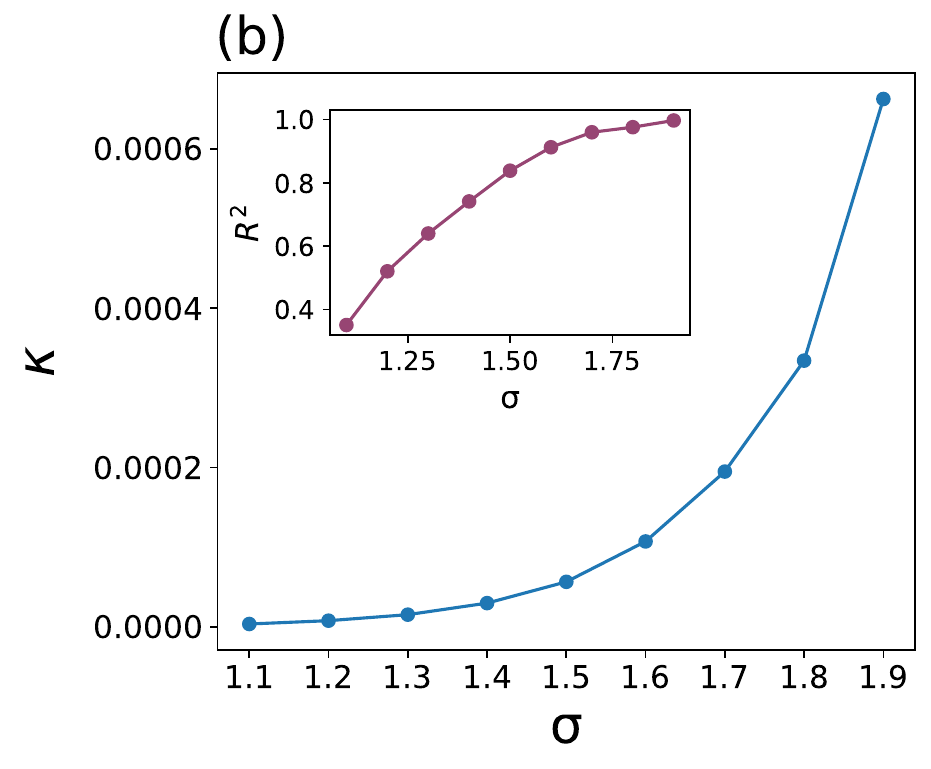}
\includegraphics[width=0.48\linewidth]{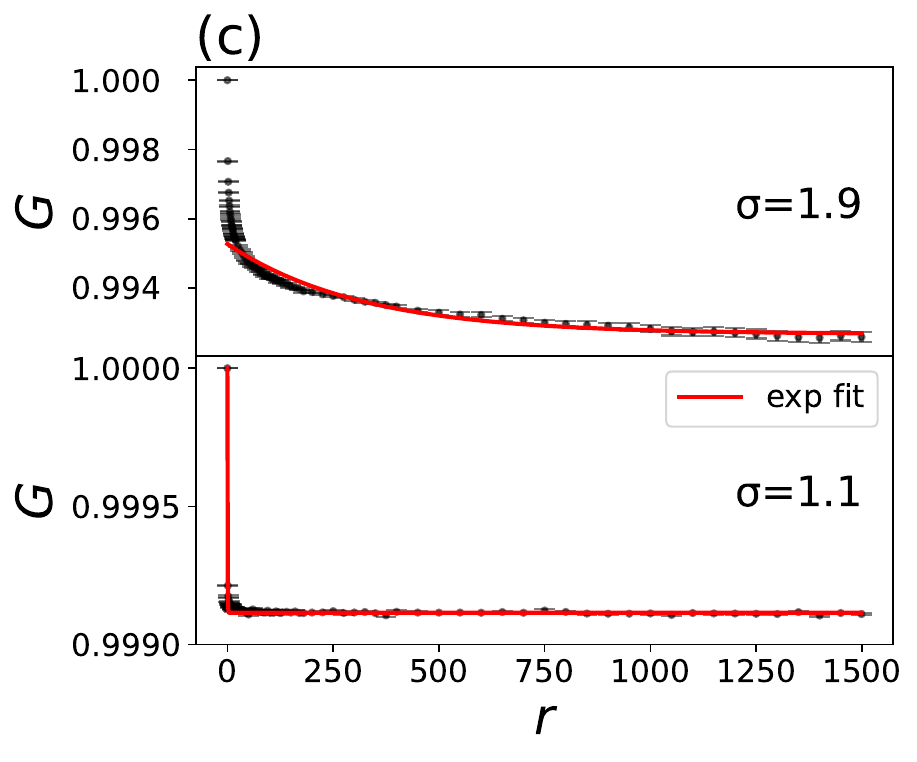}
\includegraphics[width=0.485\linewidth]{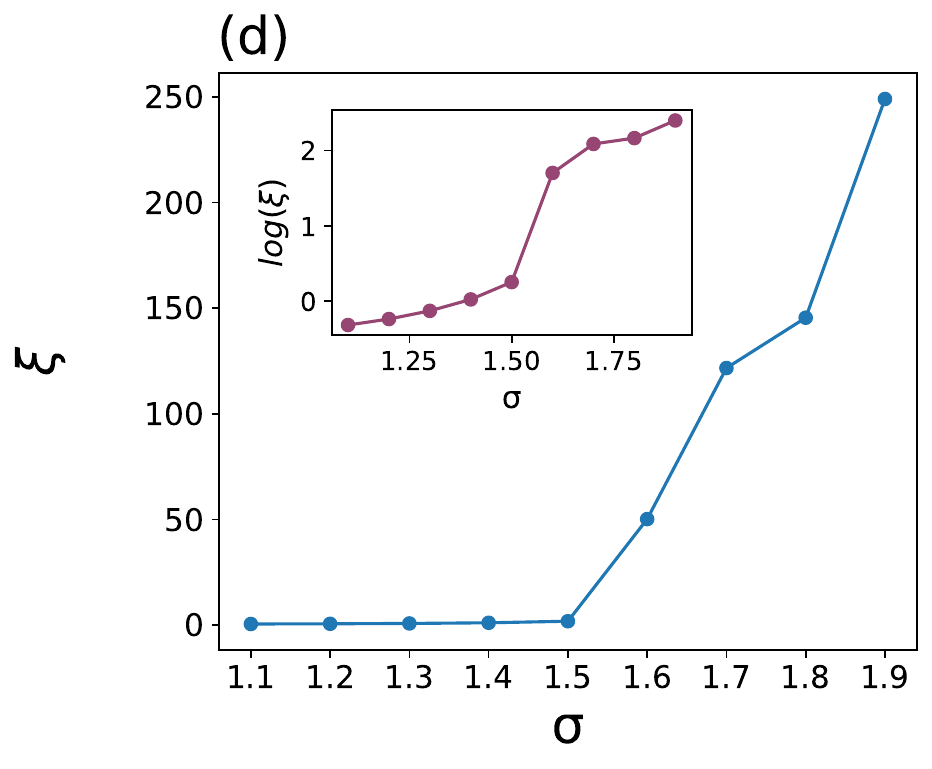}
\caption{\blue{ \red{The plots show the scaling analysis of the correlation functions for a system of \(L = 3600\) and \(T = 0.01\).} (a) Correlation functions for $\sigma = 1.9$ and $\sigma = 1.1$ on a ln-ln scale in the upper and lower panels, respectively. The blue lines represent the linear fits to the $\ln G$ vs $\ln r$. (b) The decay constant $\kappa$ extracted from the linear fit as a function of $\sigma$. The inset shows the R-squared value which quantifies the performance of the fitting. (c) Correlation functions in the original scale for $\sigma = 1.9$ and $\sigma = 1.1$ in the upper and lower panels. The red line indicates the fit using an exponential function. (d) Variation of the fitting parameter $\xi$ with different $\sigma$ values. Inset show the logarithm of $\xi$ for better visualization of the data close to zero.}}
\label{fig:fig3_2}
\end{figure}

One might observed that the magnetization presented in Fig. \ref{fig:fig2} is close to one. This is because the data is obtained for $T=0.01$, and at such a low temperature, 
\red{local defects \green{are absent}, making it easier to extract the variations in the scaling behavior induced by spin wave excitations at different $\sigma$}. 
\blue{In general, the variation in the magnetization can be large at higher temperatures. For example,} in Fig. \ref{fig:fig4}(b), we can find that at \(\sigma = 1.1\), the magnetization has significant dependence on the system size in the EnLRO region. For instance, at \( T = 3.5 \), as the system size increases from \( L = 100 \) to \( L = 4000 \), the magnetization \( M \) increases from approximately 0.36 to 0.80.
}~\\

We also estimated \(\sigma_{c}\) for temperatures $T\in[0.01,1.25]$
and found that \(\sigma_{c}\) remains almost constant at \(T \leq 0.5\).
However, as $T$ gradually increases beyond 0.5, \(\sigma_{c}\) increases,
as shown in Fig. \ref{fig:fig3}(a). In other words, as the temperature
increases, the system can maintain a EnLRO scaling behaviors under weaker long-range interactions. 
We further compared the trend of magnetization with system size at high and low temperatures when $\sigma = 1.6$. As shown in Fig. \ref{fig:fig3}(b), it is clear that the system exhibits ReLRO at $T=0.4$ and EnLRO at $T=1.2$. Figure \ref{fig:fig3}(c) further illustrates the variation of $M$ with $L$ at different temperatures. When $T<0.8$, the change in $M$ with respect to the temperature is insensitive to the system size.
On the other hand, for $T \geq 0.8$, the data points at the same temperature gradually spread out, and as $L$ increases, $M$ also increases, indicating that the system is in EnLRO in this region. \blue{We fitted the first derivative of $M$ with respect to $\ln (L)$ at different temperatures to an inverse function, i.e. \red{$\frac{\mathrm{d} M}{\mathrm{d} \ln (L)}=A (\ln L)^{-2}$}, where $A$ is a fitting parameter. This allows us to capture the scaling behavior crossover of finite systems under different $\sigma$}. Figure \ref{fig:fig3}(d) shows the fitted parameter $A$ as a function of temperature. It can be seen that $A$ changes sign at $T \approx 0.8$, \blue{which is consistent with the result shown in \ref{fig:fig3}(a) where the $T \approx 0.75$ at $\sigma =1.6$}, the result also indicating a change in the trend of the $M$-$\ln L$ curve, signifying a ReLRO-EnLRO crossover. \\

\blue{To further investigate the system's properties, we analyze the behavior of the spin-spin correlation function \red{for systems with \(L = 3600\) and \(T = 0.01\)} to distinguish between different scaling regimes \red{(the correlation functions at \(T = 0.1\) are also shown in the Appendix \ref{sec:Correlation_function_highT})}. Figure \ref{fig:fig3_2}(a) shows a ln-ln plot of the spin-spin correlation as a function of the distance $r$ between two spins. For $\sigma=1.9$, the data follows a straight line, suggesting a power-law decaying behavior of the correlation function, i.e. $G(r) \sim r^{-\kappa}$.
On the other hand, for $\sigma = 1.1$, the correlation function does not follow an algebraic decaying behavior as shown in the bottom panel of Fig. \ref{fig:fig3_2}(a). Figure \ref{fig:fig3_2}(b) shows the power-law decay exponent $\kappa$ extracted from the linear fit of the ln-ln plot for various $\sigma$, and the R-squared measuring the goodness of the fit in the inset. For large $\sigma$, $R^2$ is close to one, further evidencing the linearity of the ln-ln plot data. For small $\sigma$, the linear fit can only fit to the data with large $r$ which approches a finite constant, as reflected in a small $R^2$ and nearly vanishing $\kappa$. \\}

\red{In the quasi-long-range ordered phase or short-range ordered phase, the correlation function vanishes at infinite separation \(G(r\to\infty)=0\), so there is no constant term in the fitting function. When \(\sigma<2\), the system enters the true long-range ordered phase and a nonzero constant \(k_1\) appears, so that \(G(r)=k_1+k_2\times r^{-\kappa}\). In general, a constant plus a power law does not yield a straight line in a ln-ln plot. However, for \(\sigma=1.9\), the system is just inside the true long-range ordered phase. In the thermodynamic limit, \(k_1\) is nonzero but much smaller than one. Also, in the finite system of our rearch, the decay of \(k_2\times r^{-\kappa}\) is significantly slow, from \(r=0\) to \(r=1500\), \(G(r)\) decreases by only 0.008 (see Fig.~\ref{fig:fig3_2}(c), upper panel). Thus, over this range, \(k_1\) is negligible compared to \(k_2\times r^{-\kappa}\), so we can approximate \(G(r)\approx k_2\times r^{-\kappa}\). This approximation yields an approximately linear relation between \(\ln G\) and \(\ln r\). In Fig.~\ref{fig:fig3_2}(a) (upper panel), the data are close to a straight line, though there is a drawback in fitting performance at small \(r\). \green{As $\sigma$ decreases, the above assumption becomes less valid.}}~\\

\blue{The correlation function for small $\sigma$ in fact resemble more closely an exponential decay. We used an exponential function $G(r) = k_1 + k_2 \exp(-r/\xi)$, where  $k_1$, $k_2$, and $\xi$ are the fitting parameters, to refit the data. \red{\(k_1\) is a nonzero constant and cannot be neglected since system remains in the true long-range ordered phase when $\sigma < 2$.} As shown in Fig. \ref{fig:fig3_2}(c), the exponential function fits the correlation function for $\sigma = 1.1$ well but cannot accurately fit the correlation function for $\sigma = 1.9$.} 
\blue{Figure~\ref{fig:fig3_2}(d) displays the fitting parameter $\xi$ as a function of $\sigma$. The parameter $\xi$ exhibits a noticeable discontinuous change around $\sigma \approx 1.55$, even when viewed on a logarithmic scale, as shown in the inset. Specifically, for $\sigma \leq 1.5$, $\xi$ is approximately zero, while for $\sigma \geq 1.6$, $\xi$ becomes significantly greater than zero. This indicates a significant change in the decay trend of the correlation function, highlighting the clear difference in the scaling behavior between EnLRO and ReLRO regimes. A similar behavior is also observed in the 2D XY model with nearest-neighbor interactions. The exponent extracted from fitting an algebraic function exhibits a discontinuous change at the Berezinskii–Kosterlitz–Thouless (BKT) transition \cite{allen2021kosterlitz}. 
We suspect the differences in scaling behavior between EnLRO and ReLRO here may potentially be a non-trivial phase transition but further investigation will be needed.}~\\


\begin{figure*}[t!]
\centering
\includegraphics[width=0.239\linewidth]{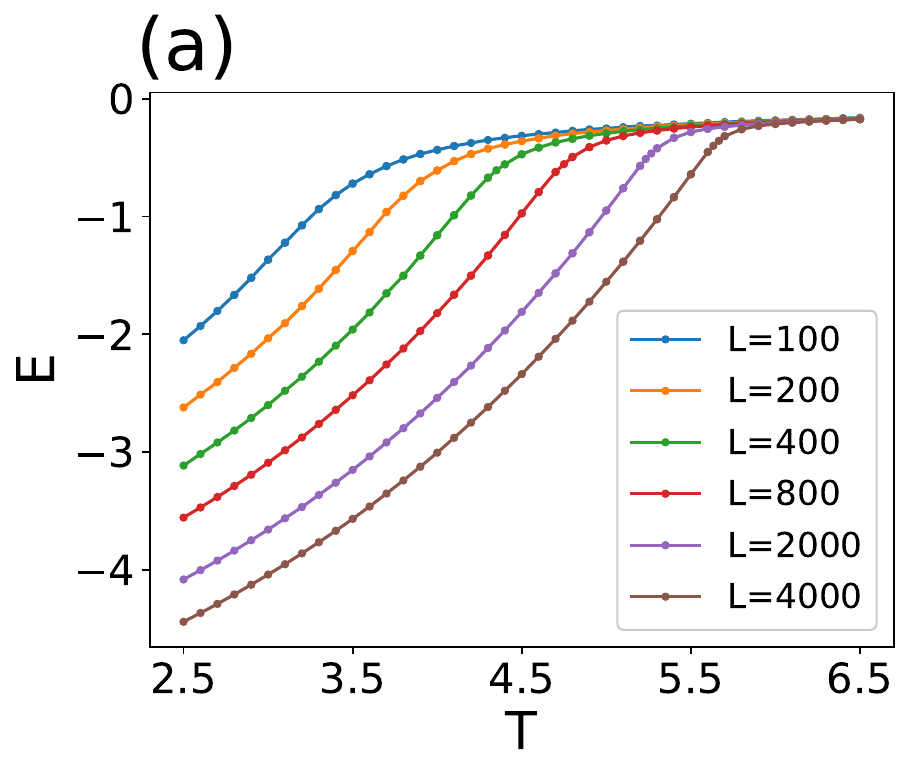}\hfill
\includegraphics[width=0.24\linewidth]{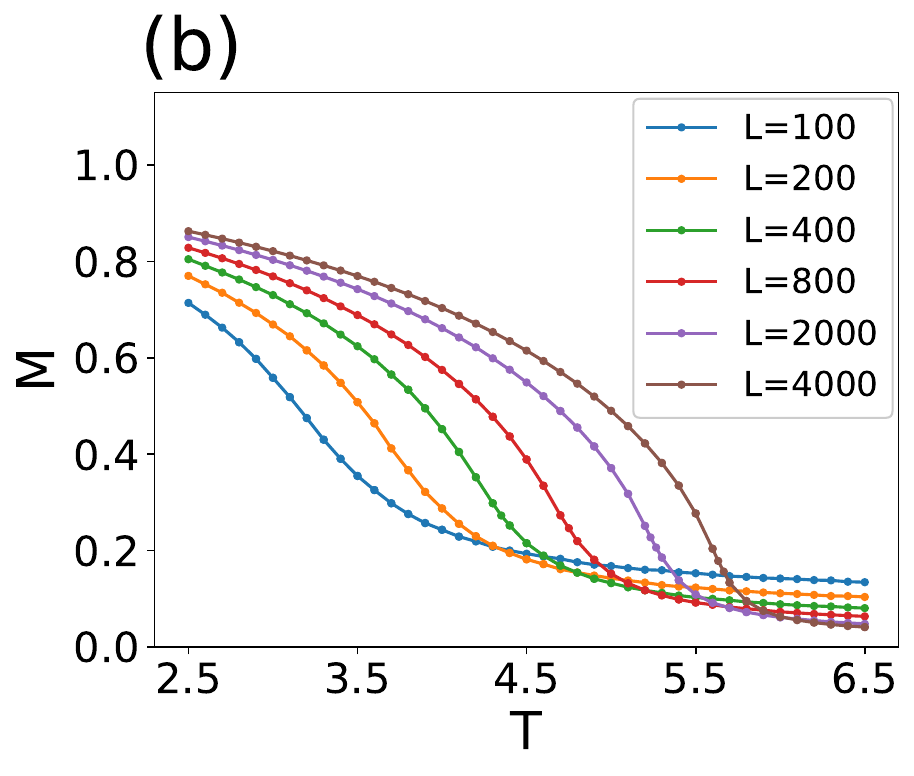}\hfill
\includegraphics[width=0.236\linewidth]{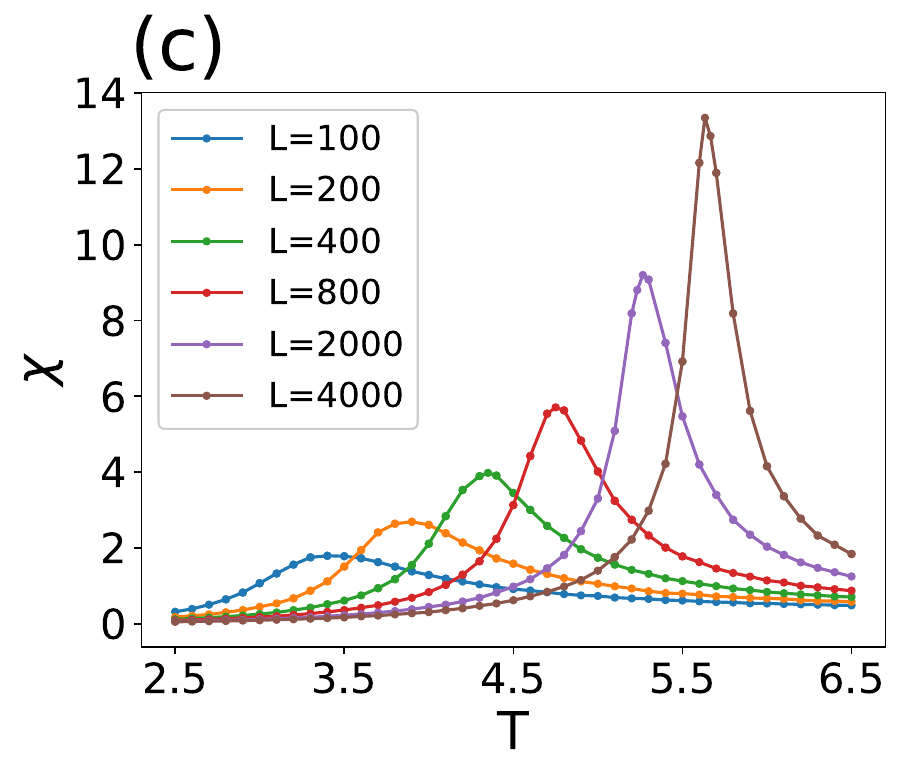}\hfill
\includegraphics[width=0.2415\linewidth]{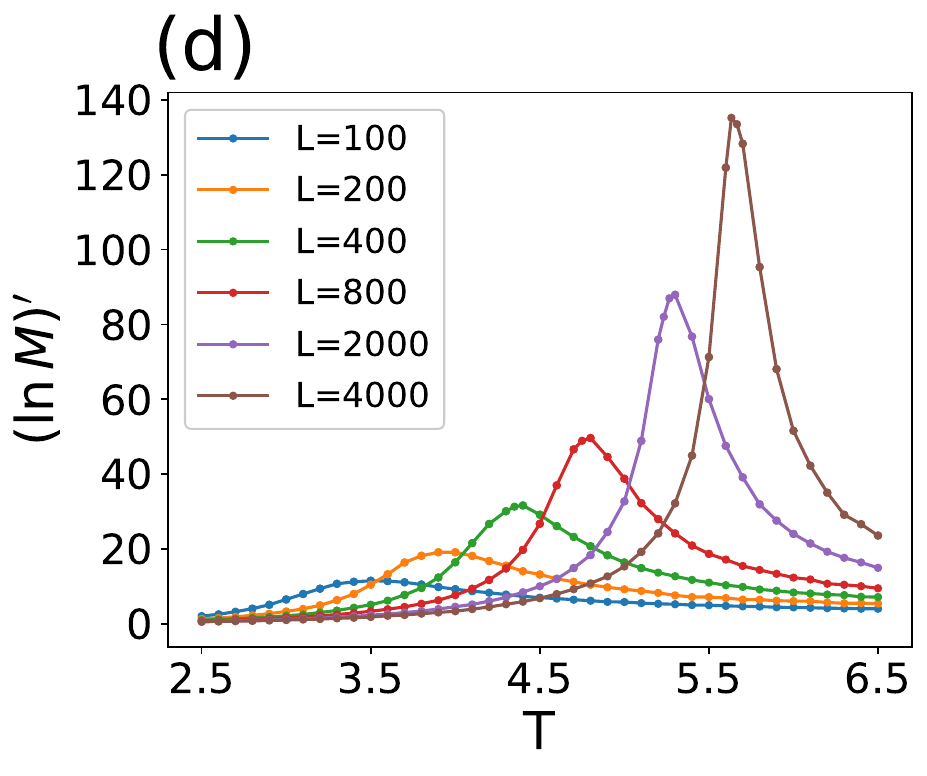}
\bigskip 
\includegraphics[width=0.245\linewidth]{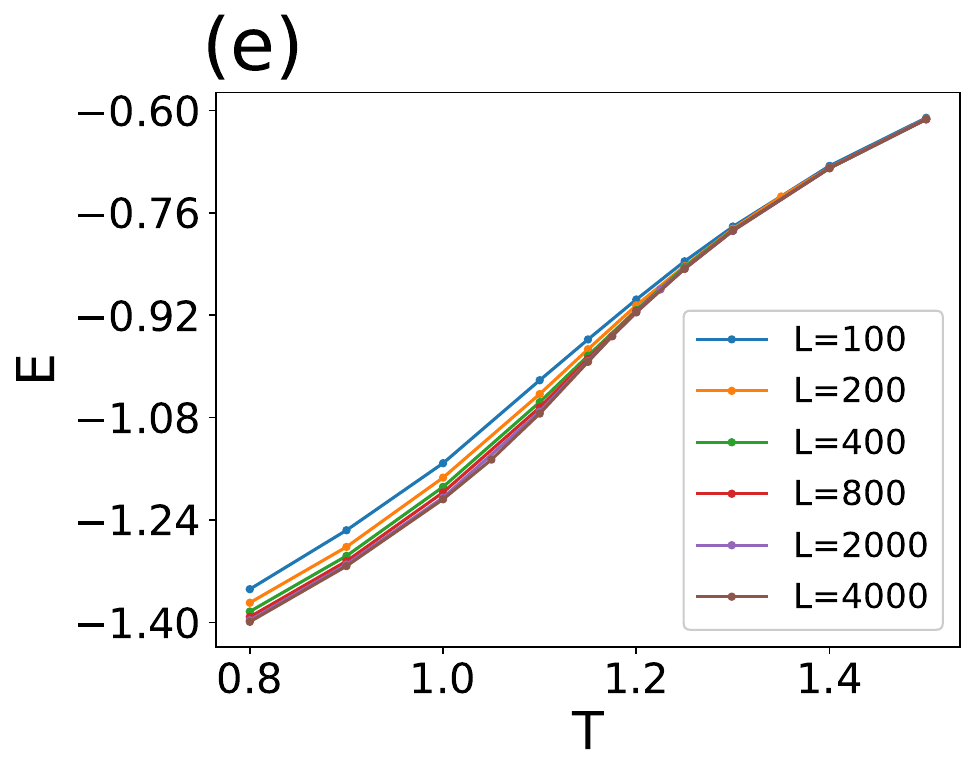}\hfill
\includegraphics[width=0.24\linewidth]{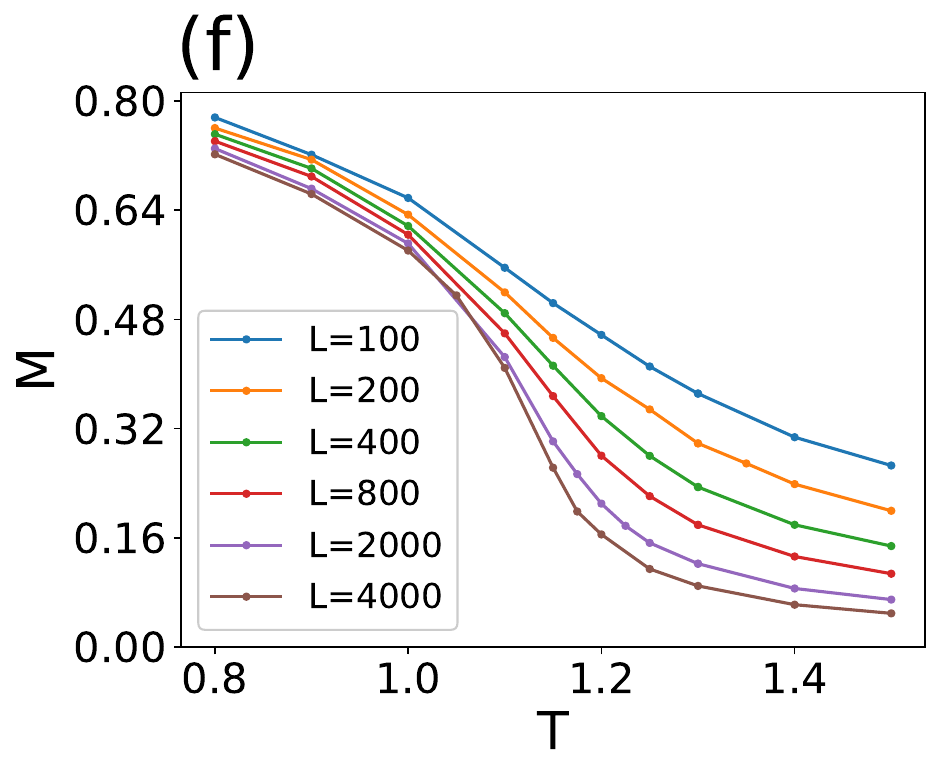}\hfill
\includegraphics[width=0.236\linewidth]{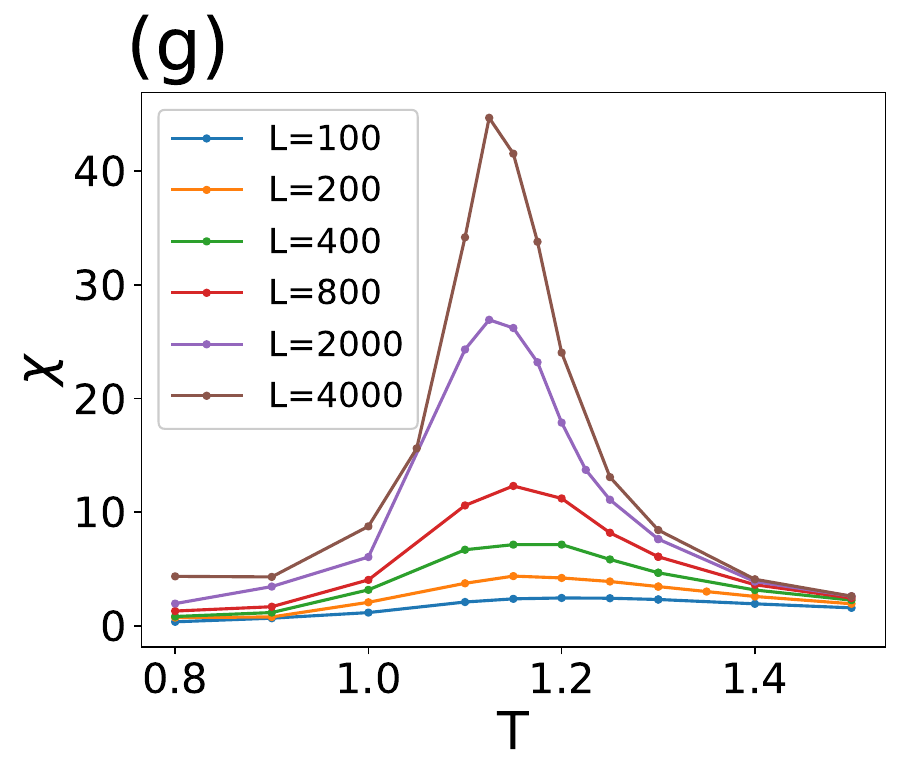}\hfill
\includegraphics[width=0.236\linewidth]{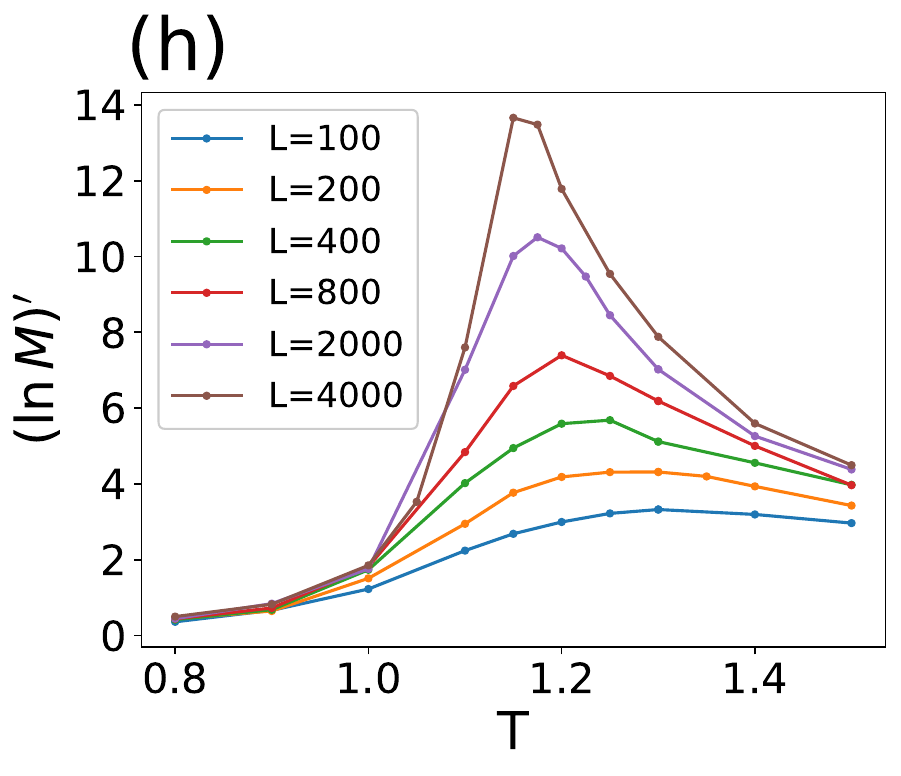}
\caption{The energy $E$, the magnetization $M$, the susceptibility $\chi$, and $(\ln M)'$ as a function of temperature for various system sizes $L$ in the 1D LR XY model. Upper and lower panels show the cases for $\sigma=1.1$ and $1.8$, respectively.}
\label{fig:fig4}
\end{figure*}

\subsubsection{1D LR XY model: \red{Order-disorder} phase transitions at high temperatures}
\label{subsubsec:1d_lr_xy_model_highT}

To obtain the complete phase diagram of the LR XY model in the non-MWH regime, we further
extracted the critical exponents \(\upsilon\) and \(\gamma\) for the EnLRO to disorder phase transition using finite size scaling analysis, and estimated the corresponding transition temperature $T_c$ (refer to Eqs. (\ref{eq:maxdUn})-(\ref{eq:Tc})). In the following, we
take the cases of \(\sigma = 1.1\) and \(\sigma = 1.8\) as examples to demonstrate the analysis procedures. The same analysis was applied to other values of $\sigma$ to obtain the phase diagram in Fig. \ref{fig:fig1}(b). ~\\

Figure \ref{fig:fig4}(a) shows that the energy at \(\sigma = 1.1\) exhibits a strong size effect. Specifically, over a broad temperature range, the energy undergoes significant changes as the system size increases. In contrast, at \(\sigma = 1.8\), the energy is much less affected by the system size as shown in Fig. \ref{fig:fig4}(e). This result is consistent with previous studies \cite{bayong1999effect}. At \(\sigma=1.1\), the interplay between long-range spin interactions and thermal fluctuations results in numerous intersection points in the magnetization curves for different system sizes near the critical temperature $T_{c}$ (Fig. \ref{fig:fig4}(b)). \blue{When the temperature is below the phase transition temperature, the system is in the EnLRO regime where the magnetization \( M \) increases monotonically with the system size \( L \).} 
On the other hand, at $\sigma=1.8$, \blue{the range of interaction is smaller and thermal fluctuations becomes more significant, the system consistently exhibits a ReLRO scaling behavior where the magnetization decreases with an increased system size.} (Fig. \ref{fig:fig4}(f)). ~\\

Figures \ref{fig:fig4} (c-d) and (g-h) display the susceptibility and $(\ln M)'$ for \(\sigma\)=1.1 and \(\sigma\)=1.8, respectively. It is evident that while the temperature at which $\chi$ and $(\ln M)'$ peak increases with the system size for $\sigma=1.1$, there are no noticeable changes for the case of $\sigma=1.8$. Similar observations for the latter case are also found in previous studies on the long-range interacting models \cite{gonzalez2021finite,scalettar1991critical,nakada2011crossover}. 
\blue{The behavior of the quantities at $\sigma = 1.8$ in Fig.~\ref{fig:fig4} is similar to that of the ordered phase in a nearest-neighbor interacting \red{2D XY model}, \red{the shift in \(T_c\) depends only weakly on the system size, and \(T_c\) decreases as the size increases.} This is consistent with the finding in the previous section that the correlation function at $\sigma = 1.9$ resembles that of the ordered phase in a \red{2D nearest-neighbor interacting XY model}. \red{In both cases, the decay of the correlation function is algebraic and cause both system have similar transition behavior, even if \green{in the 1D case, the correlation function} approaches a nonzero constant and \green{in the 2D nearest-neighbor interaction case, it} decays to zero.}} 


\begin{figure}[t!]
\centering
\begin{minipage}{\linewidth}
  \includegraphics[width=0.985\linewidth]{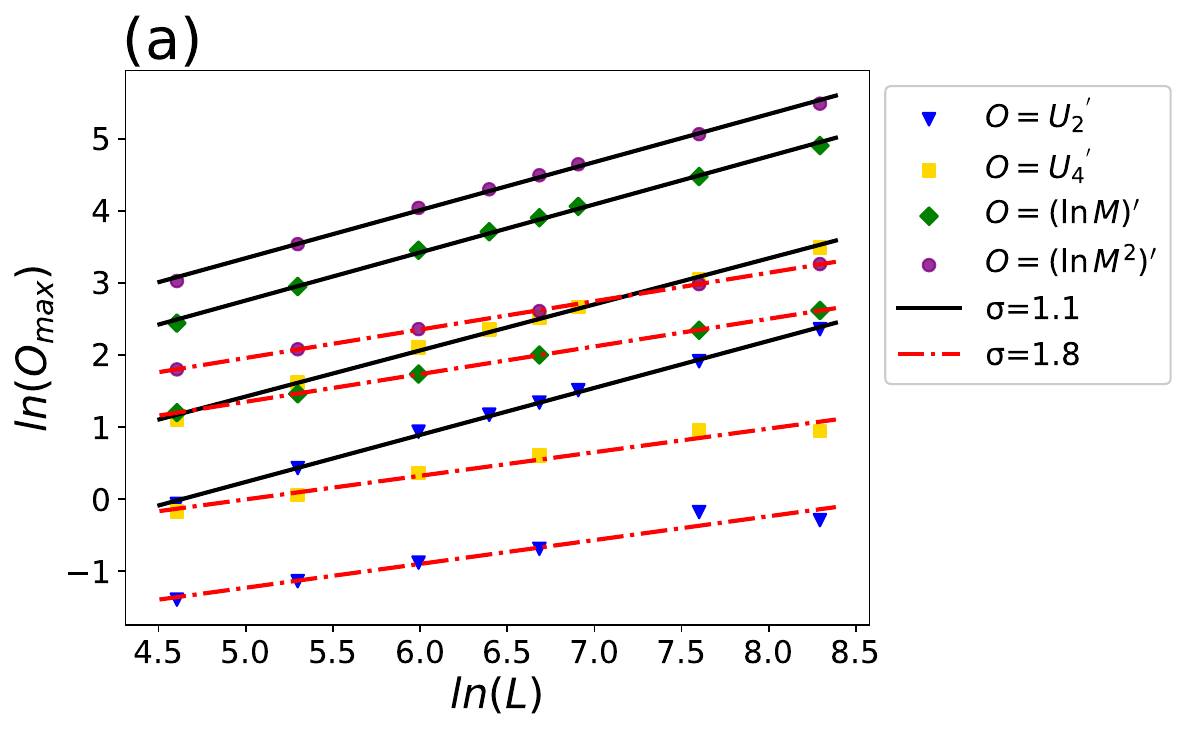}
\end{minipage}
\bigskip 
\begin{minipage}{\linewidth}
  \includegraphics[width=0.49\linewidth]{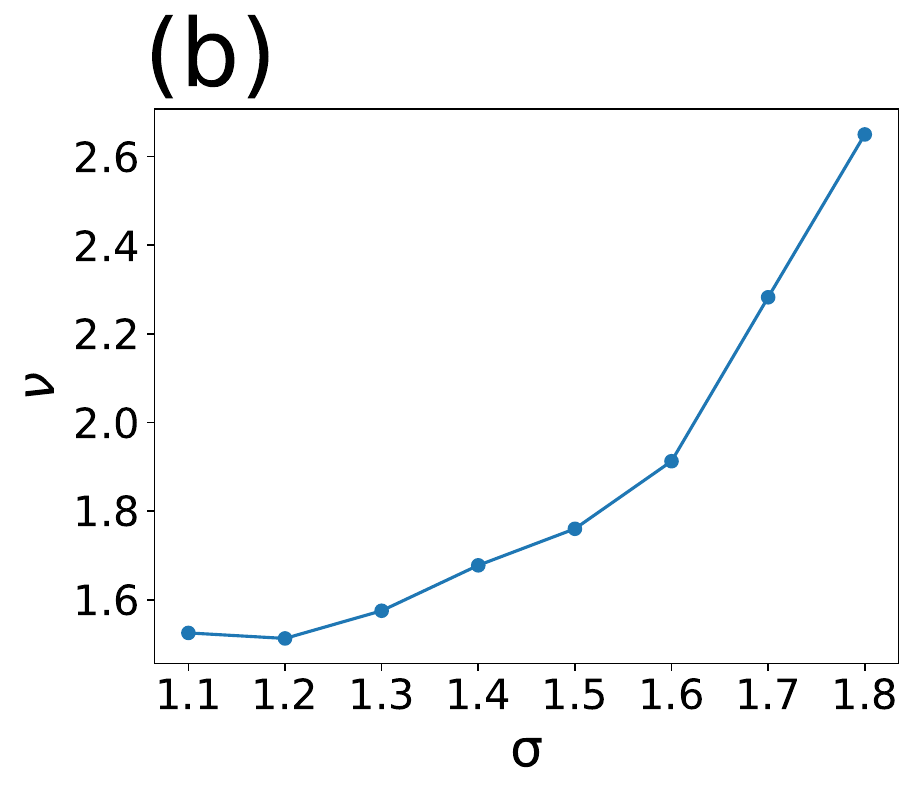}\hfill
  \includegraphics[width=0.49\linewidth]{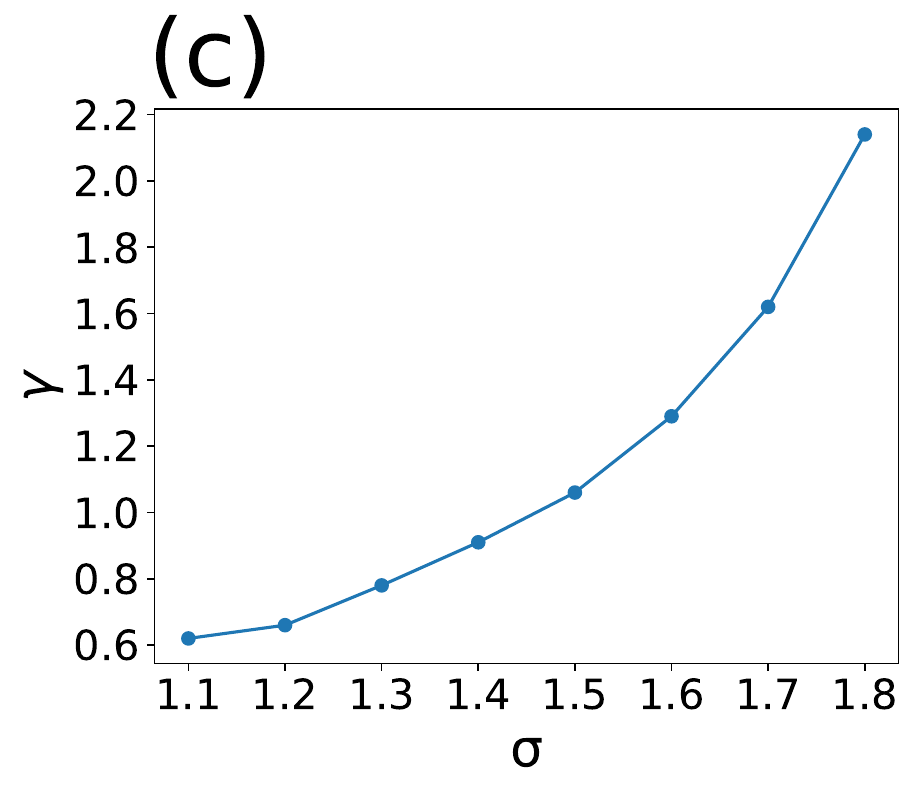}
\end{minipage}
\caption{(a) The figure shows the ln-ln plot of the relationship between the maximum values of measured quantities as a function of temperature and the system size in 1D LR XY model. The black and red straight lines represent the linear fit to each corresponding set of data for $\sigma=1.1$ and $\sigma=1.8$, respectively. 
(b, c) The figures respectively show the critical exponents \(\upsilon\) and \(\gamma\) for the transition between the true long-range ordered phase and the disordered phase.}
\label{fig:fig5}
\end{figure}

We further measured \({U_{2}}^{'}\), \({U_{4}}^{'}\), and $(\ln M^{2})'$ as a function of temperature for various system sizes. The size dependence of the maximum of these measured quantities, together with that of $(\ln M)'$, is plotted in Fig. \ref{fig:fig5}(a). The black and red lines represent the linear fits to the corresponding set of data points for \(\sigma\)=1.1 and \(\sigma\)=1.8, respectively. \blue{A near-perfect linear relationship is observed between the logarithm of the maximum values of the measured quantities and $\ln L$. The fitted lines across different measured quantities have the same slope indicates that the measured quantities share the same critical exponent $\upsilon$ as expected.} 

Utilizing Eqs. (\ref{eq:maxdUn}) and (\ref{eq:maxlnM}), we estimated the critical exponents to be \(\upsilon\)(\(\sigma\)=1.1)\(\approx\)1.525 and \(\upsilon\)(\(\sigma\)=1.8)\(\approx\)2.65. Following the same methodology, we also estimated \(\gamma\) for \(\sigma\in [1.2, 1.7]\), as shown in Fig. \ref{fig:fig5}(b). Substituting these \(\upsilon\) back to Eqs. (\ref{eq:maxchi}) and (\ref{eq:Tc}), we deduced the values of \(\gamma\) and \(T_{c}(L = \infty)\) for different \(\sigma\). The results are presented in Fig. \ref{fig:fig5}(c) and the phase diagram in Fig. \ref{fig:fig1}(b). The phase diagram in Fig. \ref{fig:fig1}(b) indicates that as \(\sigma\) approaches 1, $T_c$ shows signs of divergence. This is consistent with our expectation that for \(\sigma \leq 1\), every spin interacts with all other spins with a non-zero strength in the thermodynamic system and thus the ordered phase is maintained at all temperatures. \red{In the regime \(\sigma>2\), low‑temperature spin‑wave excitations destroy the global magnetization at any finite temperature \cite{bruno2001absence}. Consequently, in the narrow range \(1.9<\sigma<2.1\) just outside our phase diagram, there is a boundary between ReLRO and short-range order, however we do not show this boundary because it is not the focus of our study. }

\begin{figure} [t!]
\centering
\begin{minipage}{\linewidth}
  \includegraphics[width=0.79\linewidth]{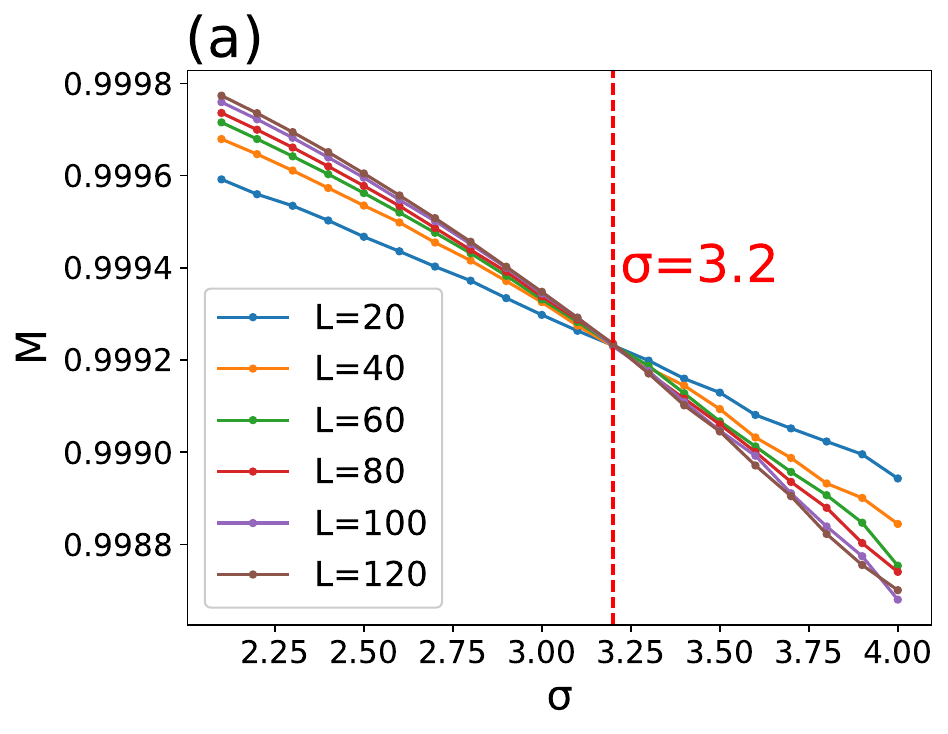}
\end{minipage}
\bigskip 
\begin{minipage}{\linewidth}
  \includegraphics[width=0.49\linewidth]{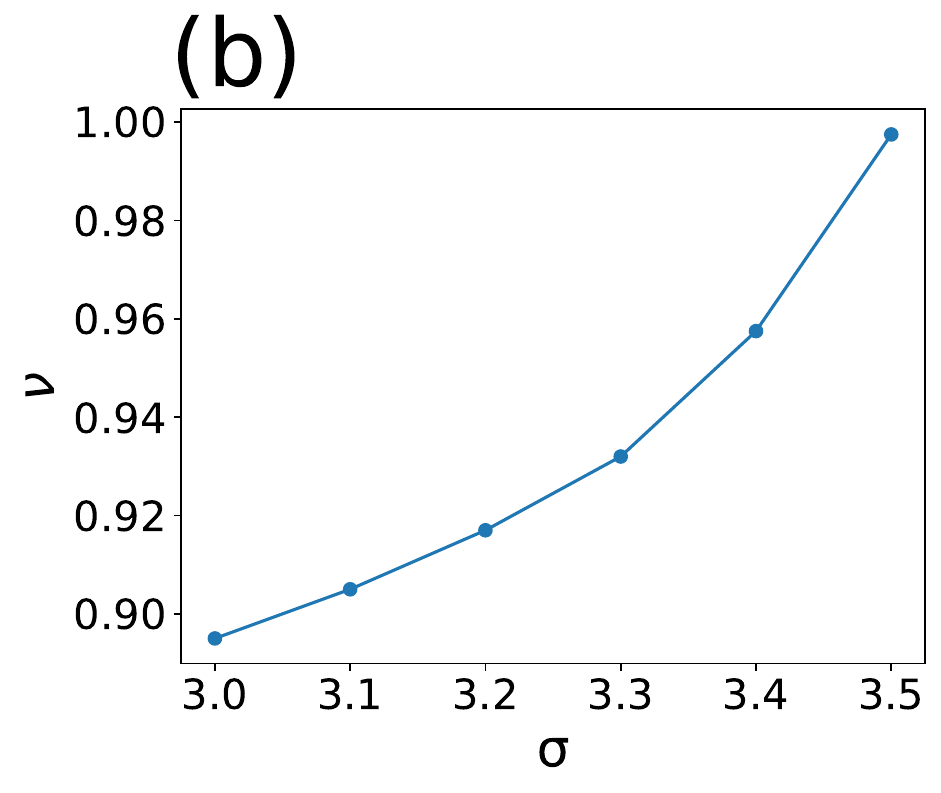}\hfill
  \includegraphics[width=0.49\linewidth]{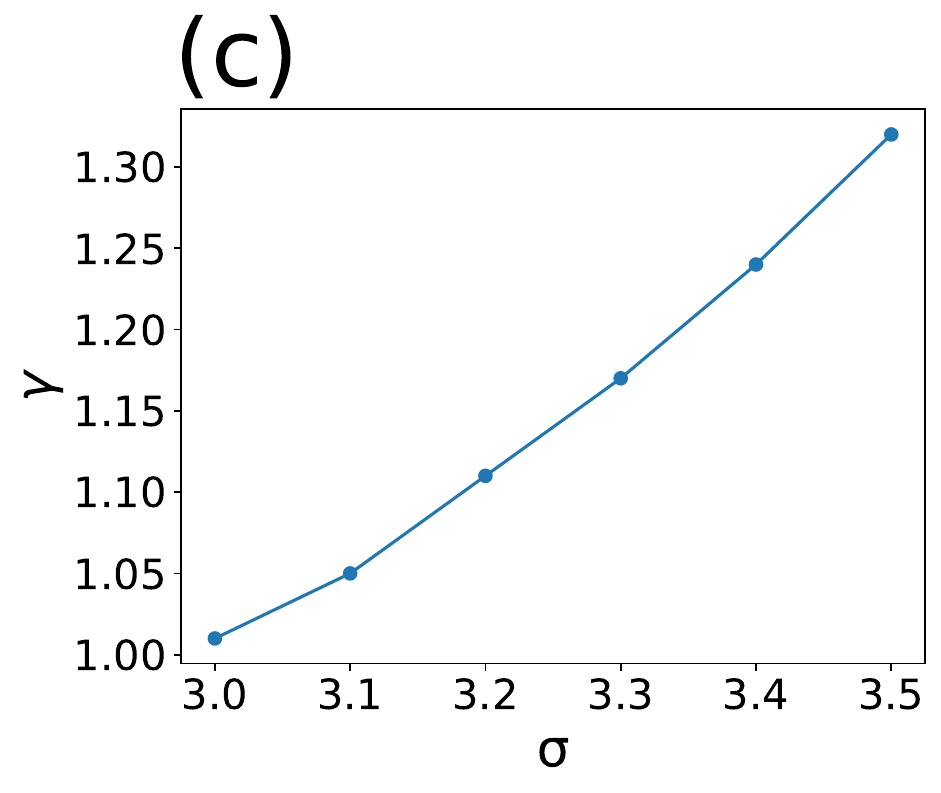}
\end{minipage}
\caption{(a) The figure shows the variation of magnetization $M$ with $\sigma$ at $T=0.01$
for different system sizes in 2D LR XY model. The red vertical dashed line identifies the
EnLRO-ReLRO crossover. (b, c) The figures respectively show the critical exponents $\nu$ and $\gamma$
for the phase transition between the true long-range ordered phase and the disordered phase.}
\label{fig:fig6}
\end{figure}

\subsubsection{2D LR XY model}
\label{subsubsec:2d_lr_xy_model}

Following the approach mentioned in the above section, we conducted similar
measurements on the LR XY model on a 2D square lattice. As depicted in Fig. \ref{fig:fig6}(a), at
low temperature ($T$=0.01), the $M$-\(\sigma\) curves for different system
sizes intersect around \(\sigma_c \approx 3.2\). Below $\sigma_c$, the magnetization increases with system size, while above $\sigma_c$, the trend is reversed. This observation is similar to the results shown in Fig. \ref{fig:fig2}(f), indicating that the 2D LR XY
model also undergoes a scaling behavior crossover from EnLRO to ReLRO. Fig. \ref{fig:fig6}(b) and (c) present the critical exponents \(\upsilon\) and \(\gamma\)
associated with the phase transition from the ordered phase (both EnLRO and ReLRO) to the disordered phase for various $\sigma$.~\\

Previous work, utilizing renormalization group analysis, has
demonstrated that the 2D LR XY model exhibits
only true long-range ordered and disordered phases for \(\sigma < 3.75\) \cite{giachetti2021berezinskii}. Integrating
this finding with our results, we constructed a phase diagram for
\(3 < \sigma < 3.5\), as shown in Fig. \ref{fig:fig1}(c). Note that the $T_c$ at $\sigma\approx3.5$ is consistent with that obtained in Ref. \cite{giachetti2021berezinskii}.
At low temperatures, the
system undergoes a scaling behavior crossover from the EnLRO to ReLRO phases at \(\sigma_{c} \approx 3.2\),
and \(\sigma_{c}\) gradually increases to 3.5 with rising temperature, mirroring the behavior observed in 1D systems. However, as \(\sigma\) decreases, \(T_{c}\) appears to increase algebraically instead of diverging, unlike that in the 1D case (Fig. \ref{fig:fig1}(b)). This could be due to the insufficiently small \(\sigma\) values being considered. 
Overall, the phase diagram for the 2D case is qualitatively similar to that in the 1D case. \red{Similar to the 1D XY model, there is a boundary between ReLRO and quasi-long-range order in the range \(3.9<\sigma<4.1\), which we also omit from our diagram \cite{bruno2001absence}.}~\\

\subsection{LR Heisenberg model}
\label{subsec:lr_heisenberg_model}
The Heisenberg model allows for spin orientation in any direction within three-dimensional space, thereby increasing the degrees of freedom and leading to more complex spin configurations. Due to this variety in spin alignment, the Heisenberg model is more susceptible to temperature variations compared to the XY model, resulting in more intense thermal disturbances. This implies that the Heisenberg model with nearest-neighbor interactions lacks quasi-long-range order even at low temperatures, exhibiting instead short-range order, as the correlation function decays exponentially with increasing $r$ \cite{shenker1980monte,holm1993critical}. This difference may affect the system behavior under long-range interactions.~\\

\begin{figure} [t!]
\includegraphics[width=0.47\linewidth]{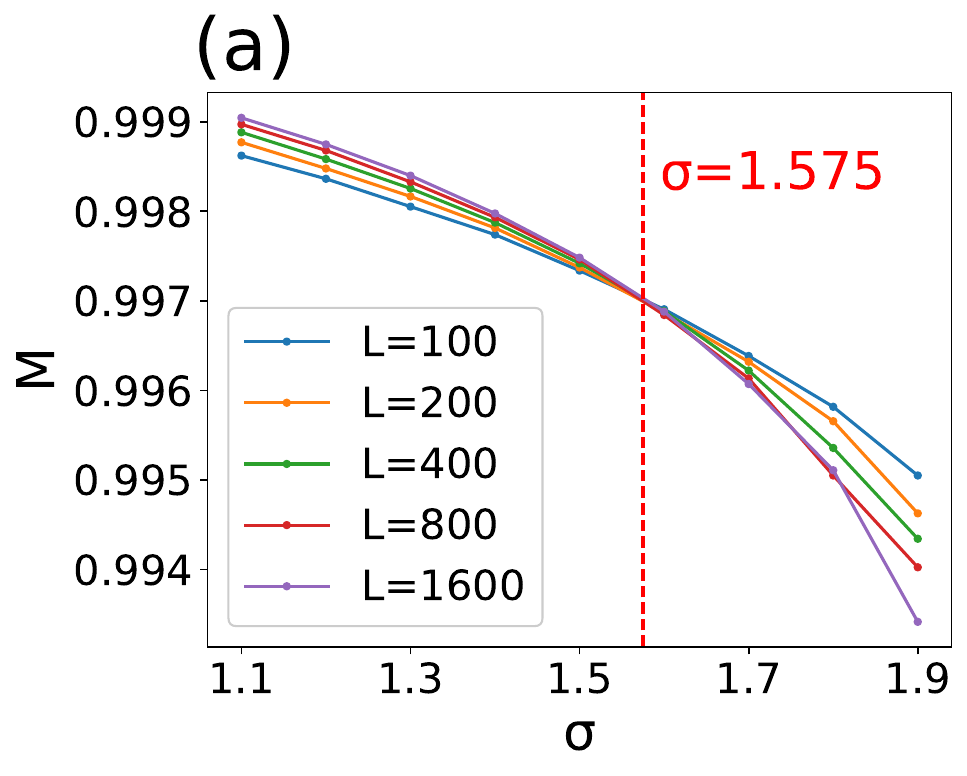}
\includegraphics[width=0.49\linewidth]{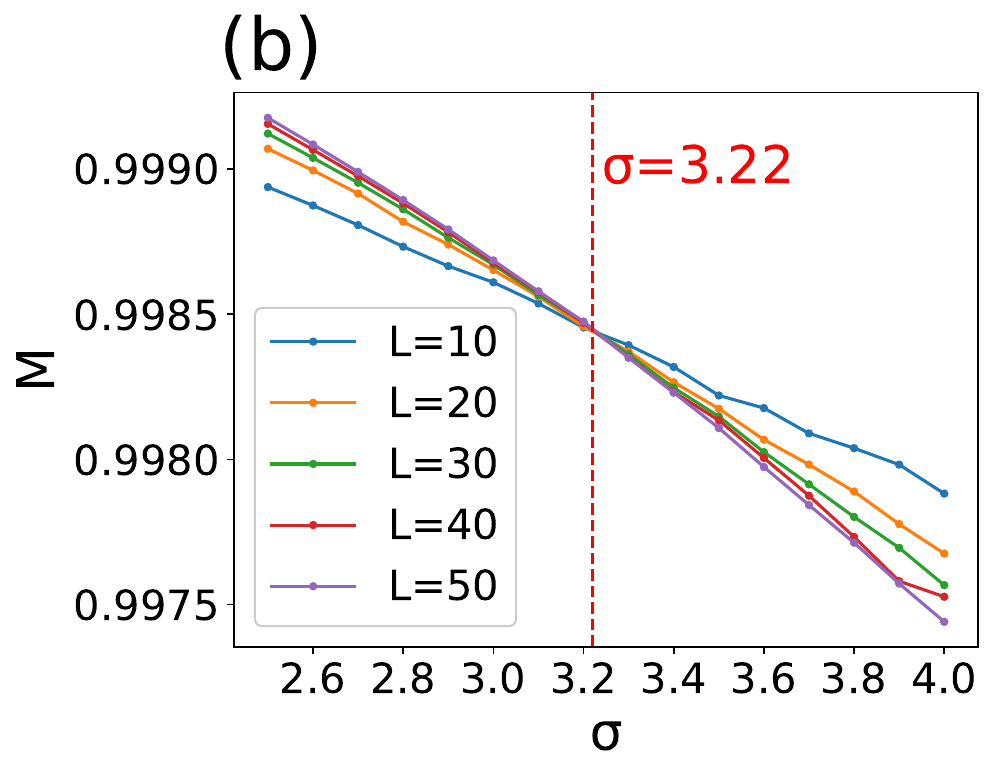}
\caption{(a, b) The variation of magnetization $M$ with $\sigma$ at $T=0.01$ for the 1D and 2D Heisenberg models, respectively. A noticeable collapse of data points is observed at $\sigma=1.575$ for the 1D model
and at $\sigma=3.22$ for the 2D model, indicating a EnLRO-ReLRO crossover.}
\label{fig:fig7}
\end{figure}

To investigate whether true long-range order exists in the Heisenberg model under long-range interactions, we analyzed the magnetization versus \(\sigma\) curves on both a 1D chain and a 2D square lattice, as illustrated in Fig. \ref{fig:fig7}. Similar to the XY model, both the 1D and the 2D LR Heisenberg models exhibit a EnLRO-ReLRO crossover within the non-MWH regime, as indicated by the intersection points of the curves at \(\sigma = 1.575\) and \(\sigma = 3.22\). We noted that \(\sigma_{c}\) of the 1D Heisenberg model is exactly the same as that of the 1D XY model, while in the 2D case, the \(\sigma_{c} = 3.22\) of the Heisenberg model is slightly larger than the \(\sigma_{c} = 3.20\) of the XY model. If we consider this small difference as a negligible numerical error, the results suggest the EnLRO-ReLRO crossover may be generic in continuous spin models and is insensitive to the change in the degrees of freedom of the spins. The similarity in the EnLRO-ReLRO crossover between the XY model and the Heisenberg model may be attributed to their shared continuous rotational symmetry. On the other hand, for a particular model with fixed spin degrees of freedom, the EnLRO-ReLRO crossover boundary is more dependent on the temperature, as shown in the phase diagram of the XY model in Fig. \ref{fig:fig1} (b) and (c).~\\

\section{Discussion}
\label{sec:discuss_results}

 \red{In Section~\ref{subsubsec:1d_lr_xy_model_lowT}, we observe that at low temperature ($T = 0.01$), varying $\sigma$ switches the one-dimensional XY model between EnLRO and ReLRO. In the large-$\sigma$ region ($1.575 \le \sigma < 1.7$), varying the temperature $T$ also induces transitions between EnLRO and ReLRO. We attribute this behavior to the competition between spin alignment and thermal fluctuations. To test this hypothesis, we divide the ordered phase in the phase diagram of Fig.~\ref{fig:fig1}(b) into four regimes: low-$T$ \green{\&} small-$\sigma$, low-$T$ \green{\&} large-$\sigma$, high-$T$ \green{\&} small-$\sigma$, and high-$T$ \green{\&} large-$\sigma$.}~\\

\subsection{Low-temperature \green{ordered phase} in 1D XY model}
\label{subsec:lowT_1D}

\begin{figure} [t!]
\includegraphics[width=0.485\linewidth]{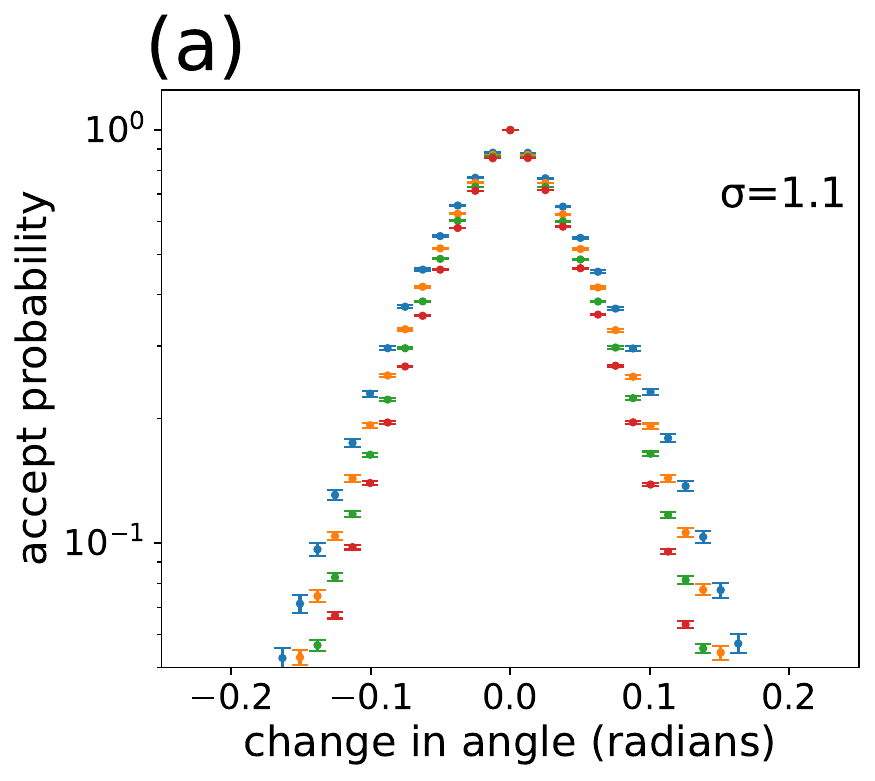}
\includegraphics[width=0.485\linewidth]{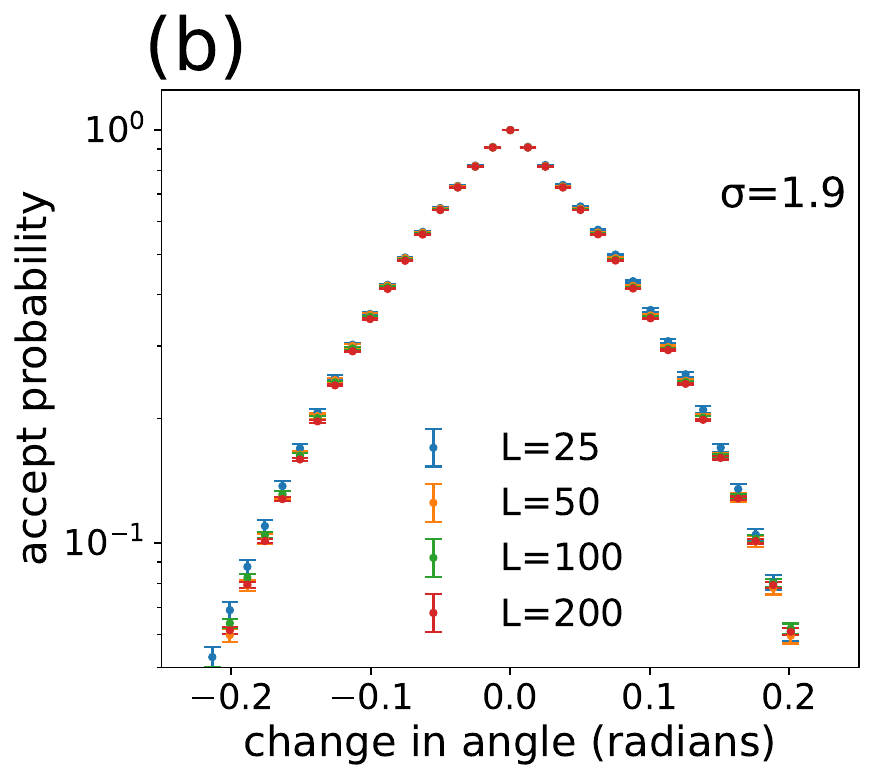}
\caption{Monte Carlo acceptance rate as a function of the flip angle in the 1D XY model with $T=0.01$ at $\sigma=1.1$ (a) and $\sigma=1.9$(b).}
\label{fig:fig3_3}
\end{figure}

\red{In Appendix~\ref{sec:1D_clock}, we analyze the 1D clock model with $q$ discrete spin angles. At $T=0.01$, Figures~\ref{fig:appendixD}(a) and (b) show that smaller $q$ suppresses spin‐wave excitations and shortens their lifetime. Since $q$ controls the strength of thermal fluctuations \green{due to spin-wave excitations} and EnLRO versus ReLRO arises from the competition between spin alignment and these fluctuations, we expect that tuning $q$ shifts the EnLRO–ReLRO crossover. Figures~\ref{fig:appendixD}(c)–(f) confirm this expectation.}~\\

\red{Returning to the 1D XY model at $T=0.01$, tuning $\sigma$ also alters the system’s response to thermal fluctuations and thus the EnLRO–ReLRO \green{crossover} appear. Consider the ground state where all spins are aligned. If one spin couples to a heat bath, thermal fluctuations deflect its angle by $\Delta\theta$. In a small-$\sigma$ system, this spin interacts with more aligned spins, so the energy cost of a flip is higher than in a large-$\sigma$ system. Therefore, the flip probability is lower and spin waves are harder to excite. Even if a large angle flip occurs on a lucky spin by chance, its energy is diluted in the long-range spin-spin coupling, in other words, a nearest-neighbor spin lowers the system energy by aligning not only with the lucky spin but also with distant spins, this reduces the lucky spin's effect on its neighbors; therefore lowering \(\sigma\) shortens the spin-wave lifetime and suppresses long-range propagation.}~\\

\red{To verify this, we measure Monte Carlo acceptance rates for different flip angles at $\sigma=1.1$ and $\sigma=1.9$ after equilibrium. Figure~\ref{fig:fig3_3} shows that for $\sigma=1.1$, the acceptance rate decays markedly as $L$ increases, consistent with suppressed excitations and propagation, while for $\sigma=1.9$ it shows little change. This demonstrates that decreasing $\sigma$ of a low temperature system strongly suppresses the system’s response to thermal fluctuations and drives the \green{crossover} from ReLRO to EnLRO.}~\\

\subsection{High-temperature \green{ordered phase} in 1D XY model}
\label{subsec:highT_1D}

\begin{figure} [t!]
\includegraphics[width=0.88\linewidth]{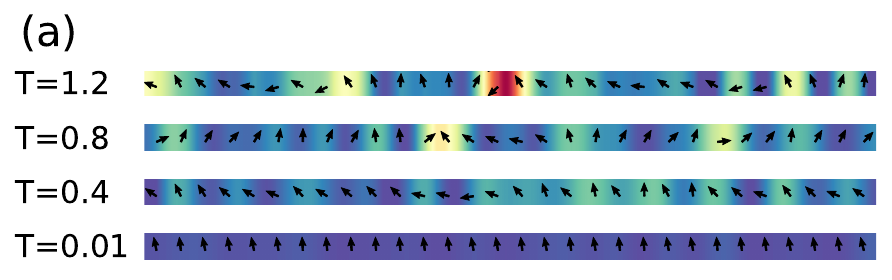}
\includegraphics[width=0.099\linewidth]{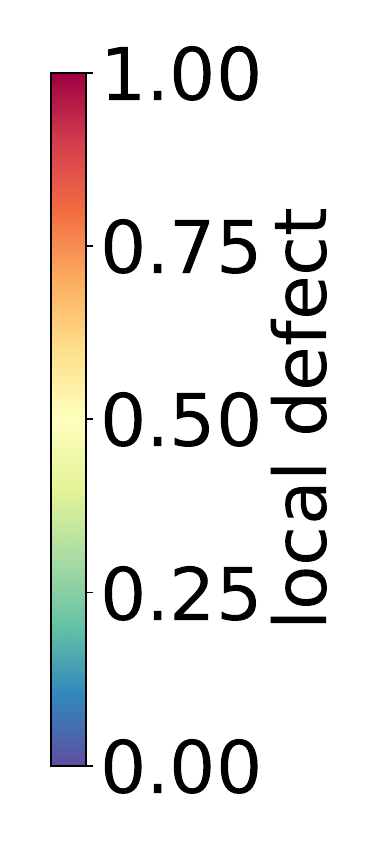}
\includegraphics[width=0.48\linewidth]{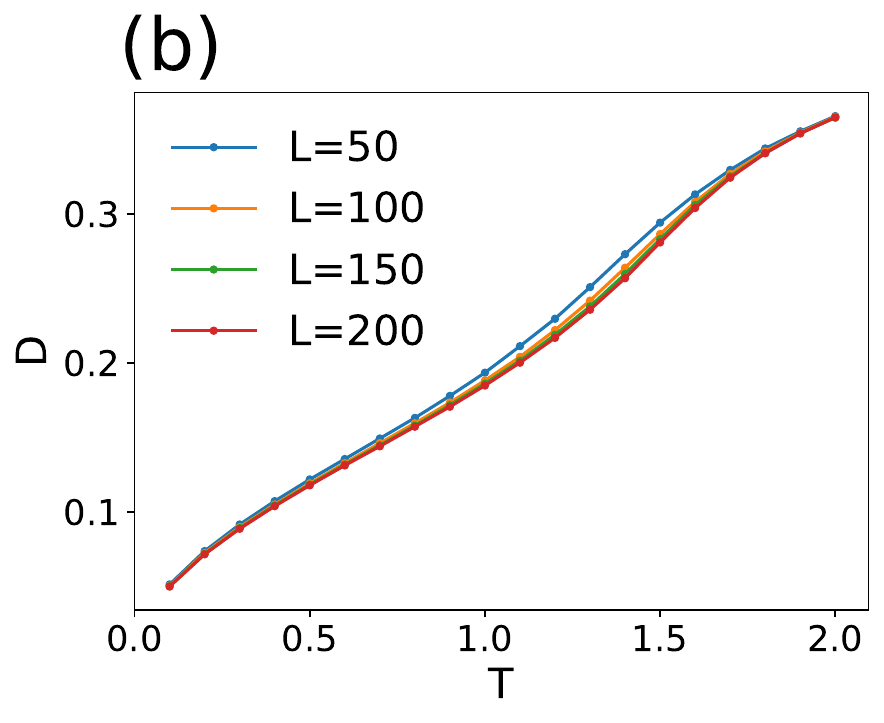}
\includegraphics[width=0.48\linewidth]{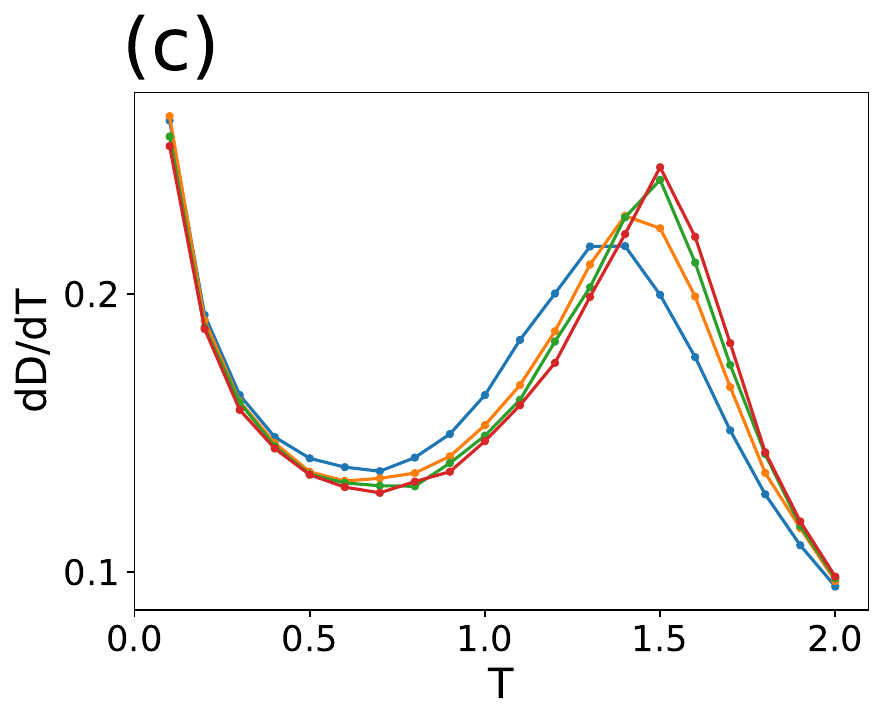}
\includegraphics[width=0.48\linewidth]{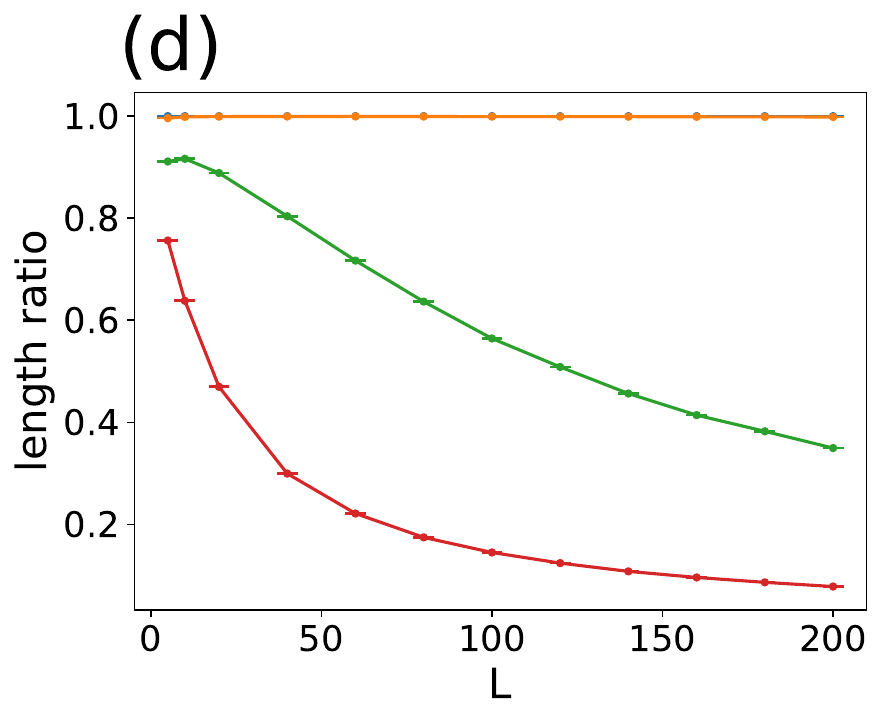}
\includegraphics[width=0.48\linewidth]{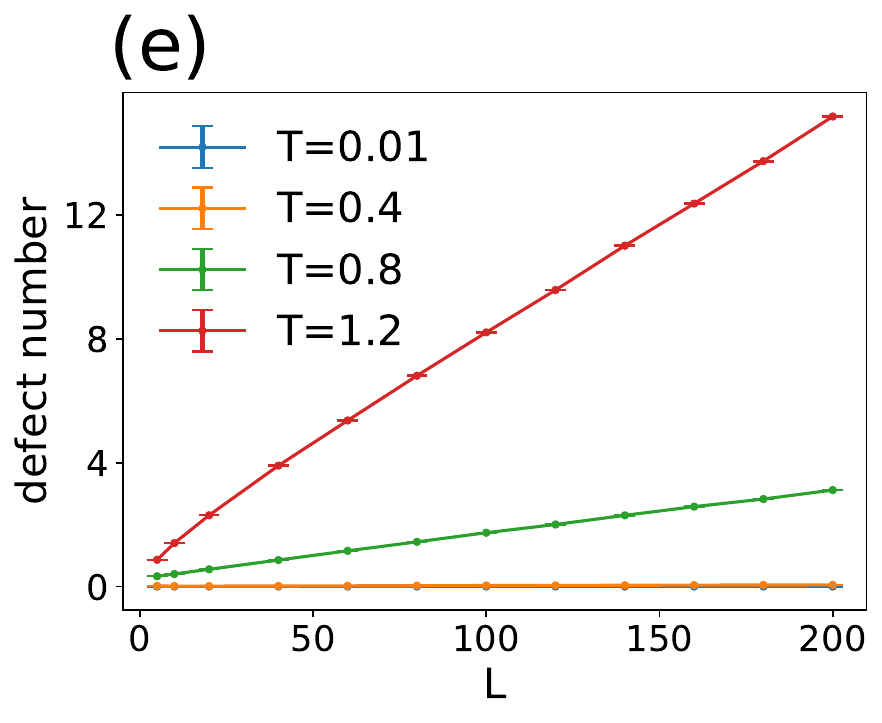}
\includegraphics[width=0.9\linewidth]{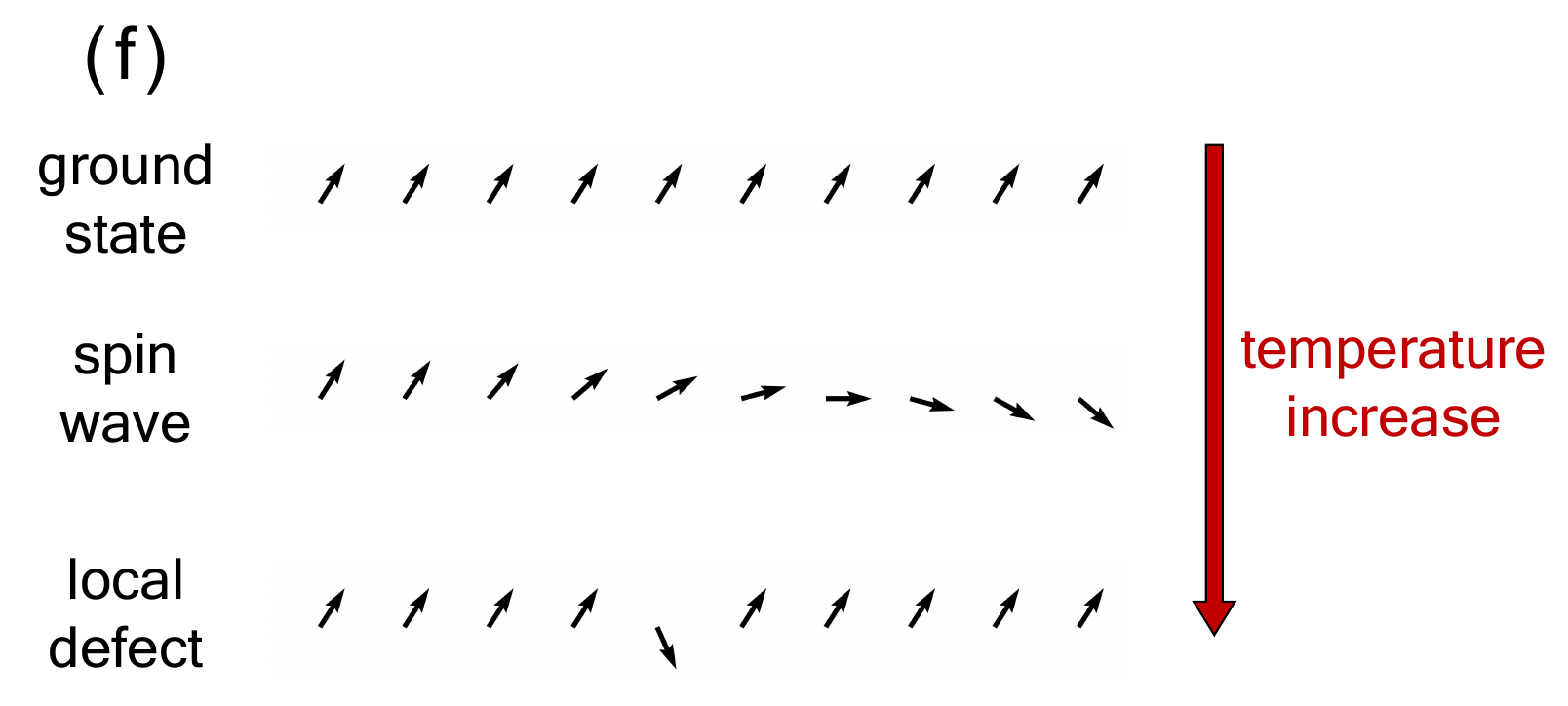}
\caption{Figure (a) shows the equilibrium spin configuration of the 1D XY model at $\sigma=1.6$ for $T=0.01$, $0.4$, $0.8$, and $1.2$, where the background color indicates the local defect $D$. Figure (b) and (c) shows the average defect $D$ and the its first derivative as a function of the temperature for different system sizes, respectively. Figures (d) and (e) show the ratio of subsystem length to total length and the number of effective defects as functions of system size at various temperatures. Figure (f) gives a schematic drawing in which the ground-state spins are disrupted by spin waves, spin waves are also disrupted by local defects as temperature increase.}
\label{fig:fig3_4}
\end{figure}

\red{In high temperatures and large-$\sigma$  ($1.57<\sigma<1.7$) region, raising the temperature induces a \green{crossover} from ReLRO to EnLRO. This cannot be explained by the low-temperature picture of competition between spin-wave \green{excitations} and spin alignment, since increasing $T$ normally strengthens spin waves and suppresses alignment.}~\\

\red{By examining equilibrium spin configurations at $\sigma=1.6$ and $L=30$ (Fig.~\ref{fig:fig3_4}(a)), we find that thermal fluctuations produce not only spin waves but also local defects at high temperature. We define the defect between neighbors as $D_{ij}=\min\bigl(|\theta_i-\theta_j|,2\pi-|\,\theta_i-\theta_j|\bigr)/\pi$. At $T=0.01$ and $0.4$, spins remain aligned and no strong defects appear. At $T=0.8$, defects (yellow-white color region in the plot) appear and break the chain into subsystems: spins to the left of a defect are still well aligned, while those to the right show strong spin-wave distortion. At $T=1.2$, more and stronger defects appear, further break the chain.  }~\\

\red{To quantify this effect, we measure the average defect per spin $D$ and its temperature derivative $dD/dT=(L/T^{2})(\langle eD\rangle-\langle e\rangle\langle D\rangle)$ at $\sigma=1.6$ (Fig.~\ref{fig:fig3_4} (b,c)). As $T$ increases, \green{$D$ increases with a slowdown in growth around the ReLRO–EnLRO crossover.}
Correspondingly, $dD/dT$ shows a local minimum \green{around ReLRO–EnLRO crossover} followed by a local maximum \green{near the order-disored transition}. As $L$ increases, the minimum tends to zero and the maximum diverges. 
Data show that for sufficiently large systems, $D$ as a function of temperature $T$ exhibits a plateau. To the left of this plateau, the system is mainly influenced by spin waves, while to the right the rapid growth of $D$ indicates the appearance of defects, suggesting both spin waves and defects affect the system's behavior.}~\\

\red{We also analyze how the subsystem length and the number of defects scale with the system size. We call a defect “effective” if $D_{ij}>0.5$, and we measure the lengths of defect‐free subsystem under periodic boundaries. Figure~\ref{fig:fig3_4} (d) shows that at low $T$, no effective defects appear, so the ratio of subsystem length and total system length remains unity; Fig.~\ref{fig:fig3_4} (e) shows the number of defect stays zero. Above a threshold $T$, the number of defect grows nearly linearly with $L$, while the subsystem length ratio decays algebraically towards zero. Small error bars indicate that the defects are uniformly distributed.  }~\\

\red{These data support the hypothesis that at high $T$, although spin-wave excitation and propagation still reduce long-range correlation, local defects block spin-wave propagation so that spin waves affect only nearby spins. As $L$ increases, defects proliferate faster than spin-wave disturbances grow, so large systems experience less disturbances and local order is enhanced, this enhancement increasing the magnetization and producing EnLRO. In other words, when a ground-state chain couples to a heat bath, spin waves break the spin alignments but the defects can block their propagation (Fig.~\ref{fig:fig3_4}f).  }~\\

\red{In the range $1.57<\sigma<1.7$, the local minimum of $\frac{dD}{dT}$ does not always coincide with the ReLRO to EnLRO \green{crossover} temperature. According to our hypothesis, this \green{crossover} arises from the competition between spin waves and defects, but $D$ and $\frac{dD}{dT}$ only reveal their combined effect and cannot show which one dominates as $T$ increases. }~\\

\red{Even for $\sigma<1.57$ or $\sigma>1.7$, defects still appear and $\frac{dD}{dT}$ presents a local minimum followed by a local maximum as $T$ rises, yet no ReLRO to EnLRO crossover appears in the phase diagram. We conjecture that for very small $\sigma$,  the spin-wave disturbances breaked by defect always weaker than in small systems, so the system stays in EnLRO throughout the ordered regime. For very large $\sigma$, the opposite holds, and the system stays in ReLRO throughout the ordered regime.}~\\

\subsection{High-temperature region in 2D XY model}
\label{subsec:highT_2D}

\begin{figure} [t!]
\includegraphics[width=0.485\linewidth]{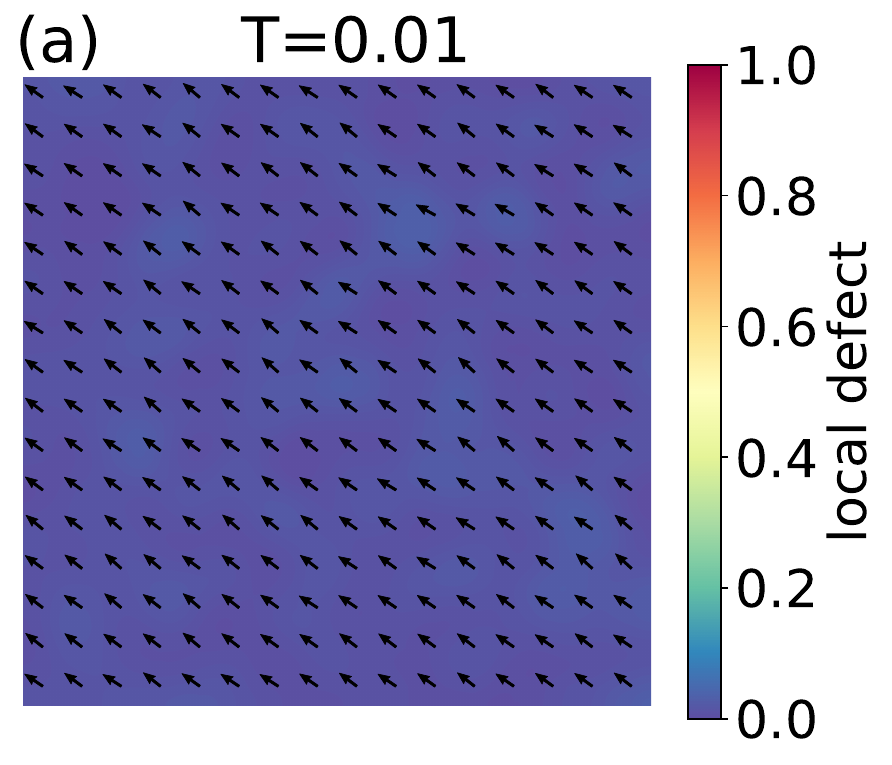}
\includegraphics[width=0.485\linewidth]{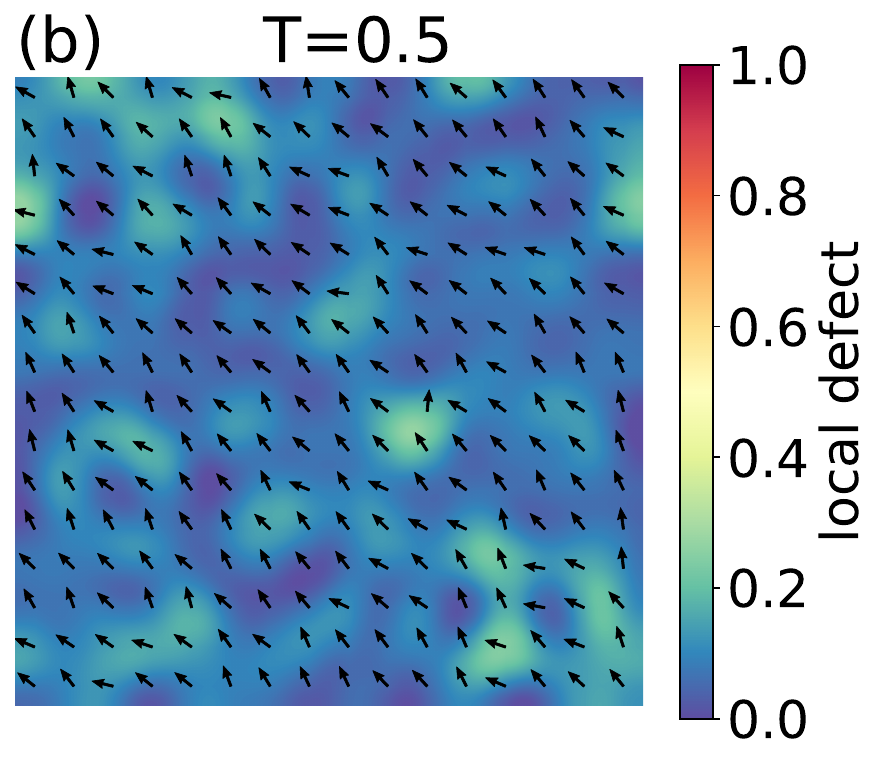}
\includegraphics[width=0.485\linewidth]{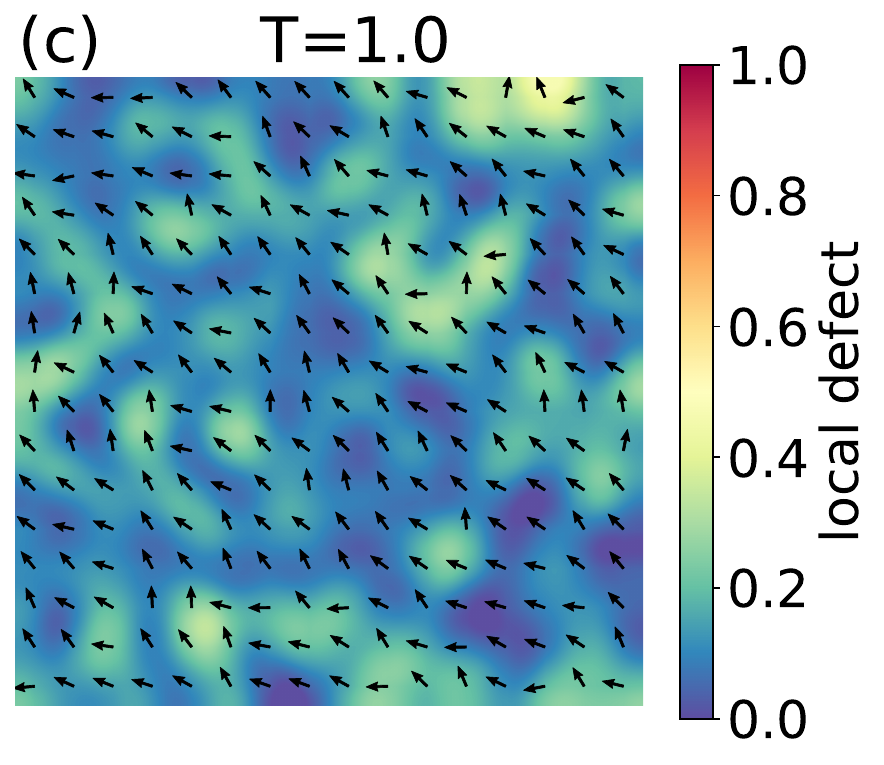}
\includegraphics[width=0.485\linewidth]{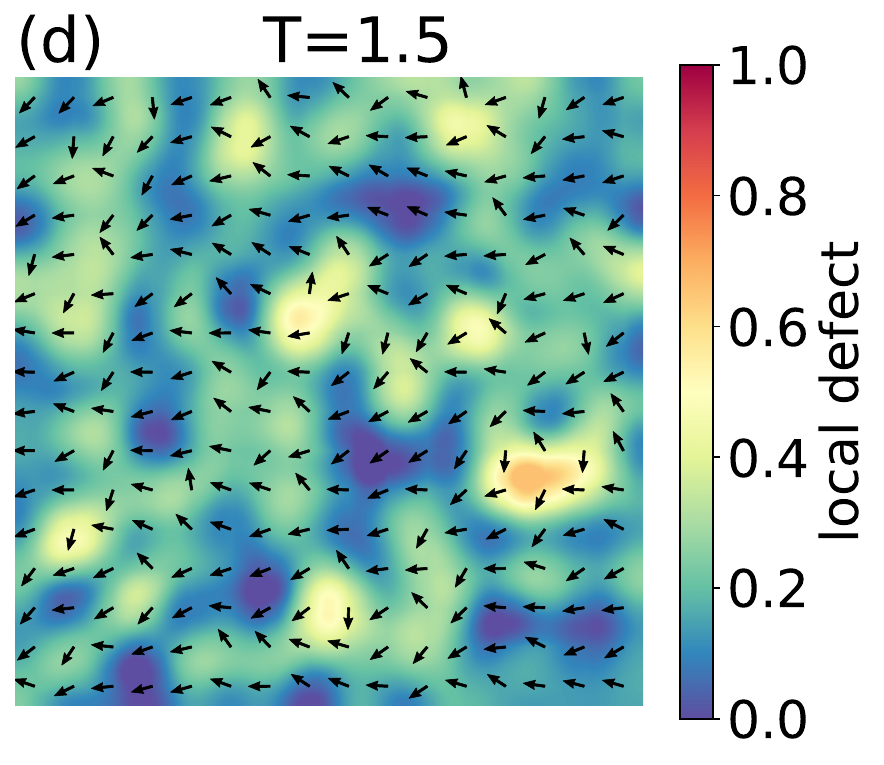}
\caption{The four figures show the equilibrium spin configurations of the 2D XY model at $\sigma=3.3$ for $T=0.01$, $0.5$, $1.0$, and $1.5$, with the background color again indicating the local defect $D$.}
\label{fig:fig6_2}
\end{figure}

\red{In the phase diagram of the 2D long-range XY model (Fig.~\ref{fig:fig1}(c)), we see that for $3.2<\sigma<3.5$, increasing temperature drives the system from ReLRO to EnLRO. This behavior is similar to that seen in the 1D model (Fig.~\ref{fig:fig1}(b)). To test whether the same mechanism applies, we define a local defect on each plaquette as the average of its four edge defects, 
$D = \bigl(D_{i,i+1} + D_{i+1,i+1+L} + D_{i+1+L,i+L} + D_{i+L,i}\bigr)/4$, 
which differs from the winding number or vortex that usually used in nearest neibough interaction XY model. Figure~\ref{fig:fig6_2} shows equilibrium spin configurations at various temperatures, with the background color indicating $D$. At $T=0.01$ and $0.5$, spins are highly aligned. At $T=1.0$, weak defects form domain walls that split the system into irregular subregions. At $T=1.5$, domain walls become clearer and strong defects appear. These results suggest that in the 2D model, raising $T$ drives a ReLRO\,$\to$\,EnLRO transition by the same competition between spin waves and local defects.}

\section{Conclusion}
\label{sec:conculsion}

We observed that both the long-range XY and LR Heisenberg models exhibit true long-range order at low temperatures in the non-MWH regime \(D < \sigma < 2D\). In the XY model, the true long-range ordered phase is further differentiated into EnLRO and ReLRO regimes with distinct system size dependence in the magnetization. \blue{Specifically, the magnetization increases with the system size in the EnLRO, whereas the magnetization decreases with the system size in the ReLRO regime. Near the EnLRO-ReLRO crossover}, the magnetization first decreases and then increases with the system size. As an example, at a low temperature \(T = 0.01\), the EnLRO-ReLRO crossover occurs at $\sigma_c = 1.575$ and 3.2 for the 1D and 2D XY model, respectively. ~\\

\blue{Furthermore, we studied the differences in the spin-spin correlation functions' scaling behaviors in different regimes. We found that in ReLRO, the correlation functions exhibit algebraic decay, whereas it is not the case in the EnLRO regime. When fitting the correlation functions with an exponential decay function, the extracted correlation length \(\xi\) shows discontinuous changes near \(\sigma_c\).}~\\

Finite-size analysis was employed to estimate the ordered-to-disordered phase transition temperatures, providing complete phase diagrams of the two systems. \blue{The phase diagrams indicate that the transition temperature from the ordered phase to the disordered phase diverges when \(\sigma \approx D\). Additionally, it was found that \blue{for $\sigma\in [1.6,1.7]$ and $\sigma\in[3.2,3.5]$ in the 1D and 2D model respectively}, as the temperature increases, the system first enters the ReLRO then the EnLRO regimes, and then undergoes a phase transition to the disordered phase.}~\\


For the Heisenberg model, our results indicate that the EnLRO-ReLRO crossover in both 1D and 2D systems within the non-MWH region is almost the same as the XY model, with \(\sigma_{c}\) values of 1.575 and 3.22 respectively at $T=0.01$. We conjecture that the continuous rotational symmetry plays an important role in the EnLRO-ReLRO crossovers in the two models. Exploring the mechanism behind and establishing a complete phase diagram of the LR Heisenberg model will be of interest in future work.~\\

\red{We also hypothesize that the distinct scaling of EnLRO and ReLRO in the 1D XY model is governed by competition among spin alignment, spin waves and local defects. At low temperature, decreasing $\sigma$ suppresses spin‐wave excitation and propagation. In the high‐temperature regime, excited defects disturb spin‐wave propagation.}~\\

This study \blue{refines} the phase diagram of classical XY models in the non-Mermin-Wagner-Hohenberg region. 
\blue{Although it remains unclear whether the difference in the scaling behavior presented above is merely a crossover or if it is a phase transition for which the order parameter has yet to be identified, studying these differences in the scaling behavior and highlighting their various characteristics may provide insights for experimental studies. Given the current experimental capabilities, Rydberg atom systems can simulate LR models typically up to systems of hundreds of atoms \cite{xiao2024two,browaeys2020many,chen2023continuous}, which is still far from the thermodynamic limit. Significant finite-size effects are expected, but these effects can serve as valuable references when comparing experimental and simulation results.}

\begin{acknowledgments}
We thank Wenlong You for helpful feedback to improve the early version of the manuscript.
We acknowledge financial support from Research Grants Council of Hong Kong (Grant No. CityU 11318722), National Natural Science Foundation of China (Grant No. 12005179, 12204130), City University of Hong Kong (Grant No. 9610438, 7005610), Harbin Institute of Technology Shenzhen (Grant No. ZX20210478, X20220001), and Shenzhen Key Laboratory of Advanced Functional CarbonMaterials Research and Comprehensive Application (No. ZDSYS20220527171407017).
\end{acknowledgments}

\nocite{*}


\begin{thebibliography}{32}%
\makeatletter
\providecommand \@ifxundefined [1]{%
 \@ifx{#1\undefined}
}%
\providecommand \@ifnum [1]{%
 \ifnum #1\expandafter \@firstoftwo
 \else \expandafter \@secondoftwo
 \fi
}%
\providecommand \@ifx [1]{%
 \ifx #1\expandafter \@firstoftwo
 \else \expandafter \@secondoftwo
 \fi
}%
\providecommand \natexlab [1]{#1}%
\providecommand \enquote  [1]{``#1''}%
\providecommand \bibnamefont  [1]{#1}%
\providecommand \bibfnamefont [1]{#1}%
\providecommand \citenamefont [1]{#1}%
\providecommand \href@noop [0]{\@secondoftwo}%
\providecommand \href [0]{\begingroup \@sanitize@url \@href}%
\providecommand \@href[1]{\@@startlink{#1}\@@href}%
\providecommand \@@href[1]{\endgroup#1\@@endlink}%
\providecommand \@sanitize@url [0]{\catcode `\\12\catcode `\$12\catcode `\&12\catcode `\#12\catcode `\^12\catcode `\_12\catcode `\%12\relax}%
\providecommand \@@startlink[1]{}%
\providecommand \@@endlink[0]{}%
\providecommand \url  [0]{\begingroup\@sanitize@url \@url }%
\providecommand \@url [1]{\endgroup\@href {#1}{\urlprefix }}%
\providecommand \urlprefix  [0]{URL }%
\providecommand \Eprint [0]{\href }%
\providecommand \doibase [0]{https://doi.org/}%
\providecommand \selectlanguage [0]{\@gobble}%
\providecommand \bibinfo  [0]{\@secondoftwo}%
\providecommand \bibfield  [0]{\@secondoftwo}%
\providecommand \translation [1]{[#1]}%
\providecommand \BibitemOpen [0]{}%
\providecommand \bibitemStop [0]{}%
\providecommand \bibitemNoStop [0]{.\EOS\space}%
\providecommand \EOS [0]{\spacefactor3000\relax}%
\providecommand \BibitemShut  [1]{\csname bibitem#1\endcsname}%
\let\auto@bib@innerbib\@empty
\bibitem [{\citenamefont {Radhakrishnan}\ \emph {et~al.}(1998)\citenamefont {Radhakrishnan}, \citenamefont {Sarma},\ and\ \citenamefont {Zacharia}}]{radhakrishnan1998modeling}%
  \BibitemOpen
  \bibfield  {author} {\bibinfo {author} {\bibfnamefont {B.}~\bibnamefont {Radhakrishnan}}, \bibinfo {author} {\bibfnamefont {G.}~\bibnamefont {Sarma}},\ and\ \bibinfo {author} {\bibfnamefont {T.}~\bibnamefont {Zacharia}},\ }\bibfield  {title} {\bibinfo {title} {Modeling the kinetics and microstructural evolution during static recrystallization—Monte Carlo simulation of recrystallization},\ }\href@noop {} {\bibfield  {journal} {\bibinfo  {journal} {Acta Mater.}\ }\textbf {\bibinfo {volume} {46}},\ \bibinfo {pages} {4415} (\bibinfo {year} {1998})}\BibitemShut {NoStop}%
\bibitem [{\citenamefont {Adam}\ \emph {et~al.}(2018)\citenamefont {Adam}, \citenamefont {Z{\"o}llner},\ and\ \citenamefont {Field}}]{adam20183d}%
  \BibitemOpen
  \bibfield  {author} {\bibinfo {author} {\bibfnamefont {K.}~\bibnamefont {Adam}}, \bibinfo {author} {\bibfnamefont {D.}~\bibnamefont {Z{\"o}llner}},\ and\ \bibinfo {author} {\bibfnamefont {D.~P.}\ \bibnamefont {Field}},\ }\bibfield  {title} {\bibinfo {title} {3d microstructural evolution of primary recrystallization and grain growth in cold rolled single-phase aluminum alloys},\ }\href@noop {} {\bibfield  {journal} {\bibinfo  {journal} {Model. Simul. Mater. Sci. Eng.}\ }\textbf {\bibinfo {volume} {26}},\ \bibinfo {pages} {035011} (\bibinfo {year} {2018})}\BibitemShut {NoStop}%
\bibitem [{\citenamefont {Graner}\ and\ \citenamefont {Glazier}(1992)}]{graner1992simulation}%
  \BibitemOpen
  \bibfield  {author} {\bibinfo {author} {\bibfnamefont {F.}~\bibnamefont {Graner}}\ and\ \bibinfo {author} {\bibfnamefont {J.~A.}\ \bibnamefont {Glazier}},\ }\bibfield  {title} {\bibinfo {title} {Simulation of biological cell sorting using a two-dimensional extended potts model},\ }\href@noop {} {\bibfield  {journal} {\bibinfo  {journal} {Phys. Rev. Lett.}\ }\textbf {\bibinfo {volume} {69}},\ \bibinfo {pages} {2013} (\bibinfo {year} {1992})}\BibitemShut {NoStop}%
\bibitem [{\citenamefont {Allena}\ \emph {et~al.}(2016)\citenamefont {Allena}, \citenamefont {Scianna},\ and\ \citenamefont {Preziosi}}]{allena2016cellular}%
  \BibitemOpen
  \bibfield  {author} {\bibinfo {author} {\bibfnamefont {R.}~\bibnamefont {Allena}}, \bibinfo {author} {\bibfnamefont {M.}~\bibnamefont {Scianna}},\ and\ \bibinfo {author} {\bibfnamefont {L.}~\bibnamefont {Preziosi}},\ }\bibfield  {title} {\bibinfo {title} {A cellular potts model of single cell migration in presence of durotaxis},\ }\href@noop {} {\bibfield  {journal} {\bibinfo  {journal} {Math. Biosci.}\ }\textbf {\bibinfo {volume} {275}},\ \bibinfo {pages} {57} (\bibinfo {year} {2016})}\BibitemShut {NoStop}%
\bibitem [{\citenamefont {Rozikov}(2021)}]{rozikov2021gibbs}%
  \BibitemOpen
  \bibfield  {author} {\bibinfo {author} {\bibfnamefont {U.}~\bibnamefont {Rozikov}},\ }\bibfield  {title} {\bibinfo {title} {Gibbs measures of potts model on cayley trees: A survey and applications},\ }\href@noop {} {\bibfield  {journal} {\bibinfo  {journal} {Rev. Math. Phys.}\ }\textbf {\bibinfo {volume} {33}},\ \bibinfo {pages} {2130007} (\bibinfo {year} {2021})}\BibitemShut {NoStop}%
\bibitem [{\citenamefont {Ohta}\ and\ \citenamefont {Jasnow}(1979)}]{ohta1979xy}%
  \BibitemOpen
  \bibfield  {author} {\bibinfo {author} {\bibfnamefont {T.}~\bibnamefont {Ohta}}\ and\ \bibinfo {author} {\bibfnamefont {D.}~\bibnamefont {Jasnow}},\ }\bibfield  {title} {\bibinfo {title} {Xy model and the superfluid density in two dimensions},\ }\href@noop {} {\bibfield  {journal} {\bibinfo  {journal} {Phys. Rev. B.}\ }\textbf {\bibinfo {volume} {20}},\ \bibinfo {pages} {139} (\bibinfo {year} {1979})}\BibitemShut {NoStop}%
\bibitem [{\citenamefont {Rosenblatt}\ \emph {et~al.}(1980)\citenamefont {Rosenblatt}, \citenamefont {Meyer}, \citenamefont {Pindak},\ and\ \citenamefont {Clark}}]{rosenblatt1980temperature}%
  \BibitemOpen
  \bibfield  {author} {\bibinfo {author} {\bibfnamefont {C.}~\bibnamefont {Rosenblatt}}, \bibinfo {author} {\bibfnamefont {R.~B.}\ \bibnamefont {Meyer}}, \bibinfo {author} {\bibfnamefont {R.}~\bibnamefont {Pindak}},\ and\ \bibinfo {author} {\bibfnamefont {N.~A.}\ \bibnamefont {Clark}},\ }\bibfield  {title} {\bibinfo {title} {Temperature behavior of ferroelectric liquid-crystal thin films: A classical xy system},\ }\href@noop {} {\bibfield  {journal} {\bibinfo  {journal} {Phys. Rev. A.}\ }\textbf {\bibinfo {volume} {21}},\ \bibinfo {pages} {140} (\bibinfo {year} {1980})}\BibitemShut {NoStop}%
\bibitem [{\citenamefont {Gingras}\ and\ \citenamefont {Huse}(1996)}]{gingras1996topological}%
  \BibitemOpen
  \bibfield  {author} {\bibinfo {author} {\bibfnamefont {M.~J. P.}\ \bibnamefont {Gingras}}\ and\ \bibinfo {author} {\bibfnamefont {D.~A.}\ \bibnamefont {Huse}},\ }\bibfield  {title} {\bibinfo {title} {Topological defects in the random-field xy model and the pinned vortex lattice to vortex glass transition in type-ii superconductors},\ }\href@noop {} {\bibfield  {journal} {\bibinfo  {journal} {Phys. Rev. B.}\ }\textbf {\bibinfo {volume} {53}},\ \bibinfo {pages} {15193} (\bibinfo {year} {1996})}\BibitemShut {NoStop}%
\bibitem [{\citenamefont {Song}\ and\ \citenamefont {Zhang}(2022)}]{song2022phase}%
  \BibitemOpen
  \bibfield  {author} {\bibinfo {author} {\bibfnamefont {F.-F.}\ \bibnamefont {Song}}\ and\ \bibinfo {author} {\bibfnamefont {G.-M.}\ \bibnamefont {Zhang}},\ }\bibfield  {title} {\bibinfo {title} {Phase coherence of pairs of cooper pairs as quasi-long-range order of half-vortex pairs in a two-dimensional bilayer system},\ }\href@noop {} {\bibfield  {journal} {\bibinfo  {journal} {Phys. Rev. Lett.}\ }\textbf {\bibinfo {volume} {128}},\ \bibinfo {pages} {195301} (\bibinfo {year} {2022})}\BibitemShut {NoStop}%
\bibitem [{\citenamefont {Zhang}\ \emph {et~al.}(2022)\citenamefont {Zhang}, \citenamefont {Ding}, \citenamefont {Deng},\ and\ \citenamefont {Zhang}}]{zhang2022surface}%
  \BibitemOpen
  \bibfield  {author} {\bibinfo {author} {\bibfnamefont {L.-R.}\ \bibnamefont {Zhang}}, \bibinfo {author} {\bibfnamefont {C.}~\bibnamefont {Ding}}, \bibinfo {author} {\bibfnamefont {Y.}~\bibnamefont {Deng}},\ and\ \bibinfo {author} {\bibfnamefont {L.}~\bibnamefont {Zhang}},\ }\bibfield  {title} {\bibinfo {title} {Surface criticality of the antiferromagnetic potts model},\ }\href@noop {} {\bibfield  {journal} {\bibinfo  {journal} {Phys. Rev. B.}\ }\textbf {\bibinfo {volume} {105}},\ \bibinfo {pages} {224415} (\bibinfo {year} {2022})}\BibitemShut {NoStop}%
\bibitem [{\citenamefont {Christiansen}\ \emph {et~al.}(2020)\citenamefont {Christiansen}, \citenamefont {Majumder}, \citenamefont {Henkel},\ and\ \citenamefont {Janke}}]{christiansen2020aging}%
  \BibitemOpen
  \bibfield  {author} {\bibinfo {author} {\bibfnamefont {H.}~\bibnamefont {Christiansen}}, \bibinfo {author} {\bibfnamefont {S.}~\bibnamefont {Majumder}}, \bibinfo {author} {\bibfnamefont {M.}~\bibnamefont {Henkel}},\ and\ \bibinfo {author} {\bibfnamefont {W.}~\bibnamefont {Janke}},\ }\bibfield  {title} {\bibinfo {title} {Aging in the long-range ising model},\ }\href@noop {} {\bibfield  {journal} {\bibinfo  {journal} {Phys. Rev. Lett.}\ }\textbf {\bibinfo {volume} {125}},\ \bibinfo {pages} {180601} (\bibinfo {year} {2020})}\BibitemShut {NoStop}%
\bibitem [{\citenamefont {Giachetti}\ \emph {et~al.}(2022)\citenamefont {Giachetti}, \citenamefont {Trombettoni}, \citenamefont {Ruffo},\ and\ \citenamefont {Defenu}}]{giachetti2022berezinskii}%
  \BibitemOpen
  \bibfield  {author} {\bibinfo {author} {\bibfnamefont {G.}~\bibnamefont {Giachetti}}, \bibinfo {author} {\bibfnamefont {A.}~\bibnamefont {Trombettoni}}, \bibinfo {author} {\bibfnamefont {S.}~\bibnamefont {Ruffo}},\ and\ \bibinfo {author} {\bibfnamefont {N.}~\bibnamefont {Defenu}},\ }\bibfield  {title} {\bibinfo {title} {Berezinskii-kosterlitz-thouless transitions in classical and quantum long-range systems},\ }\href@noop {} {\bibfield  {journal} {\bibinfo  {journal} {Phys. Rev. B.}\ }\textbf {\bibinfo {volume} {106}},\ \bibinfo {pages} {014106} (\bibinfo {year} {2022})}\BibitemShut {NoStop}%
\bibitem [{\citenamefont {Ma}(2018)}]{ma2018modern}%
  \BibitemOpen
  \bibfield  {author} {\bibinfo {author} {\bibfnamefont {S.-K.}~\bibnamefont {Ma}},\ }\bibfield  {title} {\bibinfo {title} {Modern theory of critical phenomena},\ }\bibfield {publisher} {\bibinfo {publisher} {Routledge},\ }\bibinfo {year} {2018}\BibitemShut {NoStop}%
\bibitem [{\citenamefont {Jenkins} \emph {et~al.}(2022)\citenamefont {Jenkins}, \citenamefont {R\'{o}zsa}, \citenamefont {Atxitia}, \citenamefont {Evans}, \citenamefont {Novoselov},\ and\ \citenamefont {Santos}}]{jenkins2022breaking}%
  \BibitemOpen
  \bibfield  {author} {\bibinfo {author} {\bibfnamefont {S.}~\bibnamefont {Jenkins}}, \bibinfo {author} {\bibfnamefont {L.}~\bibnamefont {R\'{o}zsa}}, \bibinfo {author} {\bibfnamefont {U.}~\bibnamefont {Atxitia}}, \bibinfo {author} {\bibfnamefont {R.~FL}~\bibnamefont {Evans}}, \bibinfo {author} {\bibfnamefont {K.~S.}~\bibnamefont {Novoselov}},\ and\ \bibinfo {author} {\bibfnamefont {E.~JG}~\bibnamefont {Santos}},\ }\bibfield  {title} {\bibinfo {title} {Breaking through the Mermin-Wagner limit in 2D van der Waals magnets},\ }\href@noop {} {\bibfield  {journal} {\bibinfo  {journal} {Nature Communications.}\ }\textbf {\bibinfo {volume} {13}},\ \bibinfo {pages} {6917} (\bibinfo {year} {2022})}\BibitemShut {NoStop}%
\bibitem [{\citenamefont {Fisher}, \citenamefont {Ma},\ and\ \citenamefont {Nickel}(1972)}]{fisher1972critical}%
  \BibitemOpen
  \bibfield  {author} {\bibinfo {author} {\bibfnamefont {M.~E.}~\bibnamefont {Fisher}}, \bibinfo {author} {\bibfnamefont {S.-K.}~\bibnamefont {Ma}}, \ and\ \bibinfo {author} {\bibfnamefont {B.~G.}~\bibnamefont {Nickel}},\ }\bibfield  {title} {\bibinfo {title} {Critical exponents for long-range interactions},\ }\href@noop {} {\bibfield  {journal} {\bibinfo  {journal} {Phys. Rev. Lett.}\ }\textbf {\bibinfo {volume} {29}},\ \bibinfo {pages} {917} (\bibinfo {year} {1972})}\BibitemShut {NoStop}%
\bibitem [{\citenamefont {Kunz and Pfister}(1976)}]{kunz1976first}%
  \BibitemOpen
  \bibfield  {author} {\bibinfo {author} {\bibfnamefont {H.}~\bibnamefont {Kunz}} \ and\ \bibinfo {author} {\bibfnamefont {C.-E.}~\bibnamefont {Pfister}},\ }\bibfield  {title} {\bibinfo {title} {First order phase transition in the plane rotator ferromagnetic model in two dimensions},\ }\href@noop {} {\bibinfo {journal} { }\textbf{\bibinfo {year}} {(1976)}}\BibitemShut {NoStop}%
\bibitem [{\citenamefont {Bruno}(2001)}]{bruno2001absence}%
  \BibitemOpen
  \bibfield  {author} {\bibinfo {author} {\bibfnamefont {P.}~\bibnamefont {Bruno}},\ }\bibfield  {title} {\bibinfo {title} {Absence of spontaneous magnetic order at nonzero temperature in one-and two-dimensional Heisenberg and XY systems with long-range interactions},\ }\href@noop {} {\bibfield  {journal} {\bibinfo  {journal} {Phys. Rev. Lett.},\ }\textbf {\bibinfo {volume} {87}},\ \bibinfo {pages} {137203} (\bibinfo {year} {2001})}\BibitemShut {NoStop}%
\bibitem [{\citenamefont {Browaeys}\ and\ \citenamefont {Lahaye}(2020)}]{browaeys2020many}%
  \BibitemOpen
  \bibfield  {author} {\bibinfo {author} {\bibfnamefont {A.}~\bibnamefont {Browaeys}}\ and\ \bibinfo {author} {\bibfnamefont {T.}~\bibnamefont {Lahaye}},\ }\bibfield  {title} {\bibinfo {title} {Many-body physics with individually controlled rydberg atoms},\ }\href@noop {} {\bibfield  {journal} {\bibinfo  {journal} {Nat. Phys.}\ }\textbf {\bibinfo {volume} {16}},\ \bibinfo {pages} {132} (\bibinfo {year} {2020})}\BibitemShut {NoStop}%
\bibitem [{\citenamefont {Chen}\ \emph {et~al.}(2023)\citenamefont {Chen}, \citenamefont {Bornet}, \citenamefont {Bintz}, \citenamefont {Emperauger}, \citenamefont {Leclerc}, \citenamefont {Liu}, \citenamefont {Scholl}, \citenamefont {Barredo}, \citenamefont {Hauschild}, \citenamefont {Chatterjee} \emph {et~al.}}]{chen2023continuous}%
  \BibitemOpen
  \bibfield  {author} {\bibinfo {author} {\bibfnamefont {C.}~\bibnamefont {Chen}}, \bibinfo {author} {\bibfnamefont {G.}~\bibnamefont {Bornet}}, \bibinfo {author} {\bibfnamefont {M.}~\bibnamefont {Bintz}}, \bibinfo {author} {\bibfnamefont {G.}~\bibnamefont {Emperauger}}, \bibinfo {author} {\bibfnamefont {L.}~\bibnamefont {Leclerc}}, \bibinfo {author} {\bibfnamefont {V.~S.}\ \bibnamefont {Liu}}, \bibinfo {author} {\bibfnamefont {P.}~\bibnamefont {Scholl}}, \bibinfo {author} {\bibfnamefont {D.}~\bibnamefont {Barredo}}, \bibinfo {author} {\bibfnamefont {J.}~\bibnamefont {Hauschild}}, \bibinfo {author} {\bibfnamefont {S.}~\bibnamefont {Chatterjee}}, \emph {et~al.},\ }\bibfield  {title} {\bibinfo {title} {Continuous symmetry breaking in a two-dimensional rydberg array},\ }\href@noop {} {\bibfield  {journal} {\bibinfo  {journal} {Nature}\ }\textbf {\bibinfo {volume} {616}},\ \bibinfo {pages} {691} (\bibinfo {year} {2023})}\BibitemShut {NoStop}%
\bibitem [{\citenamefont {Brede}\ \emph {et~al.}(2014)\citenamefont {Brede}, \citenamefont {Atodiresei}, \citenamefont {Caciuc}, \citenamefont {Bazarnik}, \citenamefont {Al-Zubi}, \citenamefont {Bl{\"u}gel},\ and\ \citenamefont {Wiesendanger}}]{brede2014long}%
  \BibitemOpen
  \bibfield  {author} {\bibinfo {author} {\bibfnamefont {J.}~\bibnamefont {Brede}}, \bibinfo {author} {\bibfnamefont {N.}~\bibnamefont {Atodiresei}}, \bibinfo {author} {\bibfnamefont {V.}~\bibnamefont {Caciuc}}, \bibinfo {author} {\bibfnamefont {M.}~\bibnamefont {Bazarnik}}, \bibinfo {author} {\bibfnamefont {A.}~\bibnamefont {Al-Zubi}}, \bibinfo {author} {\bibfnamefont {S.}~\bibnamefont {Bl{\"u}gel}},\ and\ \bibinfo {author} {\bibfnamefont {R.}~\bibnamefont {Wiesendanger}},\ }\bibfield  {title} {\bibinfo {title} {Long-range magnetic coupling between nanoscale organic--metal hybrids mediated by a nanoskyrmion lattice},\ }\href@noop {} {\bibfield  {journal} {\bibinfo  {journal} {Nat. Nanotechnol.}\ }\textbf {\bibinfo {volume} {9}},\ \bibinfo {pages} {1018} (\bibinfo {year} {2014})}\BibitemShut {NoStop}%
\bibitem [{\citenamefont {Ellis}\ and\ \citenamefont {Chantrell}(2015)}]{ellis2015switching}%
  \BibitemOpen
  \bibfield  {author} {\bibinfo {author} {\bibfnamefont {M.}~\bibnamefont {Ellis}}\ and\ \bibinfo {author} {\bibfnamefont {R.}~\bibnamefont {Chantrell}},\ }\bibfield  {title} {\bibinfo {title} {Switching times of nanoscale fept: Finite size effects on the linear reversal mechanism},\ }\href@noop {} {\bibfield  {journal} {\bibinfo  {journal} {Appl. Phys. Lett.}\ }\textbf {\bibinfo {volume} {106}} (\bibinfo {year} {2015})}\BibitemShut {NoStop}%
\bibitem [{\citenamefont {Zhao}\ \emph {et~al.}(2016)\citenamefont {Zhao}, \citenamefont {Yang}, \citenamefont {Liang}, \citenamefont {Wang}, \citenamefont {Ding}, \citenamefont {Lv}, \citenamefont {Zhang}, \citenamefont {Hu}, \citenamefont {Lu},\ and\ \citenamefont {Tang}}]{zhao2016high}%
  \BibitemOpen
  \bibfield  {author} {\bibinfo {author} {\bibfnamefont {M.}~\bibnamefont {Zhao}}, \bibinfo {author} {\bibfnamefont {F.}~\bibnamefont {Yang}}, \bibinfo {author} {\bibfnamefont {C.}~\bibnamefont {Liang}}, \bibinfo {author} {\bibfnamefont {D.}~\bibnamefont {Wang}}, \bibinfo {author} {\bibfnamefont {D.}~\bibnamefont {Ding}}, \bibinfo {author} {\bibfnamefont {J.}~\bibnamefont {Lv}}, \bibinfo {author} {\bibfnamefont {J.}~\bibnamefont {Zhang}}, \bibinfo {author} {\bibfnamefont {W.}~\bibnamefont {Hu}}, \bibinfo {author} {\bibfnamefont {C.}~\bibnamefont {Lu}},\ and\ \bibinfo {author} {\bibfnamefont {Z.}~\bibnamefont {Tang}},\ }\bibfield  {title} {\bibinfo {title} {High hole mobility in long-range ordered 2d lead sulfide nanocrystal monolayer films},\ }\href@noop {} {\bibfield  {journal} {\bibinfo  {journal} {Adv. Funct. Mater.}\ }\textbf {\bibinfo {volume} {26}},\ \bibinfo {pages} {5182} (\bibinfo {year} {2016})}\BibitemShut {NoStop}%
\bibitem [{\citenamefont {Samin}\ and\ \citenamefont {Cao}(2015)}]{samin2015monte}%
  \BibitemOpen
  \bibfield  {author} {\bibinfo {author} {\bibfnamefont {A.}~\bibnamefont {Samin}}\ and\ \bibinfo {author} {\bibfnamefont {L.}~\bibnamefont {Cao}},\ }\bibfield  {title} {\bibinfo {title} {Monte carlo study of radiation-induced demagnetization using the two-dimensional ising model},\ }\href@noop {} {\bibfield  {journal} {\bibinfo  {journal} {Nucl. Instrum. Methods Phys. Res. B: Beam Interact. Mater. At.}\ }\textbf {\bibinfo {volume} {360}},\ \bibinfo {pages} {111} (\bibinfo {year} {2015})}\BibitemShut {NoStop}%
\bibitem [{\citenamefont {Li}\ \emph {et~al.}(2023)\citenamefont {Li}, \citenamefont {Zhao}, \citenamefont {Liu}, \citenamefont {Liu}, \citenamefont {Ye}, \citenamefont {Wu}, \citenamefont {Li}, \citenamefont {Ma}, \citenamefont {Ju}, \citenamefont {Song} \emph {et~al.}}]{li2023coercivity}%
  \BibitemOpen
  \bibfield  {author} {\bibinfo {author} {\bibfnamefont {C.}~\bibnamefont {Li}}, \bibinfo {author} {\bibfnamefont {X.}~\bibnamefont {Zhao}}, \bibinfo {author} {\bibfnamefont {L.}~\bibnamefont {Liu}}, \bibinfo {author} {\bibfnamefont {W.}~\bibnamefont {Liu}}, \bibinfo {author} {\bibfnamefont {Z.}~\bibnamefont {Ye}}, \bibinfo {author} {\bibfnamefont {J.}~\bibnamefont {Wu}}, \bibinfo {author} {\bibfnamefont {Y.}~\bibnamefont {Li}}, \bibinfo {author} {\bibfnamefont {J.}~\bibnamefont {Ma}}, \bibinfo {author} {\bibfnamefont {H.}~\bibnamefont {Ju}}, \bibinfo {author} {\bibfnamefont {Y.}~\bibnamefont {Song}}, \emph {et~al.},\ }\bibfield  {title} {\bibinfo {title} {Coercivity mechanism and long-range coupling of anisotropic nd-dy-fe-co-b/fe composite thick film},\ }\href@noop {} {\bibfield  {journal} {\bibinfo  {journal} {J. Alloys Compd.}\ ,\ \bibinfo {pages} {170816}} (\bibinfo {year} {2023})}\BibitemShut {NoStop}%
\bibitem [{\citenamefont {Romano}(1997)}]{romano1997computer}%
  \BibitemOpen
  \bibfield  {author} {\bibinfo {author} {\bibfnamefont {S.}~\bibnamefont {Romano}},\ }\bibfield  {title} {\bibinfo {title} {Computer simulation evidence for berezhinski?--kosterlitz--thouless-like transitions in one dimension},\ }\href@noop {} {\bibfield  {journal} {\bibinfo  {journal} {Int. J. Mod. Phys. B.}\ }\textbf {\bibinfo {volume} {11}},\ \bibinfo {pages} {2043} (\bibinfo {year} {1997})}\BibitemShut {NoStop}%
\bibitem [{\citenamefont {M{\"u}ller}\ \emph {et~al.}(2023)\citenamefont {M{\"u}ller}, \citenamefont {Christiansen}, \citenamefont {Schnabel},\ and\ \citenamefont {Janke}}]{muller2023fast}%
  \BibitemOpen
  \bibfield  {author} {\bibinfo {author} {\bibfnamefont {F.}~\bibnamefont {M{\"u}ller}}, \bibinfo {author} {\bibfnamefont {H.}~\bibnamefont {Christiansen}}, \bibinfo {author} {\bibfnamefont {S.}~\bibnamefont {Schnabel}},\ and\ \bibinfo {author} {\bibfnamefont {W.}~\bibnamefont {Janke}},\ }\bibfield  {title} {\bibinfo {title} {Fast, hierarchical, and adaptive algorithm for metropolis monte carlo simulations of long-range interacting systems},\ }\href@noop {} {\bibfield  {journal} {\bibinfo  {journal} {Phys. Rev. X.}\ }\textbf {\bibinfo {volume} {13}},\ \bibinfo {pages} {031006} (\bibinfo {year} {2023})}\BibitemShut {NoStop}%
\bibitem [{\citenamefont {Fehske}\ \emph {et~al.}(2007)\citenamefont {Fehske}, \citenamefont {Schneider},\ and\ \citenamefont {Weisse}}]{fehske2007computational}%
  \BibitemOpen
  \bibfield  {author} {\bibinfo {author} {\bibfnamefont {H.}~\bibnamefont {Fehske}}, \bibinfo {author} {\bibfnamefont {R.}~\bibnamefont {Schneider}},\ and\ \bibinfo {author} {\bibfnamefont {A.}~\bibnamefont {Weisse}},\ }\href@noop {} {\emph {\bibinfo {title} {Computational many-particle physics}}},\ Vol.\ \bibinfo {volume} {739}\ (\bibinfo  {publisher} {Springer},\ \bibinfo {year} {2007})\BibitemShut {NoStop}%
\bibitem [{\citenamefont {Bayong}\ and\ \citenamefont {Diep}(1999)}]{bayong1999effect}%
  \BibitemOpen
  \bibfield  {author} {\bibinfo {author} {\bibfnamefont {E.}~\bibnamefont {Bayong}}\ and\ \bibinfo {author} {\bibfnamefont {H. T.}~\bibnamefont {Diep}},\ }\bibfield  {title} {\bibinfo {title} {Effect of long-range interactions on the critical behavior of the continuous ising model},\ }\href@noop {} {\bibfield  {journal} {\bibinfo  {journal} {Phys. Rev. B.}\ }\textbf {\bibinfo {volume} {59}},\ \bibinfo {pages} {11919} (\bibinfo {year} {1999})}\BibitemShut {NoStop}%
\bibitem [{\citenamefont {Bayong}\ \emph {et~al.}(1999)\citenamefont {Bayong}, \citenamefont {Diep},\ and\ \citenamefont {Dotsenko}}]{bayong1999potts}%
  \BibitemOpen
  \bibfield  {author} {\bibinfo {author} {\bibfnamefont {E.}~\bibnamefont {Bayong}}, \bibinfo {author} {\bibfnamefont {H. T.}~\bibnamefont {Diep}},\ and\ \bibinfo {author} {\bibfnamefont {V.}~\bibnamefont {Dotsenko}},\ }\bibfield  {title} {\bibinfo {title} {Potts model with long-range interactions in one dimension},\ }\href@noop {} {\bibfield  {journal} {\bibinfo  {journal} {Phys. Rev. Lett.}\ }\textbf {\bibinfo {volume} {83}},\ \bibinfo {pages} {14} (\bibinfo {year} {1999})}\BibitemShut {NoStop}%
\bibitem [{\citenamefont {Gonzalez~Lazo}\ \emph {et~al.}(2021)\citenamefont {Gonzalez~Lazo}, \citenamefont {Heyl}, \citenamefont {Dalmonte},\ and\ \citenamefont {Angelone}}]{gonzalez2021finite}%
  \BibitemOpen
  \bibfield  {author} {\bibinfo {author} {\bibfnamefont {E.}~\bibnamefont {Gonzalez~Lazo}}, \bibinfo {author} {\bibfnamefont {M.}~\bibnamefont {Heyl}}, \bibinfo {author} {\bibfnamefont {M.}~\bibnamefont {Dalmonte}},\ and\ \bibinfo {author} {\bibfnamefont {A.}~\bibnamefont {Angelone}},\ }\bibfield  {title} {\bibinfo {title} {Finite-temperature critical behavior of long-range quantum ising models},\ }\href@noop {} {\bibfield  {journal} {\bibinfo  {journal} {SciPost Phys.}\ }\textbf {\bibinfo {volume} {11}},\ \bibinfo {pages} {076} (\bibinfo {year} {2021})}\BibitemShut {NoStop}%
\bibitem [{\citenamefont {Kac}\ \emph {et~al.}(1963)\citenamefont {Kac}, \citenamefont {Uhlenbeck},\ and\ \citenamefont {Hemmer}}]{kac1963van}%
  \BibitemOpen
  \bibfield  {author} {\bibinfo {author} {\bibfnamefont {M.}~\bibnamefont {Kac}}, \bibinfo {author} {\bibfnamefont {G.}~\bibnamefont {Uhlenbeck}},\ and\ \bibinfo {author} {\bibfnamefont {P.}~\bibnamefont {Hemmer}},\ }\bibfield  {title} {\bibinfo {title} {On the van der waals theory of the vapor-liquid equilibrium. i. discussion of a one-dimensional model},\ }\href@noop {} {\bibfield  {journal} {\bibinfo  {journal} {Journal of Mathematical Physics}\ }\textbf {\bibinfo {volume} {4}},\ \bibinfo {pages} {216} (\bibinfo {year} {1963})}\BibitemShut {NoStop}%
\bibitem [{\citenamefont {Giachetti}\ \emph {et~al.}(2021)\citenamefont {Giachetti}, \citenamefont {Defenu}, \citenamefont {Ruffo},\ and\ \citenamefont {Trombettoni}}]{giachetti2021berezinskii}%
  \BibitemOpen
  \bibfield  {author} {\bibinfo {author} {\bibfnamefont {G.}~\bibnamefont {Giachetti}}, \bibinfo {author} {\bibfnamefont {N.}~\bibnamefont {Defenu}}, \bibinfo {author} {\bibfnamefont {S.}~\bibnamefont {Ruffo}},\ and\ \bibinfo {author} {\bibfnamefont {A.}~\bibnamefont {Trombettoni}},\ }\bibfield  {title} {\bibinfo {title} {Berezinskii-kosterlitz-thouless phase transitions with long-range couplings},\ }\href@noop {} {\bibfield  {journal} {\bibinfo  {journal} {Phys. Rev. Lett.}\ }\textbf {\bibinfo {volume} {127}},\ \bibinfo {pages} {156801} (\bibinfo {year} {2021})}\BibitemShut {NoStop}%
\bibitem [{\citenamefont {Xiao} \emph {et~al.}(2024)\citenamefont {Xiao}, \citenamefont {Yao}, \citenamefont {Zhang}, \citenamefont {Fan},\ and\ \citenamefont {Deng}}]{xiao2024two}%
  \BibitemOpen
  \bibfield  {author} {\bibinfo {author} {\bibfnamefont {T.}~\bibnamefont {Xiao}}, \bibinfo {author} {\bibfnamefont {D.}~\bibnamefont {Yao}}, \bibinfo {author} {\bibfnamefont {C.}~\bibnamefont {Zhang}}, \bibinfo {author} {\bibfnamefont {Z.}~\bibnamefont {Fan}},\ and\ \bibinfo {author} {\bibfnamefont {Y.}~\bibnamefont {Deng}},\ }\bibfield  {title} {\bibinfo {title} {Two-dimensional XY Ferromagnet Induced by Long-range Interaction},\ }\href@noop {} {\bibfield  {journal} {\bibinfo  {journal} {arXiv preprint arXiv:2404.08498}\ }\textbf {\bibinfo {year} {2024}}}\BibitemShut {NoStop}%
\bibitem [{\citenamefont {Allen}\ and\ \citenamefont {Rajantie}(2021)}]{allen2021kosterlitz}%
  \BibitemOpen
  \bibfield  {author} {\bibinfo {author} {\bibfnamefont {M.}~\bibnamefont {Allen}}\ and\ \bibinfo {author} {\bibfnamefont {A.}~\bibnamefont {Rajantie}},\ }\bibfield  {title} {\bibinfo {title} {The Kosterlitz-Thouless Phase Transition in Spin Models and Quantum Field Theory},\ }(\bibinfo {year} {2021})\BibitemShut {NoStop}%
\bibitem [{\citenamefont {Scalettar}(1991)}]{scalettar1991critical}%
  \BibitemOpen
  \bibfield  {author} {\bibinfo {author} {\bibfnamefont {R.}~\bibnamefont {Scalettar}},\ }\bibfield  {title} {\bibinfo {title} {Critical properties of an ising model with dilute long range interactions},\ }\href@noop {} {\bibfield  {journal} {\bibinfo  {journal} {Phys. A: Stat. Mech. Appl.}\ }\textbf {\bibinfo {volume} {170}},\ \bibinfo {pages} {282} (\bibinfo {year} {1991})}\BibitemShut {NoStop}%
\bibitem [{\citenamefont {Nakada}\ \emph {et~al.}(2011)\citenamefont {Nakada}, \citenamefont {Rikvold}, \citenamefont {Mori}, \citenamefont {Nishino},\ and\ \citenamefont {Miyashita}}]{nakada2011crossover}%
  \BibitemOpen
  \bibfield  {author} {\bibinfo {author} {\bibfnamefont {T.}~\bibnamefont {Nakada}}, \bibinfo {author} {\bibfnamefont {P.~A.}\ \bibnamefont {Rikvold}}, \bibinfo {author} {\bibfnamefont {T.}~\bibnamefont {Mori}}, \bibinfo {author} {\bibfnamefont {M.}~\bibnamefont {Nishino}},\ and\ \bibinfo {author} {\bibfnamefont {S.}~\bibnamefont {Miyashita}},\ }\bibfield  {title} {\bibinfo {title} {Crossover between a short-range and a long-range ising model},\ }\href@noop {} {\bibfield  {journal} {\bibinfo  {journal} {Phys. Rev. B.}\ }\textbf {\bibinfo {volume} {84}},\ \bibinfo {pages} {054433} (\bibinfo {year} {2011})}\BibitemShut {NoStop}%
\bibitem [{\citenamefont {Shenker}\ and\ \citenamefont {Tobochnik}(1980)}]{shenker1980monte}%
  \BibitemOpen
  \bibfield  {author} {\bibinfo {author} {\bibfnamefont {S.~H.}\ \bibnamefont {Shenker}}\ and\ \bibinfo {author} {\bibfnamefont {J.}~\bibnamefont {Tobochnik}},\ }\bibfield  {title} {\bibinfo {title} {Monte carlo renormalization-group analysis of the classical heisenberg model in two dimensions},\ }\href@noop {} {\bibfield  {journal} {\bibinfo  {journal} {Phys. Rev. B.}\ }\textbf {\bibinfo {volume} {22}},\ \bibinfo {pages} {4462} (\bibinfo {year} {1980})}\BibitemShut {NoStop}%
\bibitem [{\citenamefont {Holm}\ and\ \citenamefont {Janke}(1993)}]{holm1993critical}%
  \BibitemOpen
  \bibfield  {author} {\bibinfo {author} {\bibfnamefont {C.}~\bibnamefont {Holm}}\ and\ \bibinfo {author} {\bibfnamefont {W.}~\bibnamefont {Janke}},\ }\bibfield  {title} {\bibinfo {title} {Critical exponents of the classical three-dimensional heisenberg model: A single-cluster monte carlo study},\ }\href@noop {} {\bibfield  {journal} {\bibinfo  {journal} {Phys. Rev. B.}\ }\textbf {\bibinfo {volume} {48}},\ \bibinfo {pages} {936} (\bibinfo {year} {1993})}\BibitemShut {NoStop}%
\end{thebibliography}

%

~\\

\clearpage
\appendix
\section{Detailed derivation of Eqs. (10) and Eqs. (11)}
\label{sec:Detailed_derivation}

Consider Eqs. (\ref{eq:Z}) and (\ref{eq:O}), the expectation value of \( M^{n} \) reads
\begin{equation}
\langle M^{n} \rangle = \frac{1}{Z} \sum_{i} M_{i}^{n} e^{-\beta N E_i}.
\end{equation}
The derivative of \(\langle M^{n} \rangle\) with respect to \(\beta\), which is used in the derivations of Eqs. (\ref{eq:dUn}) and (\ref{eq:dlnM^n}), is given by
\begin{equation}
\begin{aligned}
\frac{d \langle M^{n} \rangle}{d\beta} 
&= \frac{d}{d\beta} \left( \frac{1}{Z} \sum_{i} M_{i}^{n} e^{-\beta N E_{i}} \right ),  \\
&= \frac{d (1/Z)}{d\beta} \sum_{i} M_{i}^{n} e^{-\beta N E_{i}} \\ 
&\quad + \frac{1}{Z}
         \frac{d}{d\beta}
         \Bigl(\sum_{i} M_{i}^{n} e^{-\beta N E_{i}}\Bigr),\\
&= -\frac{1}{Z^{2}} \frac{dZ}{d\beta} \sum_{i} M_{i}^{n} e^{-\beta N E_{i}} \\
&\quad + \frac{1}{Z} \sum_{i} M_{i}^{n} (-N E_{i}) e^{-\beta N E_{i}}, \\
&= \frac{1}{Z^{2}} \sum_{i} N E_{i} e^{-\beta N E_{i}} \sum_{j} M_{j}^{n} e^{-\beta N E_{j}} \\
&\quad - \frac{1}{Z} \sum_{i} M_{i}^{n} N E_{i} e^{-\beta N E_{i}},  \\
&= N\langle E \rangle \langle M^{n} \rangle - N\langle M^{n} E \rangle.
\end{aligned}
\end{equation}
~\\

Taking the derivative of \(U_{n}\) in Eq. (\ref{eq:Un}) with respect to \(\beta\) and applying the chain rule, we obtain
\begin{equation}
\begin{aligned}
\frac{dU_n}{d\beta} &= -\frac{d}{d\beta} \left( \frac{\langle M^n \rangle}{3 \langle M^{n/2} \rangle^2} \right), \\
&= -\frac{1}{3} \left[ \frac{d\langle m^n \rangle}{d\beta} \frac{1}{\langle m^{n/2} \rangle^2} + \langle m^n \rangle \frac{d \langle m^{n/2} \rangle^{-2}}{d\beta} \right], \\
&= -\frac{1}{3}
   \left[\,\frac{d\langle m^n \rangle}{d\beta}\,
              \frac{1}{\langle m^{n/2}\rangle^2}
   \right. \\ 
&\quad \left. 
   -\,2\,\langle m^n \rangle\,
      \frac{1}{\langle m^{n/2}\rangle^3}\,
      \frac{d \langle m^{n/2}\rangle}{d\beta}
   \right], \\
&= -\frac{1}{3 \langle m^{n/2} \rangle^2} \left[ \frac{d\langle m^n \rangle}{d\beta} - 2 \langle m^n \rangle \frac{1}{\langle m^{n/2} \rangle} \frac{d \langle m^{n/2} \rangle}{d\beta} \right], \\
&= \frac{N}{3 \langle m^{n/2} \rangle^2} \left[ \langle m^n E \rangle - \langle m^n \rangle \langle E \rangle \right. \\
&\quad \left. - 2 \langle m^n \rangle \frac{\langle m^{n/2} E \rangle}{\langle m^{n/2} \rangle} + 2 \langle m^n \rangle \langle E \rangle \right], \\
&= \frac{N}{3 \langle m^{n/2} \rangle^2}
   \left[ \frac{\langle m^n \rangle \langle m^n E \rangle}{\langle m^n \rangle} - 2 \langle m^n \rangle \frac{\langle m^{n/2} E \rangle}
    {\langle m^{n/2} \rangle} + \right. \\   
&\quad \left. \langle m^n \rangle \langle E \rangle
   \right], \\
&= \frac{N \langle m^n \rangle}{3 \langle m^{n/2} \rangle^2} \left[ \langle E \rangle - 2 \frac{\langle m^{n/2} E \rangle}{\langle m^{n/2} \rangle} + \frac{\langle m^n E \rangle}{\langle m^n \rangle} \right], \\
&= N(1 - U_n) \left[ \langle E \rangle - 2 \frac{\langle m^{n/2} E \rangle}{\langle m^{n/2} \rangle} + \frac{\langle m^n E \rangle}{\langle m^n \rangle} \right],
\end{aligned}
\end{equation}
which is Eq. (\ref{eq:dUn}) in the main text.

Next, we provide the derivation of Eq. (\ref{eq:dlnM^n}). Taking the derivative of the logarithm of \(\langle M^{n} \rangle\) with respect to \(\beta\), we have
\begin{equation}
\begin{aligned}
\frac{d \ln(\langle M^{n} \rangle)}{d\beta} 
&= \frac{1}{\langle M^{n} \rangle} \frac{d \langle M^{n} \rangle}{d\beta} \\
&= \frac{1}{\langle M^{n} \rangle} \left( N\langle E \rangle \langle M^{n} \rangle - N\langle M^{n} E \rangle \right ) \\
&= N \left( \langle E \rangle - \frac{ \langle M^{n} E \rangle }{ \langle M^{n} \rangle } \right ).
\end{aligned}
\end{equation}

\section{Correlation function of 1D XY model at different system size}
\label{sec:Correlation_function_diffL}
\red{In the main text, we conjecture that in the region $\sigma<2D$ the system has both EnLRO and ReLRO. This conjecture assumes that true long‐range order exists when $\sigma<2D$. To support this, we here provide indirect evidence through the finite‐size \green{scaling} analysis of the correlation functions. Figure ~\ref{fig:fig3_2}(a,c) (upper panel) and Fig.~\ref{fig:appendixC} (upper panel) shows that for $\sigma=1.9$ at \green{$T=0.01$ and $T=0.1$ respectively}, the correlation function decays algebraically. We expect the correlation function converges to a finite constant in the \green{thermodynamic} limit, $G(r)=k_{1}+k_2\times r^{-\kappa}$ with $k_{1}>0$ in the thermodynamic limit. \green{To verify this,} one may fit $G(r)$ for different system sizes $L$ to the above power-law decaying form directly, and then track how the fitted parameter $k_{1}$ changes with $L$. If $k_{1}$ grows with $L$ to a clear finite value, one can deduce that $k_{1}>0$ in the thermodynamic limit. However, we found that such a fitting process to the Monte Carlo data is unstable. \green{The fit is sensitive to the exponent $\kappa$. There can be multiple $(k_{1},k_{2},\kappa)$'s for $k_1\in[0,1]$ that gives equally good fit, making the extracted $k_{1}$ unreliable.}}\\

\begin{figure} [t!]
\includegraphics[width=0.49\linewidth]{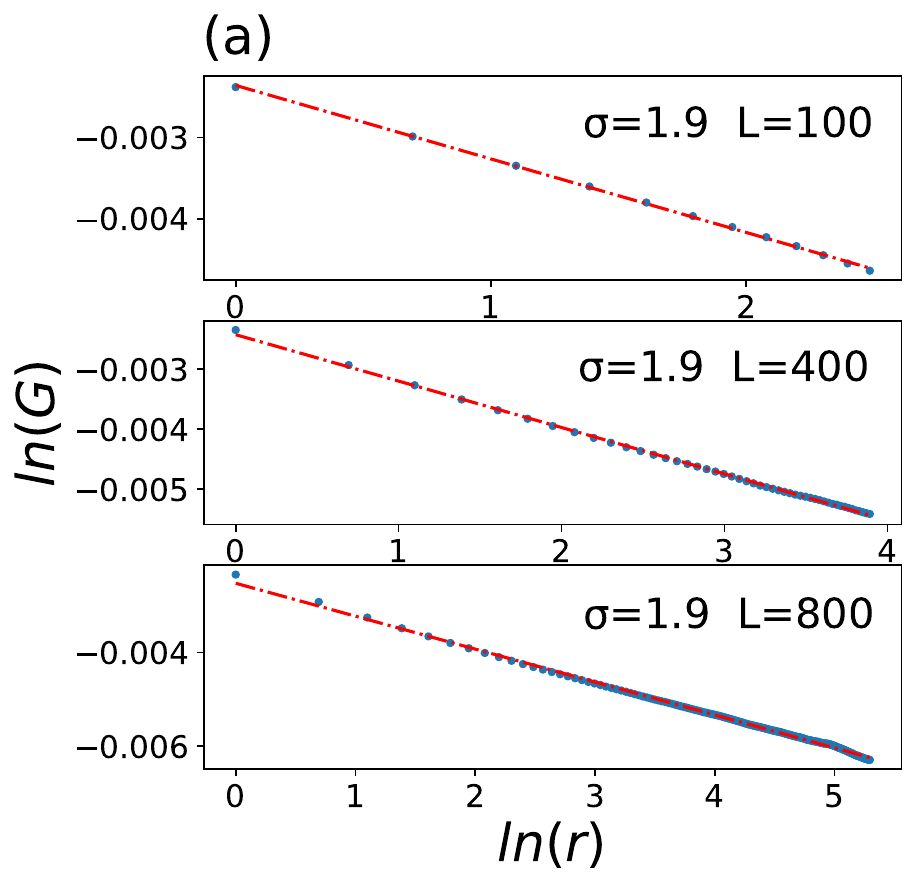}
\includegraphics[width=0.49\linewidth]{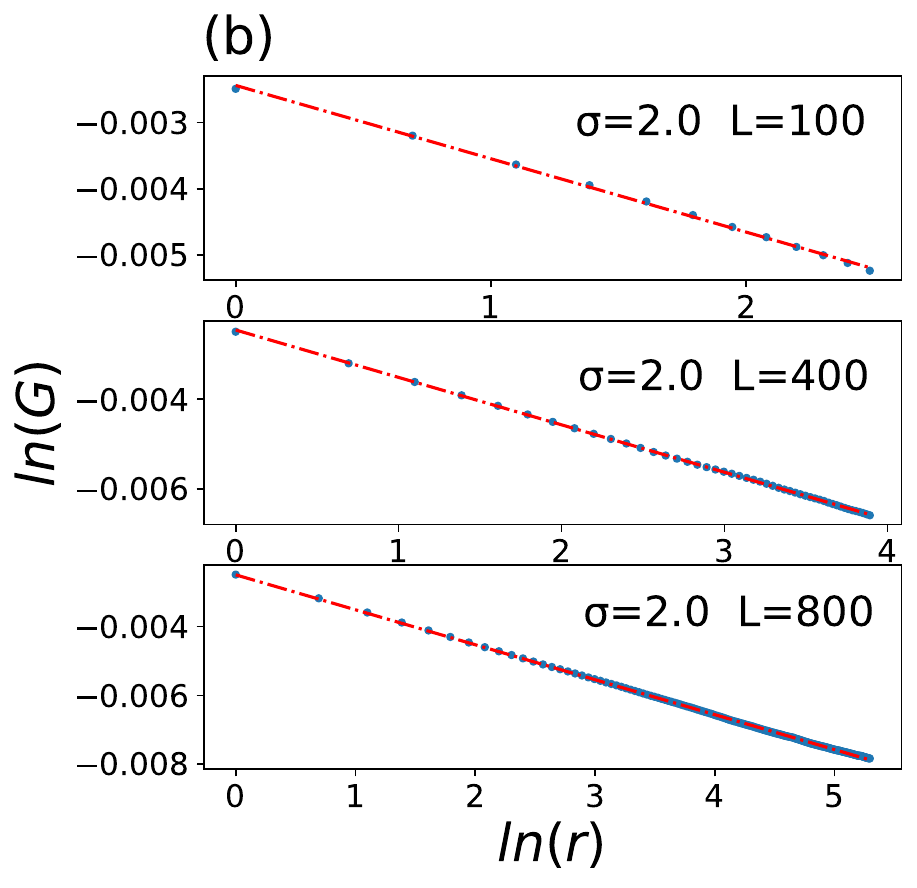}
\includegraphics[width=0.49\linewidth]{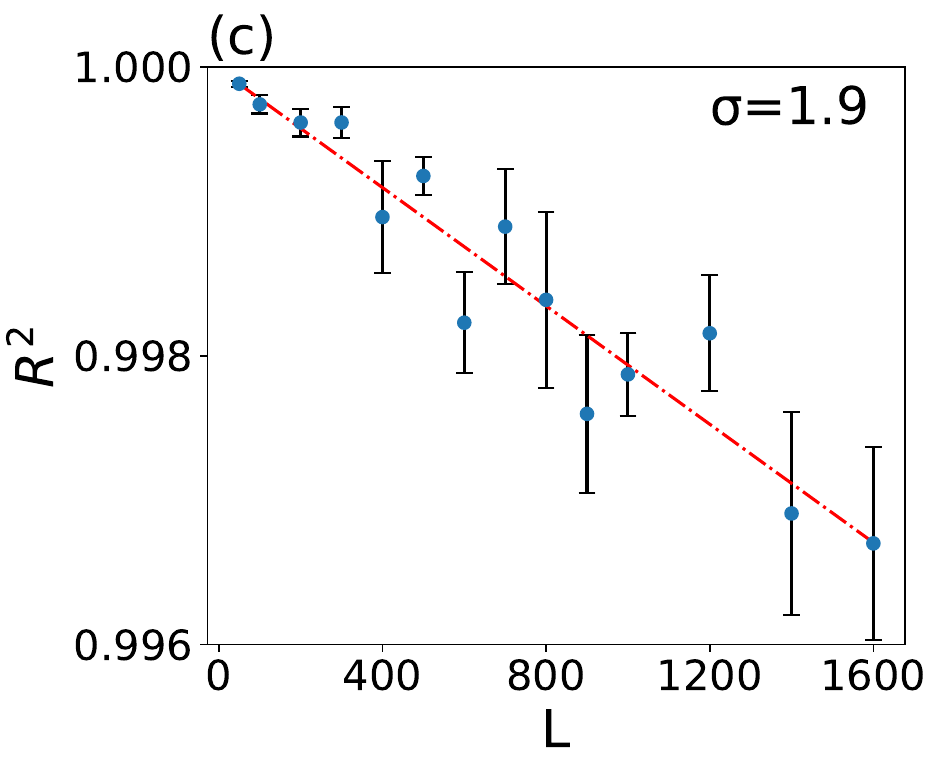}
\includegraphics[width=0.49\linewidth]{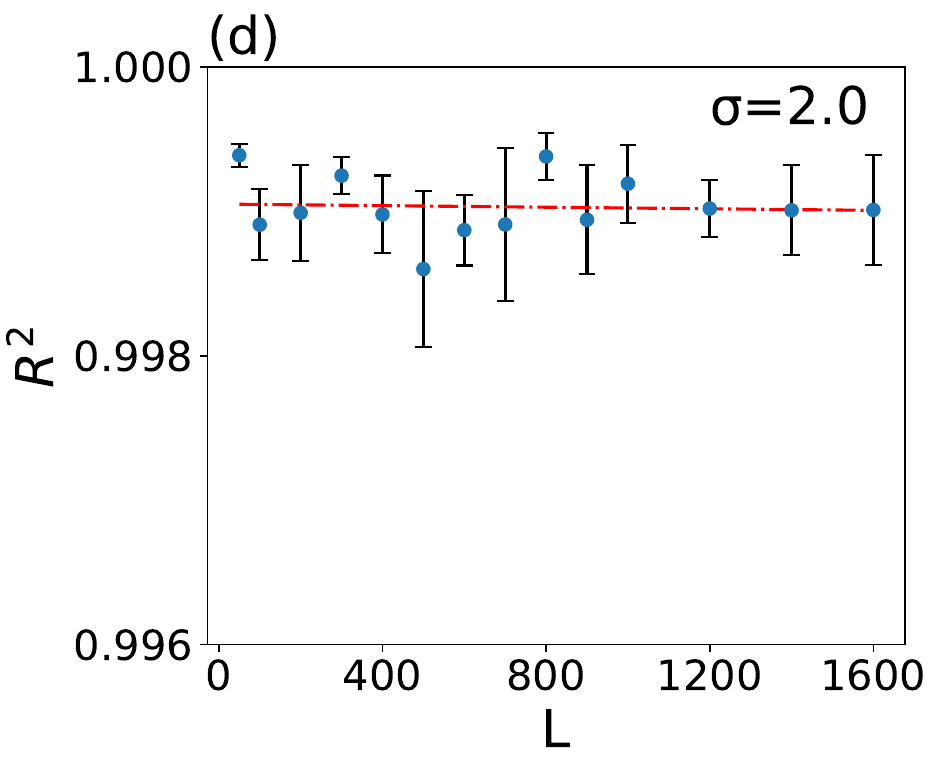}
\caption{Figures (a) and (b) show the linearity of \(\ln G\) versus \(\ln r\) for \(\sigma=1.9\) and \(\sigma=2.0\), respectively. In each panel, from top to bottom, the system size is \(L=100\), \(200\), and \(800\), the red dash–dot line denotes the linear fit. Figures (c) and (d) show the R-squared value of the linear fits as a function of system size for \(\sigma=1.9\) and \(\sigma=2.0\), respectively. The red dash–dot line denotes the linear fit.}
\label{fig:appendixB}
\end{figure}

\green{Alternatively, we can first assume $k_1=0$, i.e. $G(r)\approx k_{2}\times r^{-\kappa}$ and perform a linear fit to $\ln G$ versus $\ln r$. If the fitting performance $R^2$ is good, that suggests $k_1=0$. Vice versa, a poor fitting performance suggests a non-zero $k_1$. By tracking $R^2$ as a function of $L$, we would expect it to decrease if there exists true long-range order.}
\red{Figures~\ref{fig:appendixB}(a) and (b) show data and linear fits of $\ln G$ versus ~$\ln r$ for $\sigma=1.9<2D$ and $\sigma=2.0>2D$, respectively, at $L=\{100,400,800\}$. For $\sigma=2.0$, the data lies on a straight line for all $L$. However, for $\sigma=1.9$, the fit is good at $L=100$ but fails at larger $L$. 
Figures~\ref{fig:appendixB}(c) and (d) plot $R^{2}$ as a function of~$L$. For $\sigma=1.9$, $R^{2}$ clearly drops with $L$; for $\sigma=2.0$, $R^{2}$ stays near one. This suggests that $k_{1}>0$ for $\sigma=1.9$ and the system possesses true long‐range order exists, whereas for $\sigma=2.0$, $k_{1}=0$ and only quasi-long‐range order exists. The slightly lower $R^{2}$ values for $\sigma = 2.0$ as compared to $\sigma = 1.9$ at small $L$ is attributed to the broader range of the measured $\ln G$ values for $\sigma = 2.0$.}~\\

\section{Correlation function of 1D XY model at \(T = 0.1\)}
\label{sec:Correlation_function_highT}

\begin{figure} [t!]
\includegraphics[width=0.485\linewidth]{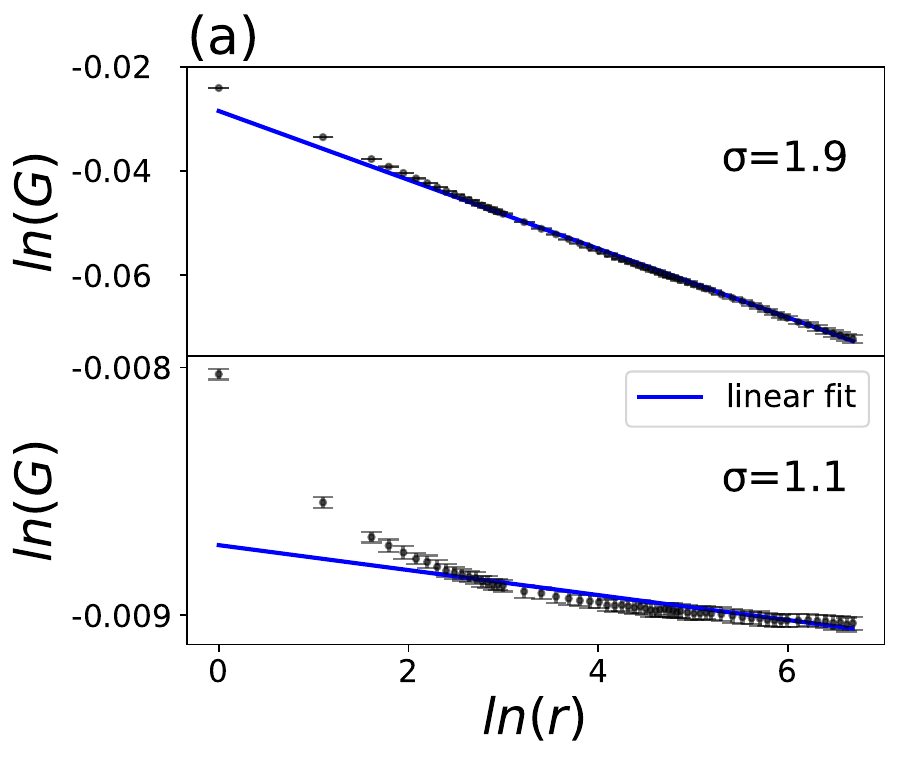}
\includegraphics[width=0.485\linewidth]{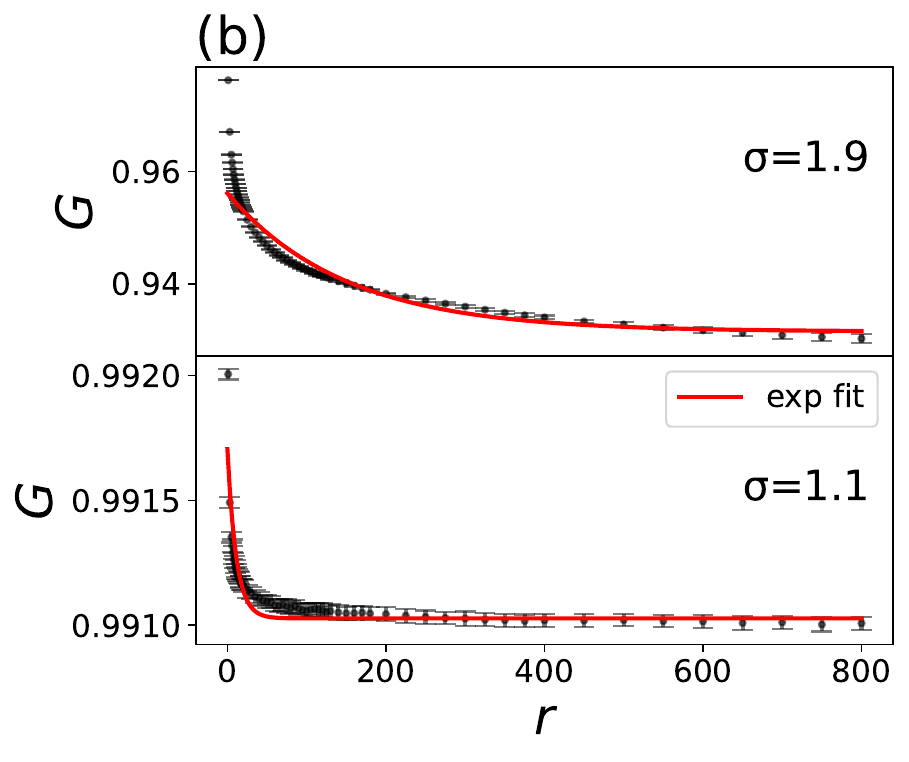}
\caption{The correlation functions of the 1D XY model on a ln-ln scale (left) and a linear scale (right) for $\sigma=1.9$ and $\sigma=1.1$. Here \(L = 3600\) and \(T = 0.1\). The blue straight lines in the left panel show the linear fits and the red curves in the right panel show the exponential fits to the correlation functions.} 
\label{fig:appendixC}
\end{figure}

\red{To ensure that our conclusions do not depend on the very low temperature (\(T = 0.01\)) used in the main text, we compute the correlation function of a one-dimensional XY model with size \(L = 3600\) at \(T = 0.1\).
Figure~\ref{fig:appendixC}(a) shows the correlation function \(G(r)\) versus distance \(r\) on a ln--ln scale for \(\sigma = 1.9\) (top panel) and \(\sigma = 1.1\) (bottom panel).
Figure~\ref{fig:appendixC}(b) presents the same data on a linear scale.
The main features of these plots are similar to those in Figure~\ref{fig:fig3_2}(a,c).
On the ln--ln scale, the data for \(\sigma = 1.9\) exhibit a linear trend, while the data for \(\sigma = 1.1\) do not.
On the linear scale, the data for \(\sigma = 1.1\) are well fitted by an exponential function, but the exponential fit fails for \(\sigma = 1.9\).
This outcome supports our conclusion in the main text that the correlation function decays differently in the EnLRO and ReLRO regions.}~\\

\red{We also note that, as temperature increases, thermal fluctuations grow and push the \(\sigma = 1.9\) data toward algebraic decay.
Consequently, the exponential fit in the lower panel of Figure~\ref{fig:appendixC}(b) is slightly worse than that in the lower panel of Figure~\ref{fig:fig3_2}(c).}

\section{Thermal perturbation in 1D Clock model at \(T = 0.01\)}
\label{sec:1D_clock}

\begin{figure} [t!]
\includegraphics[width=0.49\linewidth]{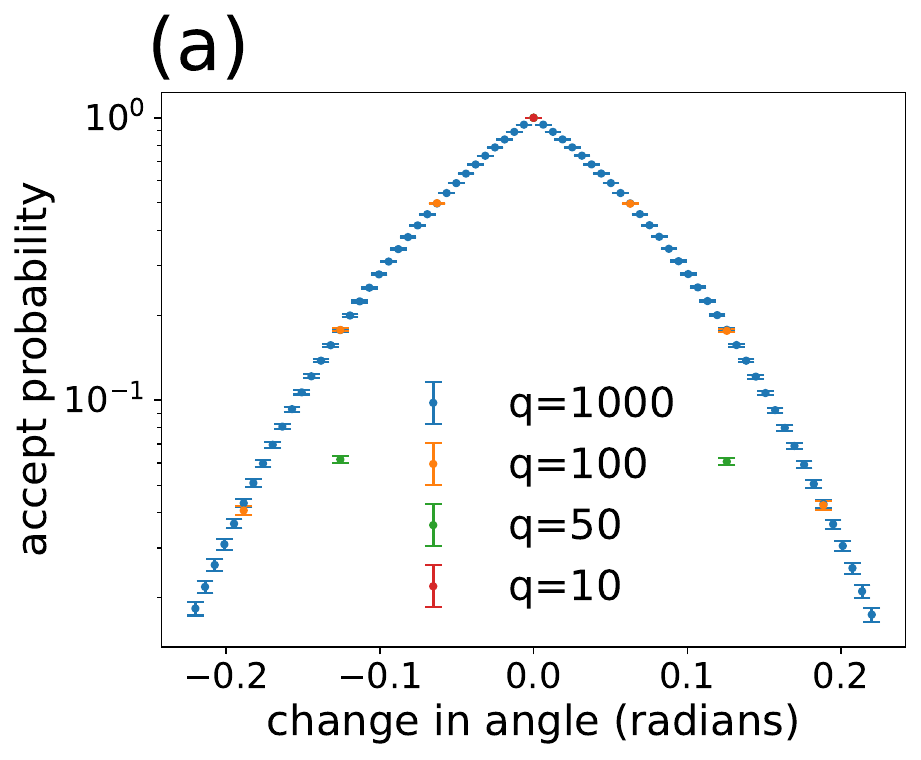}
\includegraphics[width=0.49\linewidth]{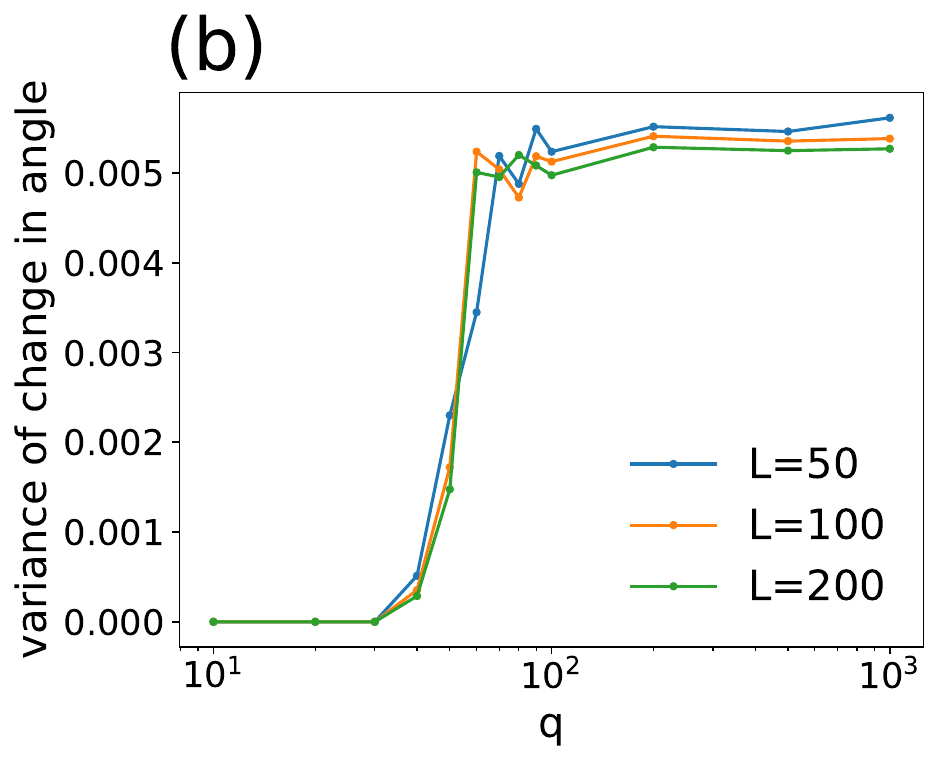}
\includegraphics[width=0.49\linewidth]{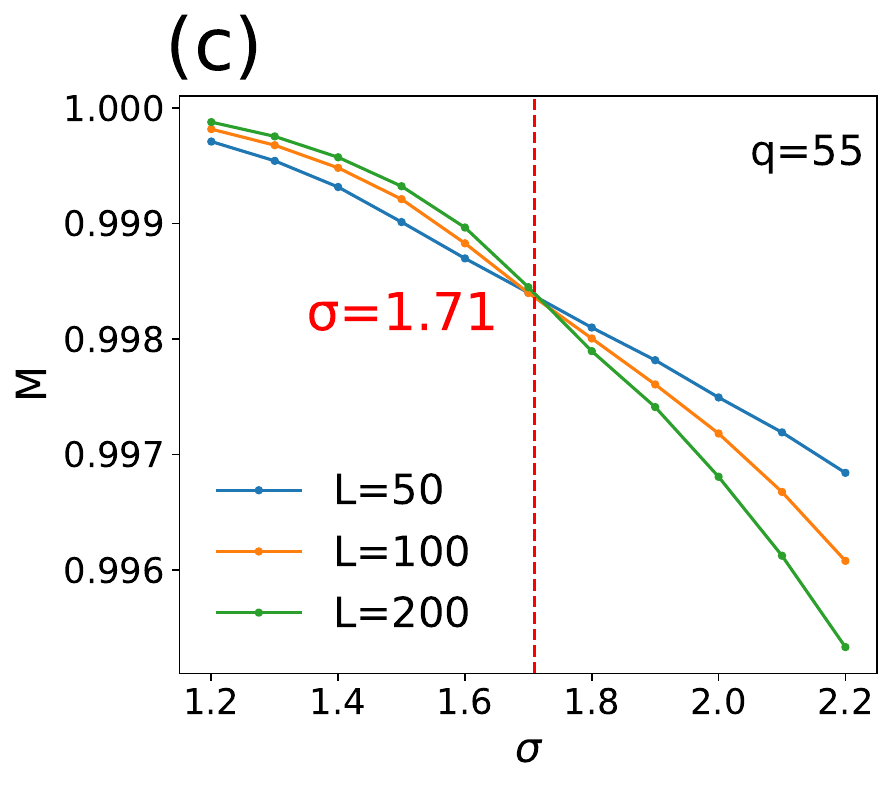}
\includegraphics[width=0.49\linewidth]{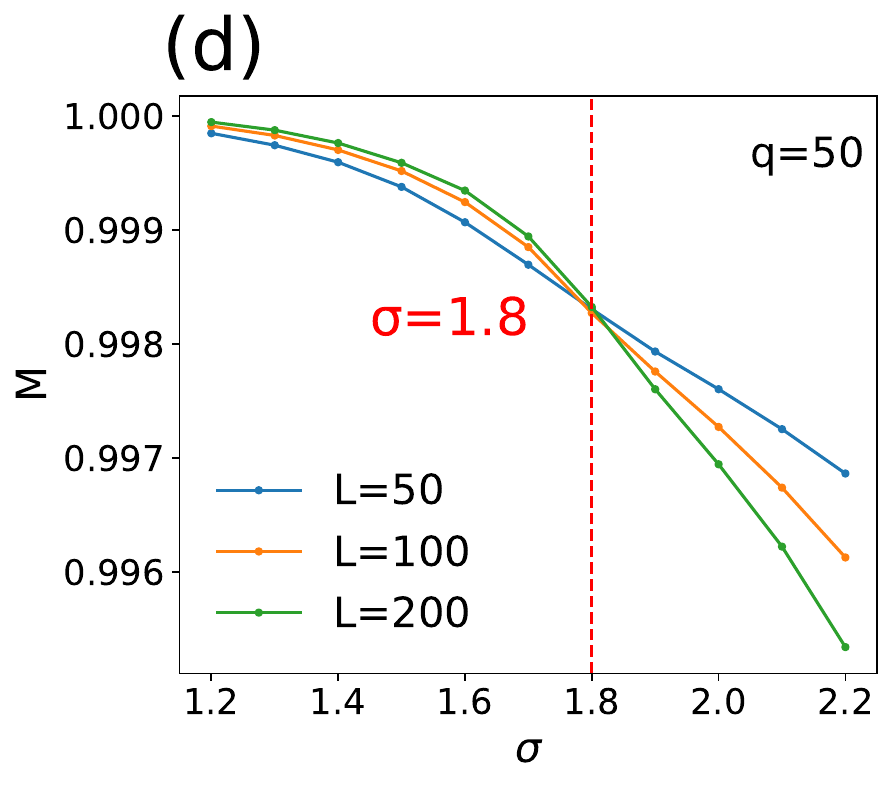}
\includegraphics[width=0.49\linewidth]{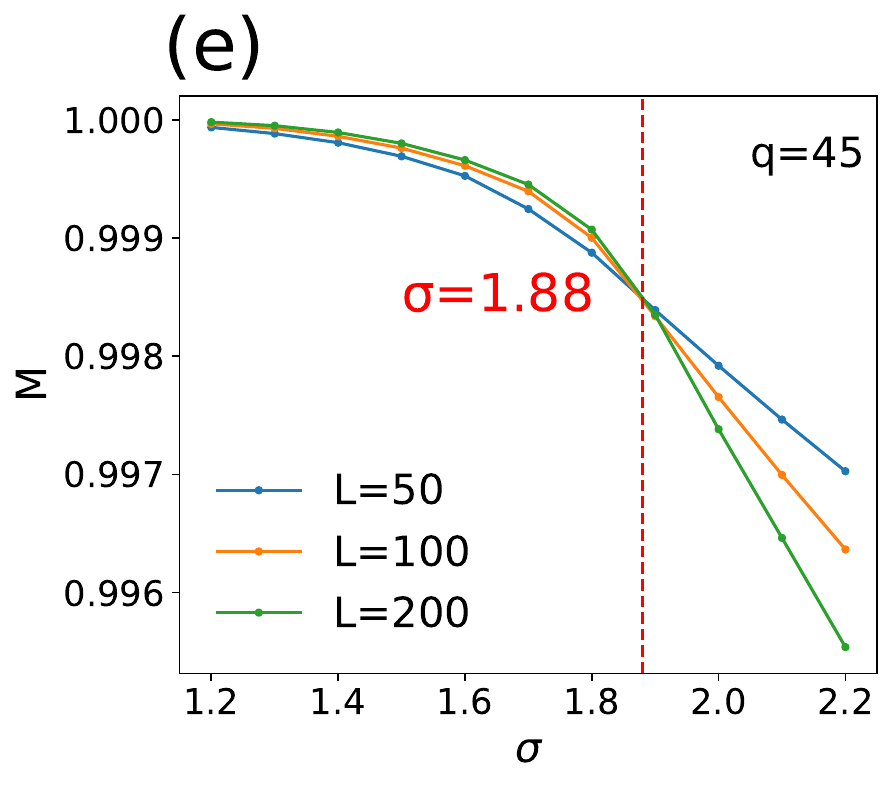}
\includegraphics[width=0.49\linewidth]{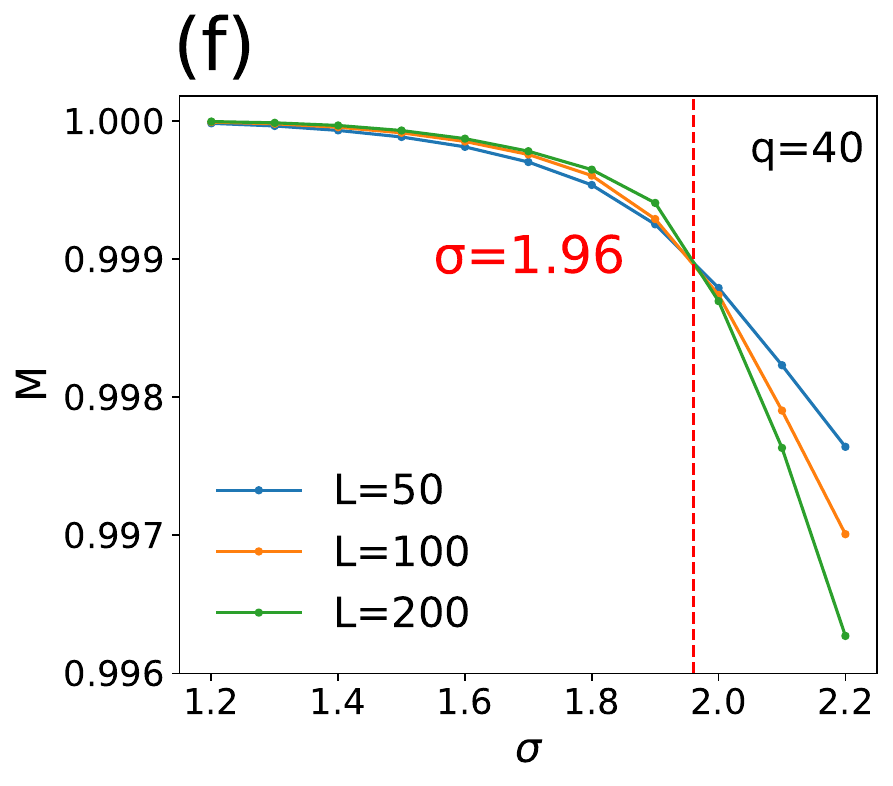}
\caption{Figure (a) shows the relation between flip angle and Monte Carlo acceptance rate for the long-range clock model with $L=100$, $\sigma=1.5$, and $T=0.01$ at different $q$. Figure (b) shows the variance of the flip angles changes with $q$ for various system sizes. Figures (c–f) show the average magnetization $M$ versus $\sigma$ at $T=0.01$ for $q=55$, $50$, $45$, and $40$, respectively. The red vertical lines mark the crossover point $\sigma_c$.}
\label{fig:appendixD}
\end{figure}

\red{In order to understand how spin alignment and thermal fluctuations affect the system and lead to EnLRO and ReLRO regions, we analyze the long-range interacting clock model. The clock model is similar to the classical XY model but with discrete spin angles. In this model, each spin can only take $q$ equally spaced angles $\theta_i=2\pi k/q$ for $k=0,1,\dots,q-1$. The Hamiltonian is the same as that of the XY model, but the rotation symmetry is reduced from continuous to discrete. When $q=2$, it reduces to the Ising model, while the phase transition behavior approaches that of the XY model as $q$ increases and tend to infinity.}~\\

\red{First, we study the effect of $q$ on spin flips. We set the system size to $L=100$, $\sigma=1.5$, temperature $T=0.01$, and choose $q\in\{10,50,100,1000\}$. For each $q$, we run 100 Monte Carlo simulations starting from high-temperature states. We measure the Monte Carlo acceptance rates for different flip angles and compute their mean and standard error after each Markov chain reaching equilibrium. Figure~\ref{fig:appendixD}(a) shows that when $q$ decreases from 1000 to 100, the acceptance rate changes little except for the discretization of flip angles. However, when $q=50$, the acceptance rate drops by about one order of magnitude, and when $q=10$, almost all flips are rejected.}~\\

\red{From these results, we propose two hypotheses. First, lowering $q$ reduces the effect of thermal fluctuations on the system. At the same temperature, large flips become less probable when $q$ is small, thermal fluctuations cannot produce large random flips for many spins. This suppresses exitation of spin waves and keeps most spins aligned. In another view point, in a long-range interacting system with small $q$, a flip with large angle costs more energy than at large $q$, reducing its probability. Second, thermal fluctuations on one spin do not transfer easily to its neighbors. In other words, the lifetime of spin waves is limited. If one lucky spin makes a large flip by chance, its neighbors still rarely flip, so they cannot respond to this disturbance.}~\\

\red{We also analyze how the variance of flip angles changes with $q$ for $L\in\{50,100,200\}$, other settings of the system remain the same as that in Figure~\ref{fig:appendixD}(a). Here the variance is computed from the normalized acceptance probabilities of each flip angle. Figure~\ref{fig:appendixD}(b) shows that for all three system sizes, the variance decreases to zero as $q$ decreases, meaning no flips occur.}~\\

\red{In summary, lowering $q$ reduces the response to thermal fluctuations by suppressing spin wave excitations and shortening their lifetime. In the main text, we argued that EnLRO and ReLRO in the 1D XY model arise from the competition between spin alignment and thermal fluctuations. Lowering $q$ reduces thermal fluctuations, allowing spin alignment to dominate and expanding the EnLRO region in $\sigma$. Figures~\ref{fig:appendixD}(c)-(f) show that for $q\in\{55,50,45,40\}$ with $L=50,100,200$, the average magnetization versus \ $\sigma$ indicates that the crossover point $\sigma_c$ increases toward 2 as $q$ decreases. This trend matches our hypothesis and supports that the competition between spin alignment and thermal fluctuations leads to EnLRO and ReLRO.}~\\

\end{document}